\documentclass[journal]{IEEEtran}
\usepackage{ifpdf}

\usepackage{cite}

\ifCLASSINFOpdf
  \usepackage[pdftex]{graphicx}
 \graphicspath{{figures/}}  
   \DeclareGraphicsExtensions{.pdf,.jpeg,.png}
\else
   \usepackage[dvips]{graphicx}
 \graphicspath{{figures/}}  
   \DeclareGraphicsExtensions{.eps}
\fi
\usepackage{amsmath}
\usepackage[cmintegrals]{newtxmath}
\usepackage{algorithmic}

\usepackage{array}

\ifCLASSOPTIONcompsoc
  \usepackage[caption=false,font=normalsize,labelfont=sf,textfont=sf]{subfig}
\else
  \usepackage[caption=false,font=footnotesize]{subfig}
\fi
\usepackage{fixltx2e}
 \usepackage{dblfloatfix}

\usepackage{overpic}
\usepackage{booktabs}

\DeclareMathOperator*{\argmin}{arg\,min}
\usepackage{comment}

\begin{document}
%
\title{A subspace pursuit method to infer refractivity in the marine atmospheric boundary layer}
%
%
%

\author{Marc~Aur\`ele~Gilles,~Christopher~Earls~and~David~Bindel
\thanks{This work was supported
by the Office for Naval Research through grant N00014-16-1-2077}
\thanks{M. A. Gilles is with the Center for Applied Mathematics, Cornell University, Ithaca, NY, 14850 USA e-mail: mtg79@cornell.edu}
\thanks{C. Earls is with the School of Civil and Environmental Engineering and the Center for Applied Mathematics, Cornell University, Ithaca, NY, 14850 USA}
\thanks{D. Bindel is with the Department of Computer Science and the Center for Applied Mathematics, Cornell University, Ithaca, NY, 14850 USA}
\thanks{Manuscript received September 27, 2018; revised January 11, 2019.}}

\maketitle

\begin{abstract}
Inferring electromagnetic propagation characteristics within the marine atmospheric boundary layer (MABL) from data in real time is crucial for modern maritime navigation and communications. The propagation of electromagnetic waves is well modeled by a partial differential equation (PDE): a Helmholtz equation. A natural way to solve the MABL characterization inverse problem is to minimize what is observed and what is predicted by the PDE. However, this optimization is difficult because it has many local minima. We propose an alternative solution that relies on the properties of the PDE but does not involve solving the full forward model. Ducted environments result in an EM field which can be decomposed into a few propagating, trapped modes. These modes are a subset of the solutions to a Sturm-Liouville eigenvalue problem. We design a new objective function that measures the distance from the observations to a subspace spanned by these eigenvectors. The resulting optimization problem is much easier than the one that arises in the standard approach, and we show how to solve the associated nonlinear eigenvalue problem efficiently, leading to a real-time method.
\end{abstract}


%
\IEEEpeerreviewmaketitle

\section{Introduction}
\begin{figure}[!t]
\centering
\includegraphics[width=2.5in]{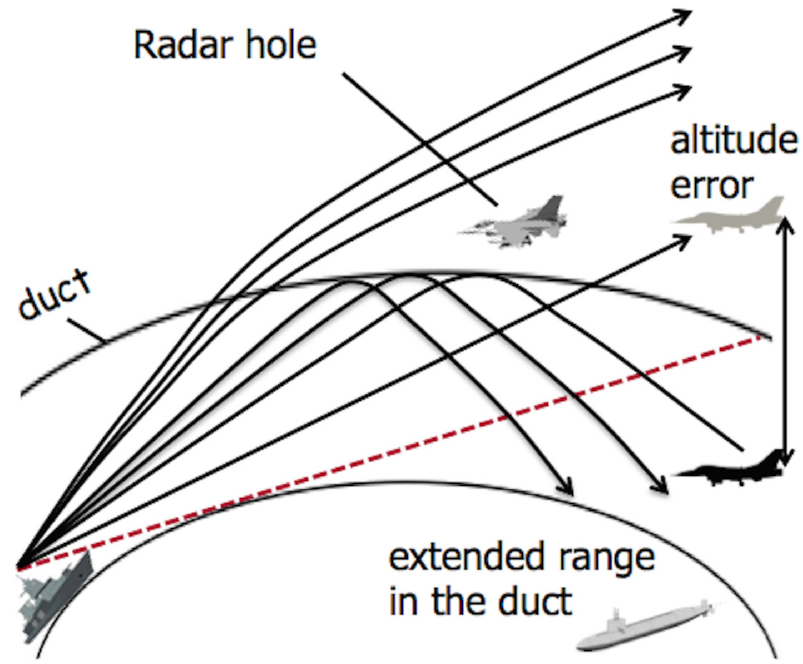}
  \caption{Effects of atmospheric ducting on EM waves \cite{vasilis}.}
  \label{fig:ducteffects}
\end{figure}

%
%
%
%
\IEEEPARstart{T}{he} 
marine atmospheric boundary layer (MABL) is the part of the lower troposphere in direct contact with the ocean. This contact creates a zone of particularly high inhomogeneity due to the exchanges of heat, moisture, and momentum between the atmosphere and the ocean \cite{sikora_ufermann_2004}. 
Within the lower MABL, the index of refraction - the speed of light in the medium relative to that in a vacuum - may change rapidly with height above the ocean surface; this causes ducting, i.e., bending of EM waves to the surface.
 Atmospheric ducting greatly changes the behavior of EM propagation within the MABL from what is expected in a ``standard" atmosphere \cite{sirkova_2012}. 
Ducting impacts radio communication, and it is also detrimental to maritime radars: it creates radar holes where no EM wave can travel, increases sea surface clutter, and changes the maximal operating range (illustrated in Fig. \ref{fig:ducteffects}).
It is therefore of great interest to be able to identify and characterize the presence of ducts in real time.

A variety of methods have been developed to characterize EM ducts. A few methods link refractive index profiles and weather conditions \cite{bean_dutton_1966} to estimate the MABL characteristics. Such estimates of meteorological conditions may come from numerical weather predictions \cite{lefurjah_marshall_casey_haack_boyer_2010,haack_wang_garrett_glazer_mailhot_marshall_2010}. Although these numerical weather predictions are successful for long-term, general characterization, their accuracy is unsatisfactory for characterizing ducting in a local sense and in real time. Another way to estimate meteorological conditions conducive to ducting is by using radiosondes or rocket sondes \cite{rowland1987fine} but those methods are costly, slow to deploy and very local. Yet another way to estimate ducting conditions is by lidar \cite{willitsford2005lidar} but this approach is sensitive to clouds, fog, and aerosols.

Other methods that rely on global positioning system (GPS)  satellites have been proposed which use information about the distortion of the GPS signal to estimate refractivity profiles \cite{lowry2002vertical, xie2006approach}. These methods rely on the GPS being placed over the horizon with respect to the receiver, which renders this method impractical for other contexts than targets of opportunity.

In the last decade, refractivity from clutter (RFC) methods have received a lot of attention in the literature.
RFC methods use a radar to estimate the refractivity profile by emitting radiation and measuring the backscattered signal from the rough ocean surface (also called clutter). RFC methods typically use either a forward model of EM propagation or a database to ``predict" the clutter under some EM condition, and compare the measurements to the prediction to infer the refractivity profiles. For a review of RFC methods, see \cite{review}.

The most popular forward model in RFC applications is the parabolic equation (PE) specialization of Maxwell's equations, briefly discussed in section \ref{Background}. A wealth of inverse solution methods for characterizing MABL refractivity have been developed that combine the accurate and relatively fast solvers inherited by the PE with a statistical or machine-learning method. For example, \cite{bayesian} uses a recursive Bayesian approach, \cite{kalman} uses Kalman filters, \cite{svm} uses support vector machine, and \cite{MCMC} uses Bayesian Monte Carlo analysis. An example of an RFC method that learns from a database is \cite{fountoulakis_earls_2016}. It uses the proper orthogonal bases of collected data to form an approximate forward model to be used in an inversion aimed at characterizing the EM duct itself.

In the current paper, we are interested in a different sampling approach: the bistatic case~\cite{gingras1997electromagnetic, gerstoft2000estimation, penton2018rough, pozderac2018x, wagner2016estimating, zhang2018study, zhao2012evaporation, zhao2011theoretical}. An example situation could involve two separate phased arrays, one transmitting and one receiving down range, with the receiver able to sample at different heights. Our method is similar to RFC methods in that it uses a forward model to predict clutter, as it also relies on a forward model of EM propagation. However, the proposed method does not involve actually solving the associated differential equation to predict the signal. Instead, we exploit a structure that is present in the partial differential equation which governs the physics: namely the approximately low-rank structure of the field within specific parts of the domain. The low-rank structure arises because only a few eigenvectors are needed to reconstruct the PDE when it is solved through separation of variables. This allows us to design an algorithm that seeks a refractivity profile associated with eigenvectors that best fit the data. The method presented in this paper is close in spirit to the idea presented in \cite{vasilis} and \cite{fountoulakis_earls_2016}. However, while they used a basis induced by the data, we use a basis induced by the forward model. 

The rest of the paper is organized as follows: in section \ref{Background} we state some background on the problem; in section \ref{motivation} we motivate and describe our algorithm; in section \ref{implementation} we give details needed for a fast implementation; and in section \ref{experiments} we present our numerical results.

\section{Background} \label{Background}
\subsection{Forward problem: propagation}
The physics that govern electromagnetic wave propagation are described by Maxwell's equations. Assuming a horizontal polarization and suppressing a time dependence of the form $\exp\left(-i\omega t\right)$, Maxwell's equation can be transformed into the Helmholtz equation (cf. (\ref{eq:Helmholtz}), along with useful boundary conditions (\ref{eq:leontovich}), (\ref{eq:source}), (\ref{eq:radiation})), by means of an exact earth flattening transformation \cite{vtrpe} :

\begin{equation} \label{eq:Helmholtz}
\frac{\partial^2 f \left(x,z \right)}{\partial x^2}
+ \frac{\partial^2 f\left(x,z\right) }{\partial z^2}+k_0^2n(x,z)^2f\left(x,z\right) = 0 
\end{equation}

\begin{equation} \label{eq:leontovich}
\frac{\partial f(x,z)}{\partial z} \bigg\rvert_{z = 0 }= -\left( \frac{1}{2a_e} + ik_0\sqrt{\epsilon _s -1}\right) f(x,0) 
\end{equation}
\begin{equation} \label{eq:source}
f(0,z)=F_0(z) 
\end{equation}

\begin{equation} \label{eq:radiation}
\begin{aligned}
\lim_{r \rightarrow \infty}  r \left( \frac{\partial }{\partial r }-i k_0 \right)f(x,z)    =  0  \\ 
\end{aligned}
\end{equation}

These equations are in 2D cartesian coordinates, where $x$ denotes the horizontal range, $z$ denotes the vertical altitude (the direction of invariance), $r=| (x,z )| $, $k_0 = 2\pi / \lambda$ is the free-space wavenumber, $\lambda$ is the wavelength, $\epsilon_s$ is the complex dielectric constant at the ocean free surface, $a_e$ is the radius of the earth, $n(x,z)$ is the index of refraction, and $f$ denotes the electric field in horizontal polarization. 
This Helmholtz equation is equipped with boundary conditions.
In the case of the MABL, this is achieved by imposing continuity of the tangential field components by modeling the sea surface as a locally homogeneous dielectric and specifying a surface boundary condition \cite{vtrpe}. 
This surface boundary condition is implemented via the Leontovich surface impedance condition, which for horizontal polarization is expressed as (\ref{eq:leontovich}).
Equation (\ref{eq:source}) is the boundary at $x=0$ and represents the source, i.e. the transmitter antenna. 
The domain is semi-infinite in both the $x$ and $z$ direction, for these boundaries radiating boundary conditions of the form of (\ref{eq:radiation}) are appropriate \cite{vtrpe}.

In the particular case of the MABL, the index of refraction is often approximated to be horizontally constant \cite{MCMC,svm}. This assumption seems to be approximately valid for open-sea for a small region (less than $100$~km) \cite{kerr_1951} but may not hold within coastal regions. In the argument that follows, we will make this assumption, and thus fix $n(x,z) := n(z)$. In section \ref{experiments}, we relax this assumption. That is, we allow for some horizontal change in the refractivity and attempt to characterize the mean refractive index, where the mean is taken over the downrange distance.

\subsection{Index of refraction}

\begin{figure}[!t]

  \centering
\begin{overpic}[trim={5cm 1.5cm 3cm 1cm}, width=3.5in]{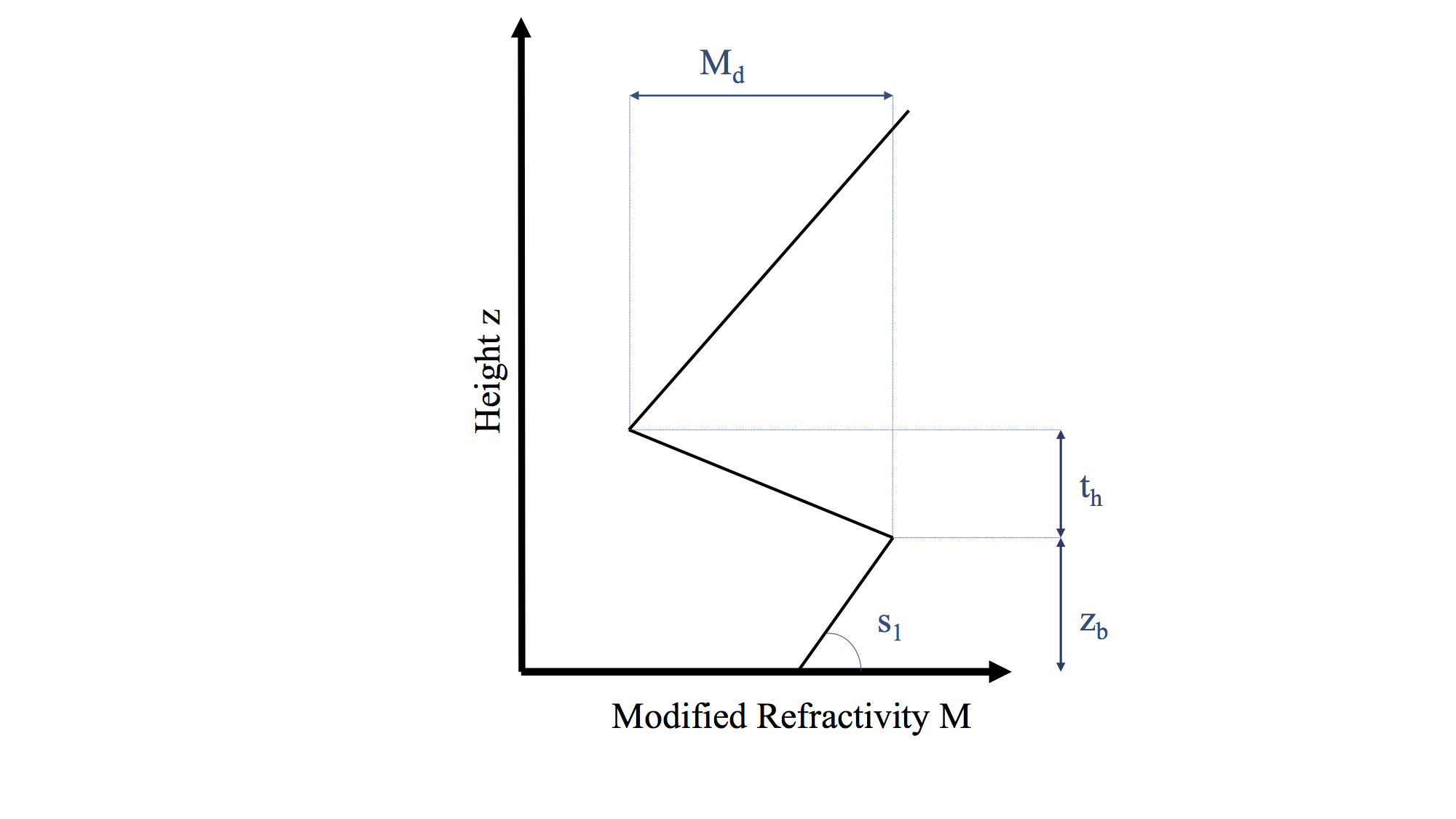}
\put(22,41) { \rotatebox{90}{\scriptsize (m)}}
\put(70,2) {\scriptsize (M-unit)}
\end{overpic}
 \caption{Example of a modified refractivity profile of a surface based duct. The modified refractivity profile is modeled as a tri-linear function represented by four coefficients: $z_b$ is the height of the base of the duct in meters, $t_h$ is the thickness of the duct in meters, ${M_d}$ is the M-deficit in M-units, and $s_1$ is the slope of the lowest linear portion in M-unit/meter.}
  \label{fig:Mprofile}
  
\end{figure}

The literature suggests that in the MABL the refraction can be well approximated by employing a modified refractivity $M(z)$, defined as

\begin{equation} \label{eq: define_M}
 n =  M \cdot 10^{-6} - z / a_e + 1
\end{equation}

where $a_e = 6370$~km is the radius of the earth. $ M(z)$ is modeled as a tri-linear function represented by four coefficients~\cite{gerstoft2003inversion}: $z_b$ is the height of the base of the duct in meters, $t_h$ is the thickness of the duct in meters, ${M_d}$ is the M-deficit in M-units, and $s_1$ is the slope of the lowest linear portion in M-unit/meter. This parameterization is typically used to represent surface-base ducts~\cite{gerstoft2000estimation} for which the height of the ducts is a few tens of meters. Figure \ref{fig:Mprofile} displays graphically the parametrization of $M$, and (\ref{eq:trilinear}) shows the analytical form. The slope $0.118$ M-unit/m of the upper part of the refractivity profile is consistent with the mean over the whole of the United States and is not a sensitive parameter in the inversion~\cite{gerstoft2000estimation}.
\begin{equation}\label{eq:trilinear}
 M(z) = \begin{cases} 
      M_0 + s_1z & z \leq z_b \\
      M(z_b) - \frac{ M_d }{t_h}(z-z_b) & z_b < z \leq z_b + t_h \\
      M(z_b + t_h) + 0.118 z & z_b +t_h < z 
   \end{cases}
\end{equation}

For the rest of the paper, we will refer to the parametrization of  $n$, through $M$  as $\gamma   = (s_1, z_b, t_h, M_d)$. We note that $M_0$, the modified refractivity at the mean free surface of the ocean, could be included in the parametrization, but propagation of EM waves have been found to be insensitive to this parameter \cite{bayesian}; therefore following the example of the authors in \cite{bayesian}, we fix it to a typical value of $M_0 = 340$ . We note that the method described in section \ref{motivation} does not rely on this parametrization, and any other parametrization could be used.

\subsection{SSFPE }

The most used method for solving the PDE in (\ref{eq:Helmholtz}) together with boundary conditions of the form of  (\ref{eq:leontovich}), (\ref{eq:source}), (\ref{eq:radiation}) is to use the split step Fourier transform for the parabolic equation method (SSFPE) \cite{vtrpe, dockery88, petool, sirkova_2012}. This method relies on a parabolic equation approximation of the Helmholtz equation. It is accurate, stable, relatively fast, and fairly easy to implement. Our surrogate field data is obtained by this method. In particular, our code is based on the software PETOOL, described in \cite{petool}.

\subsection{Modal solution}
For simple boundary conditions and geometries, the Helmholtz equation can be solved exactly by separation of variables \cite{COA}; that is,

\begin{equation*} 
f^{\gamma}(x,z) = \sum_{m=1}^{ \infty } \Phi_m^{\gamma}(x)  \Psi_m^{\gamma}(z) \ ,
\end{equation*}
where the eigenpair $(\Psi_m^{\gamma}(z), k^{\gamma}_{rm})$\footnote{We use $\gamma$ superscripts to emphasize quantities that depend on the refractivity profile parametrized by $\gamma$.} are solutions to a Sturm-Liouville (SL) eigenvalue problem:
 \begin{equation*} 
\frac{d^2 \Psi_m^{\gamma}(z)}{dz^2}+\left[ k_0^2n(z)^2-\left(k_{rm}^{\gamma}\right)^2 \right] \Psi_m^{\gamma}(z)=0
 \end{equation*} 
together with the associated boundary conditions. The functions $\Psi_m^{\gamma}(z)$ are called eigenfunctions or modes, and the scalars $(k_{m}^{\gamma})^2$ are the associated eigenvalues ($k_m^{\gamma} $ are also called associated wavenumbers). Throughout, we assume that $\Psi_m^{\gamma}(z)$ are normalized so that $\|\Psi_m^{\gamma}(z)\|_{L_2} = 1$.
 
For example, in the case where the source is modeled through a boundary condition at $x=0$ (as a point source at height $z_s$) and  the boundary condition is homogeneous Dirichlet at $z=0$ along with homogeneous Neumann at $z=D$, the electric field solution is \cite{COA}:
\begin{equation*}
f^{\gamma}(x,z) = \frac{i}{4}\sum_{m=1}^{ \infty } \Psi_m^{\gamma}(z_s) \Psi_m^{\gamma}(z)\exp \left( i k_m^{\gamma} z \right) \ .
\end{equation*}

When we consider an infinite domain and more complicated boundary conditions, such as (\ref{eq:leontovich}) and (\ref{eq:radiation}), 
separation of variables does not provide an exact solution. 
In this case, we use contour integration to obtain a solution involving a linear combination of modes from the discrete part of the spectrum and an integral term from the continuous spectrum.
In practice, the integral term can be neglected if we are sufficiently far from the source \cite{COA}. The modes can be further divided into two categories:
 \begin{enumerate}
 \item Leaky modes which are not observed in range.
 \item Trapped modes which propagate in range. 
 \end{enumerate}
As noted in \cite{COA} for most long-range propagation, only the trapped modes whose wavenumber is within a certain interval of interest are important. 
In our case, the interval of interest contains the admissible speeds of propagation of the modes. These admissible speeds of the propagating modes are bounded by the minimum and maximum speed induced by the refractivity in the domain. 
In the ducting case, we are concerned primarily with the energy emitted, propagated, and received in the MABL.
Therefore, by restricting our domain of dependence to the MABL, we get heuristics bounds on the set of relevant eigenvalues. Formally: we say that a solution $\Psi_m^{\gamma}$ of the SL eigenvalue problem is one of $K$ propagating modes if $\text{Im}(k_m^{\gamma}) = 0 $ and
\begin{equation*}
\text{Re}(k_m^{\gamma})\in \left[ \min_z \left\lbrace k_0n(z) \right\rbrace, \max_z \left\lbrace k_0 n(z) \right\rbrace \right] \ ,
\end{equation*}
where the maximum and minimum are taken over a domain of dependence: $0<z<z_{max}$. In our case, we define $z_{max}$ to be the maximal height at which we consider non-standard refractivity. In our numerical experiments in section \ref{experiments}, we take $z_{max} = 60$ m.

\begin{figure}[!t]
\centering

 \subfloat[mode 1 ]{
    \includegraphics[width=0.5in,keepaspectratio]{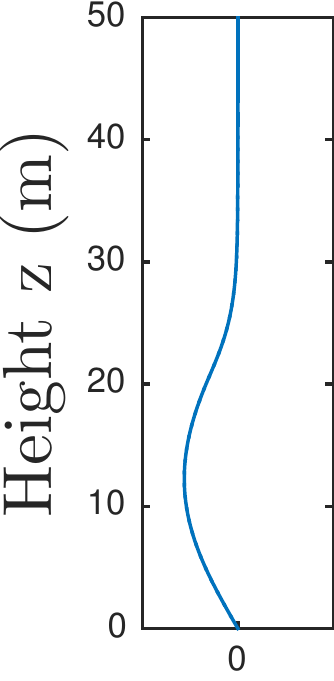}
    \label{fig:subfig1}}
    \hspace{0cm} 
    \subfloat[mode 2]{
    \includegraphics[width=0.5in,keepaspectratio]{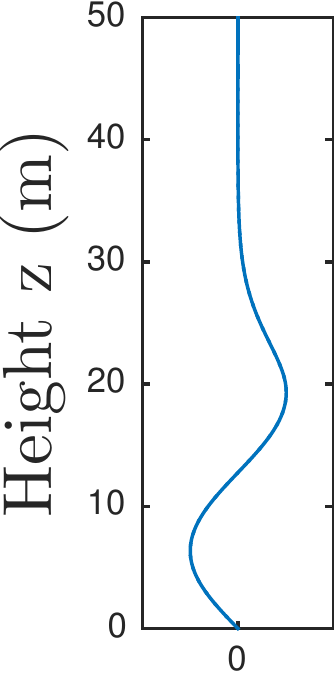}
        \label{fig:subfig2}}
    \hspace{0cm} 
 \subfloat[mode 3]{
    \includegraphics[width=0.5in,keepaspectratio]{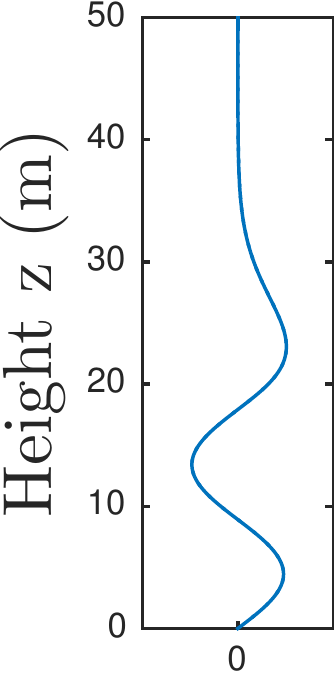}
        \label{fig:subfig3}}
    \hspace{0cm}
       \subfloat[mode 4]{
    \includegraphics[width=0.5in,keepaspectratio]{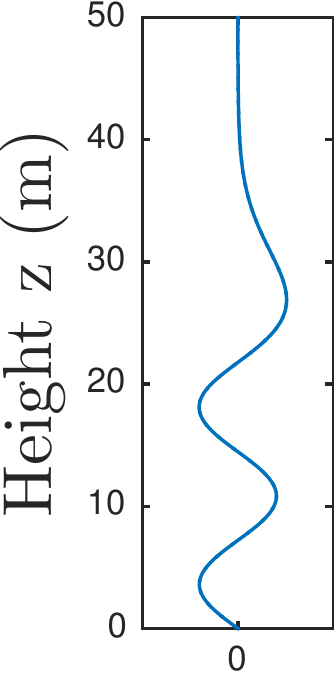}
        \label{fig:subfig4}}
    \hspace{0cm} 
     \subfloat[mode 5]{
    \includegraphics[width=0.5in,keepaspectratio]{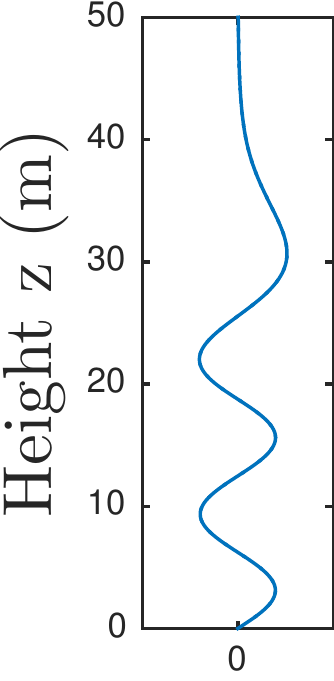}
        \label{fig:subfig5}}

  \caption{Plots of the first 5 propagating vertical modes induced by a particular refraction index parametrized by $\gamma = (0.118, 5,40,30)$.
    \label{fig:modes}}

\end{figure}

\begin{figure}[!h]
  \centering
\includegraphics[width = 2.5in]{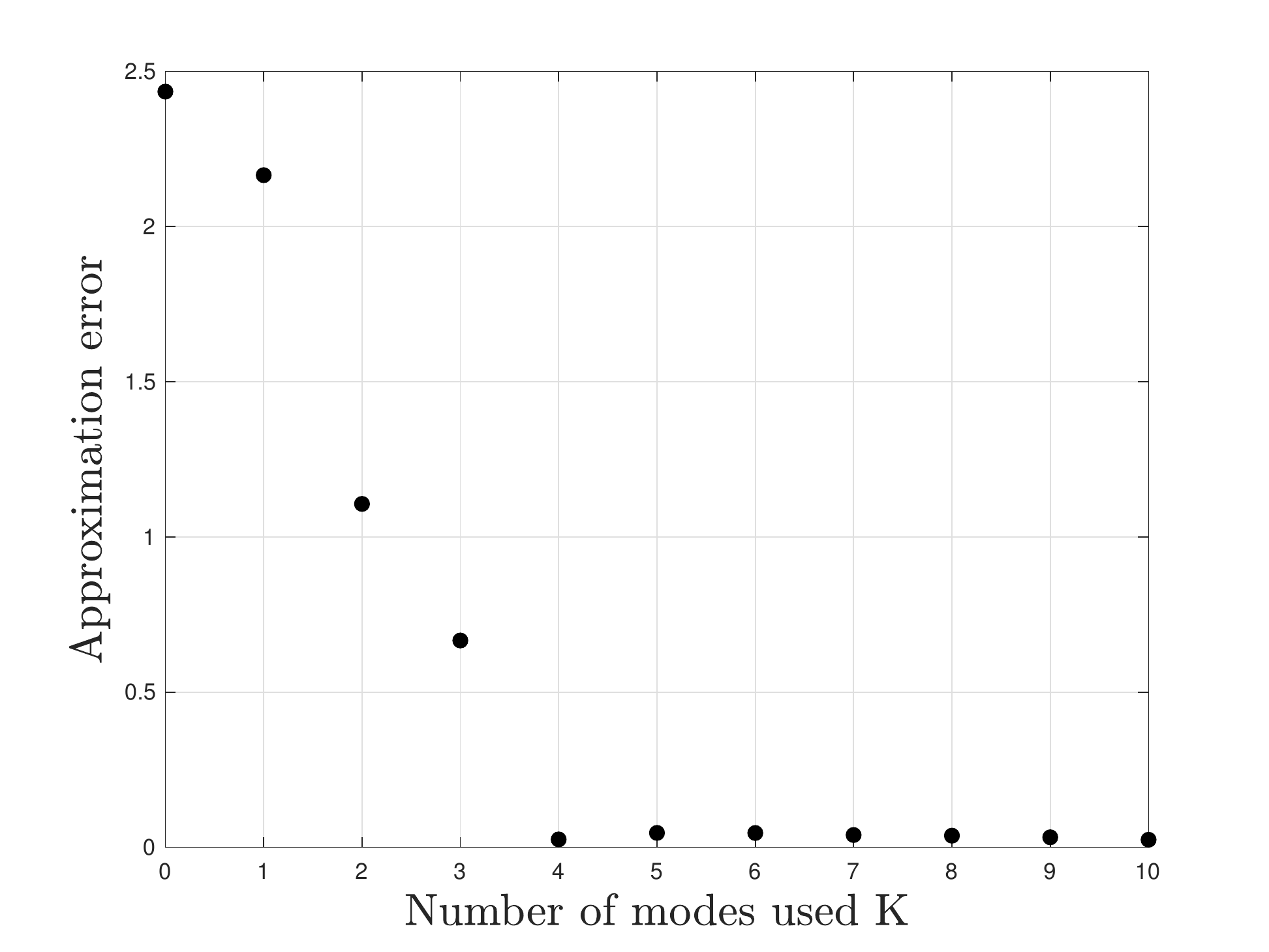}
  \caption{Plot of error: $ \displaystyle \|f_K(x_{\text{obs}},z_{\text{obs}}) - f(x_{\text{obs}},z_{\text{obs}})\|_2$ as a function of the number of modes retained $K$, where $x_{\text{obs}}=50$ km . } 
\label{fig:residual_of_field}
\end{figure}

Formally, we have for $x $ large (on the order of $50$ km) and $z<z_{max}$:

\begin{equation} \label{eq:approx}
f^{\gamma}_K(x,z) \approx \overset{K}{\underset{m=1}{\sum }}a_m^{\gamma} \Psi_m^{\gamma}(z) \exp \left( ik_m^{\gamma}x \right) \ ,
\end{equation}

 where $\Psi_m^{\gamma}(z)$ is a propagating mode and $a_m^{\gamma} = \int \bar{\Psi}^{\gamma}_m(z) F_0(z) dz$.
In the case where we use the boundary conditions in (\ref{eq:leontovich}) \& (\ref{eq:radiation}) along with a trilinear refractivity profile parametrized with $\gamma = (0.118,5,40,30)$, the first five propagating modes are shown in Fig. \ref{fig:modes}. Fig. \ref{fig:fieldsformodes} shows the approximation of the field with different numbers of modes used, and the field obtained by using the SSFPE solution. Figure \ref{fig:residual_of_field} shows the norm difference of $f_K(x,z)$ and $f(x,z)$ at $x=50 $ km and  for $z \in [ 0; 30] $. We observe that after 4 modes, the field is well reconstructed.

In the rest of the paper, we will assume that the observations are collected at a fixed range, and at multiple fixed altitudes: that is we fix $x=x_{\text{obs}} \in \mathbb{R}$ and  $z = z_{\text{obs}}\in \mathbb{R}^{v_{\text{obs}}}$. We denote 
\begin{align}\label{eq:discretized_modal_sol}
F(\gamma) = \begin{bmatrix} f_K^{\gamma}(x_{\text{obs}},z_{\text{obs},1}) \\ \vdots \\ f_K^{\gamma}(x_{\text{obs}},z_{\text{obs},v_{\text{obs}}}) \end{bmatrix}  \in \mathbb{R}^{v_{\text{obs}}}  
\end{align}
where $\gamma$ is the parametrization of $n(z)$.

\begin{figure}[!t]
\centering

 \subfloat[1 mode used]{
    \includegraphics[width=2.5in]{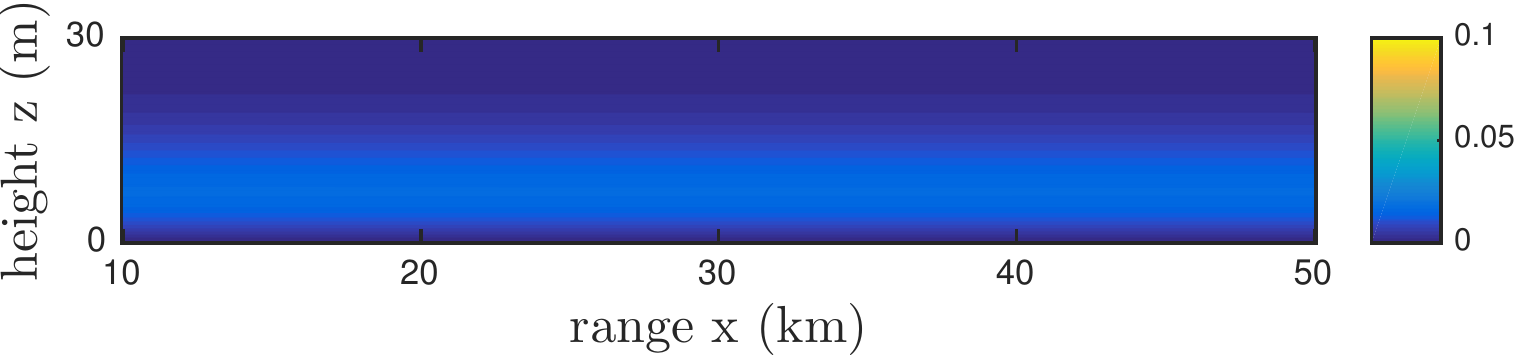}
    \label{fig:subfig1}}
 
    \subfloat[2 modes used]{
    \includegraphics[width=2.5in]{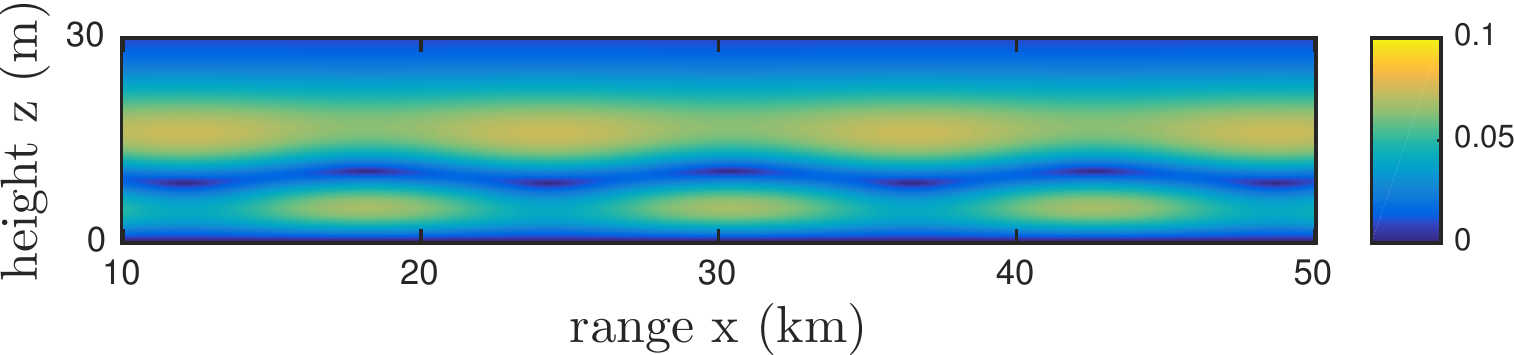}
        \label{fig:subfig2}}

 \subfloat[3 modes used]{
    \includegraphics[width=2.5in]{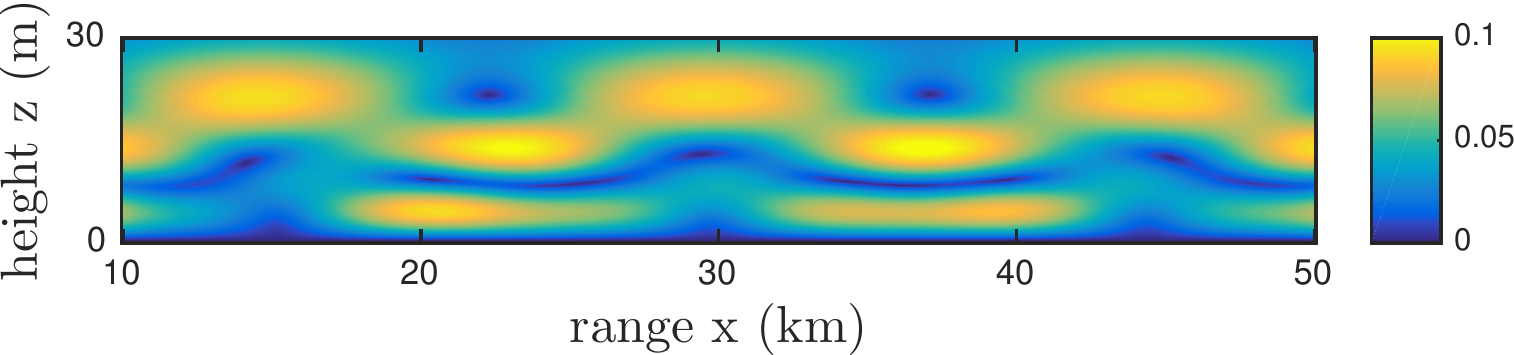}
        \label{fig:subfig3}}

       \subfloat[4 modes used]{
    \includegraphics[width=2.5in]{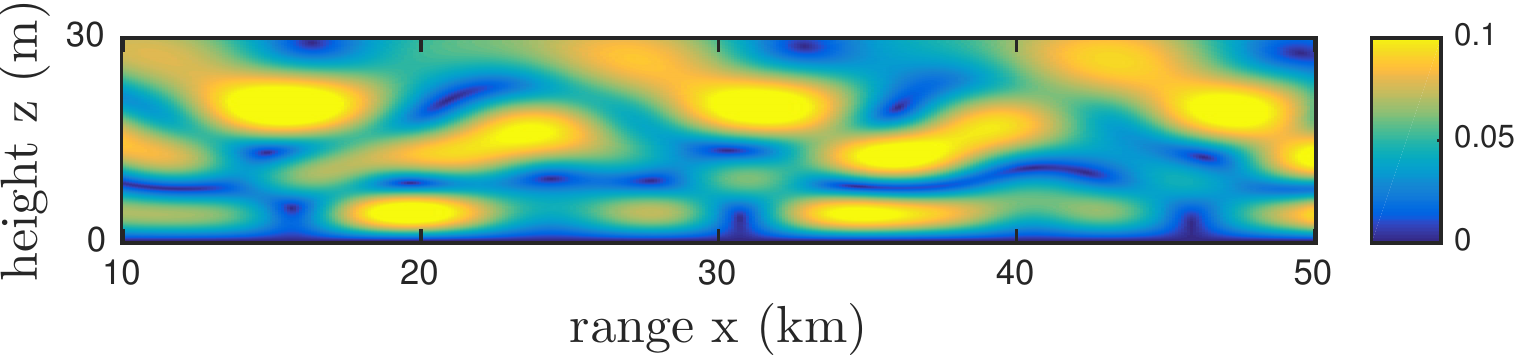}
        \label{fig:subfig4}}

     \subfloat[5 modes used]{
    \includegraphics[width=2.5in]{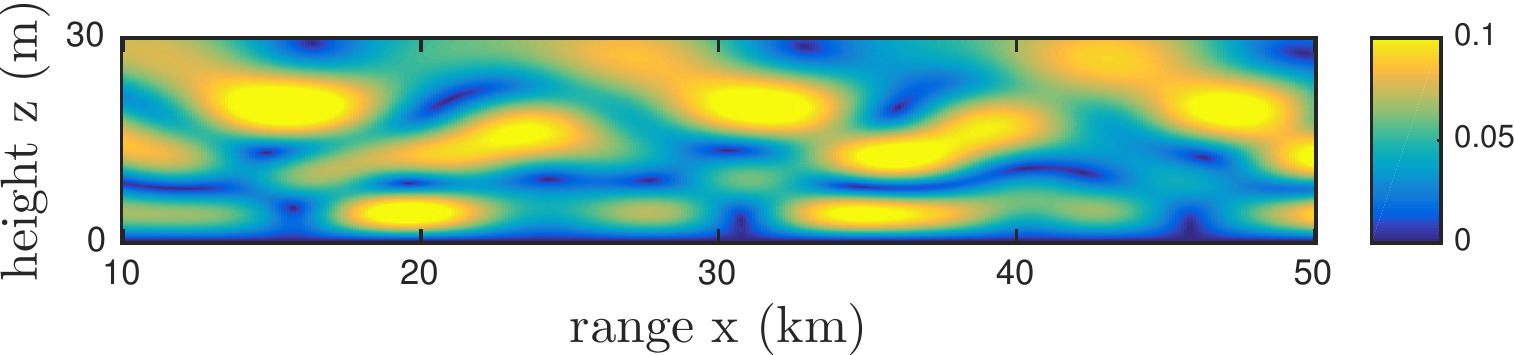}
        \label{fig:subfig5}}

     \subfloat[SSFPE solution]{
    \includegraphics[width=2.5in]{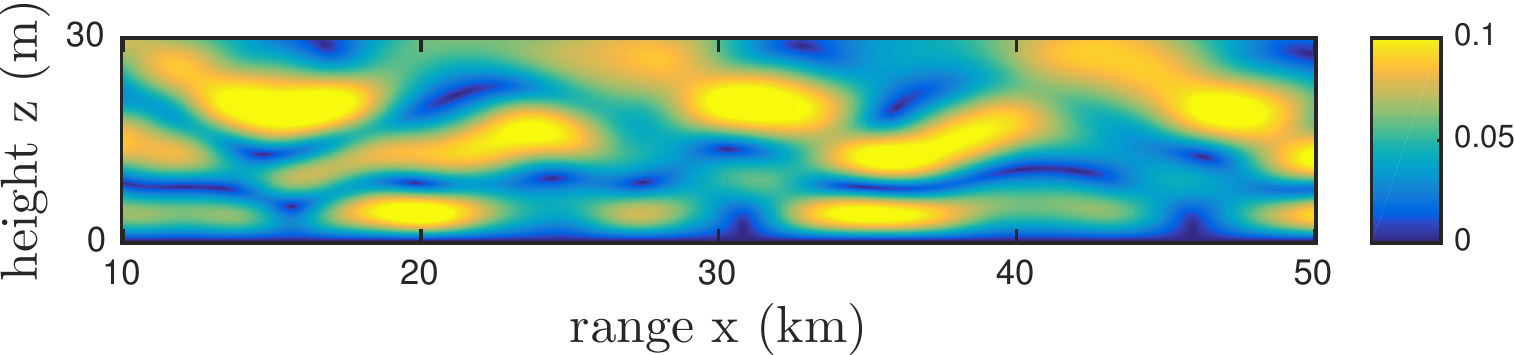}
        \label{fig:subfig6}}

      \label{fig:fieldsformodes}

      \caption{ Low order approximation of the field: $f_K(x,z)$ for increasing numbers of retained modes ((a) - (e)) and also the true field computed by SSFPE (f)   \label{fig:fieldsformodes} }
\end{figure}

\section{Inverse problem: characterizing refractivity} \label{motivation}

Our problem can be described in the following way: given observations of the EM response within the MABL in the form of observed data at a fixed range, $x_{\text{obs}}$, and different heights given by the vector $z_{\text{obs}}$, identify the prevailing vertical profile of the index of refraction $n(z)$. The BVP described in section \ref{Background} provides a forward model, which given an index of refraction, $n(z)$, lets us compute the field at any points in the domain. We would like to solve the inverse problem: find the refractivity profile given some observations of the field. 
A natural approach is to seek 
\begin{equation} \label{eq:minl2}
\gamma^{\text{inv}} = \argmin_{\gamma} ||F(\gamma) - F_{\text{obs}}|| 
\end{equation}
where $\gamma$ is a parametrization of the refractivity profile, and $F(\gamma)$ is the predicted observation under the forward model, e.g., as computed by an SSFP solver. 

\subsection{Analysis of the inverse problem properties}
\label{sec:analysis}
\begin{figure}[t!]
  \centering
\begin{overpic}[width=2.5in]{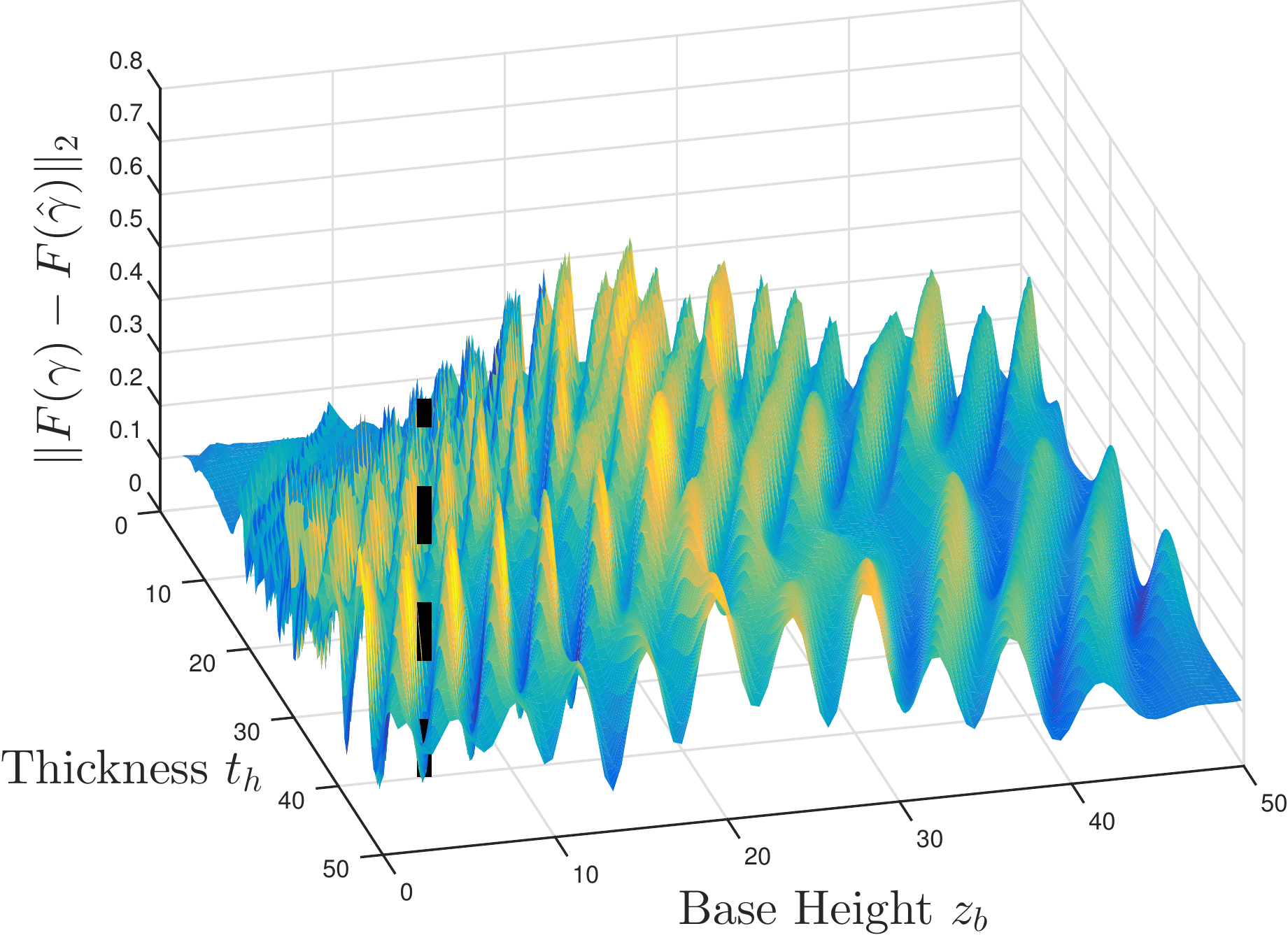}
\put(78,1) {\scriptsize (m)}
\put(6,8) {\scriptsize (m)}
\end{overpic}
    \caption{2-dimensional slice of objective function \ref{eq:minl2}. The black dotted line indicates the value of $\gamma = \hat{ \gamma}$, which by definition of the function is the global minimum.
   }
  \label{fig:ssfpe}
\end{figure}

The objective function in (\ref{eq:minl2}) has thousands of local minima, which makes global minimization very difficult. For demonstration purposes, Fig. \ref{fig:ssfpe} shows a plot of a two-dimensional cut of the objective function (\ref{eq:minl2}), with $F_{\text{obs}} = F \left(\hat{\gamma}\right)$ for a fixed refractivity profile parameterization $\hat{\gamma } = (0.118, 5,40, 30)$.

\begin{figure}[!t]
  \centering
\begin{overpic}[width=2.5in]{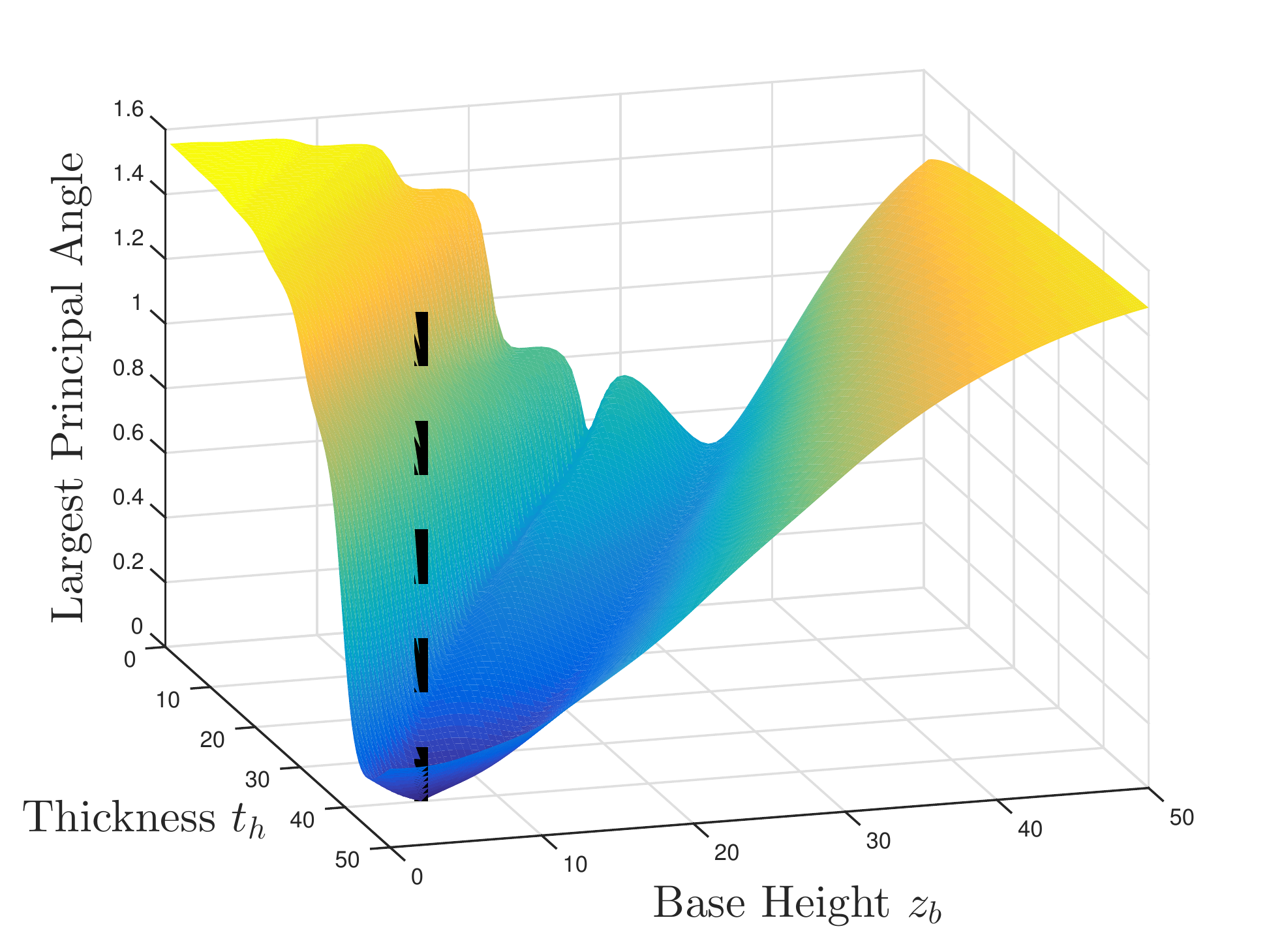}
\put(76,3) {\scriptsize (m)}
\put(7,6) {\scriptsize (m)}
\end{overpic}  \caption{Principal angle between two subspaces: one constant, and the other induced by different refractivity profiles. The dotted line indicates the value where $\gamma = \hat{ \gamma}$. }
  \label{fig:subspaceangle}
\end{figure}

\begin{figure}[!t]
  \centering
\begin{overpic}[width=2.5in]{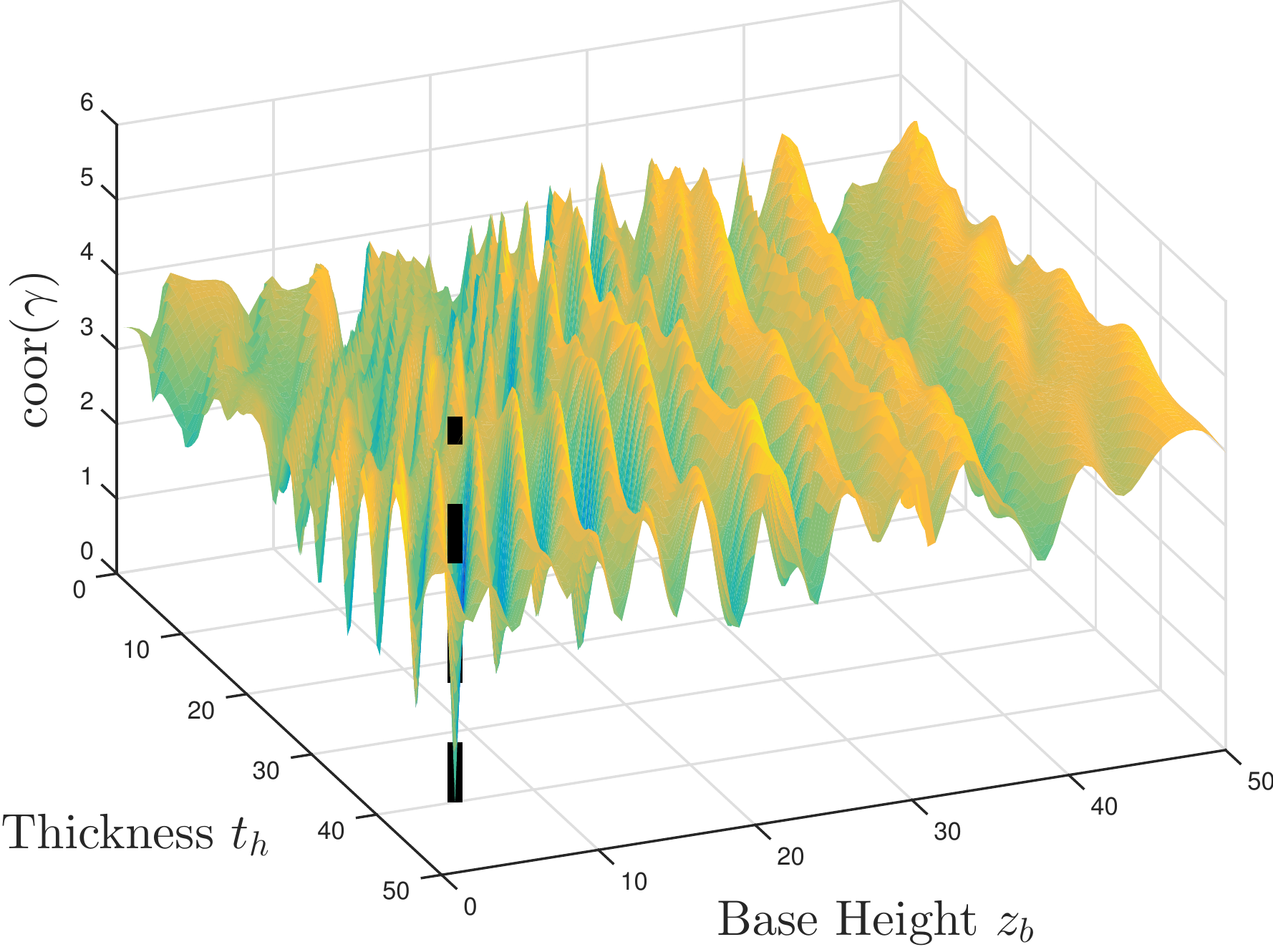}
\put(83,1) {\scriptsize (m)}
\put(7,5) {\scriptsize (m)}
\end{overpic}
  \caption{Difference of norm of ``coordinates" induced by different refractivity profiles. The dotted line indicates the value where $\gamma = \hat{ \gamma}$.}
  \label{fig:coordinateplot}
\end{figure}

Although the complex function behavior is a property of the solution, and not of the modal approximation, it is illuminating to reason about this behavior in terms of the modal approximation. Using the modal approximation in (\ref{eq:discretized_modal_sol}), we see that in the region of interest:
  \begin{align*}
 F(\gamma) &\approx
\begin{bmatrix} f_K^{\gamma}(x_{\text{obs}},z_{\text{obs},1}) \\ \vdots \\ f_K^{\gamma}(x_{\text{obs}},z_{\text{obs},v_{\text{obs}}}) \end{bmatrix} \\
& = 
 \begin{bmatrix}  \overset{K}{\underset{m=1}{\sum }} \Psi_m^{\gamma}(z_{\text{obs},1}) a_m \exp \left( ik_m^{\gamma}x_{\text{obs}} \right) \ , \\ \vdots \\  \overset{K}{\underset{m=1}{\sum }} \Psi_K^{\gamma}(z_{\text{obs},v_{\text{obs}}}) a_K  \exp \left( ik_K^{\gamma}x_{\text{obs}} \right) \end{bmatrix}  {\mathrel{\mathop=}:} U(\gamma) c(\gamma),
 \end{align*}
where  
 \begin{align*}
 & U(\gamma) = 
\begin{bmatrix}
\Psi^{\gamma}_{1} ( z_{\text{obs},1})& \dots& \Psi^{\gamma}_{K} ( z_{\text{obs},1})  \\
\vdots & \ddots & \vdots   \\
\Psi^{\gamma}_{1} ( z_{\text{obs},v_{\text{obs}}})& \dots &\Psi^{\gamma}_{K} ( z_{\text{obs},v_{\text{obs}}})  \\
\end{bmatrix},  \\ 
& c(\gamma) = \begin{bmatrix} 
a_{1}^{\gamma} \exp\left({ik_{1}^{\gamma}x_{\text{obs}}}\right) \\
\vdots \\
a_{v_{\text{obs}}}^{\gamma} \exp \left( {ik_{v_{\text{obs}}}^{\gamma}x_{\text{obs}}} \right)
\end{bmatrix} .
  \end{align*}
The $U(\gamma)$  matrix spans a basis for a subspace in which the approximation is expressed and the vector $ c(\gamma)$ represents the coordinates of the approximation within that subspace. Armed with this representation of the forward model, we would like to determine which of the two components causes the highly oscillatory behavior observed in Fig.~\ref{fig:ssfpe}. Fig.~\ref{fig:subspaceangle} shows how the subspace changes as we vary the refractivity profile $\gamma$ from a reference value of $\hat{\gamma} =  \left(0.118, 5, 20, 40 \right)$ by plotting the largest principal angle \cite{matrixcomp} between $U(\gamma)$ and $U(\hat{\gamma})$. 
 Fig. \ref{fig:coordinateplot} shows how the coordinate part of the function evolves as we change $\gamma$ in the discrete 2-norm, i.e. it is a plot of:
  \begin{equation*} \label{eq:coor}
\mbox{coor}(\gamma) = \|c(\gamma) - c(\hat{\gamma} )\|_2^2
  \end{equation*}
  
\begin{figure}[!t]
  \centering
\begin{overpic}[width=2.5in]{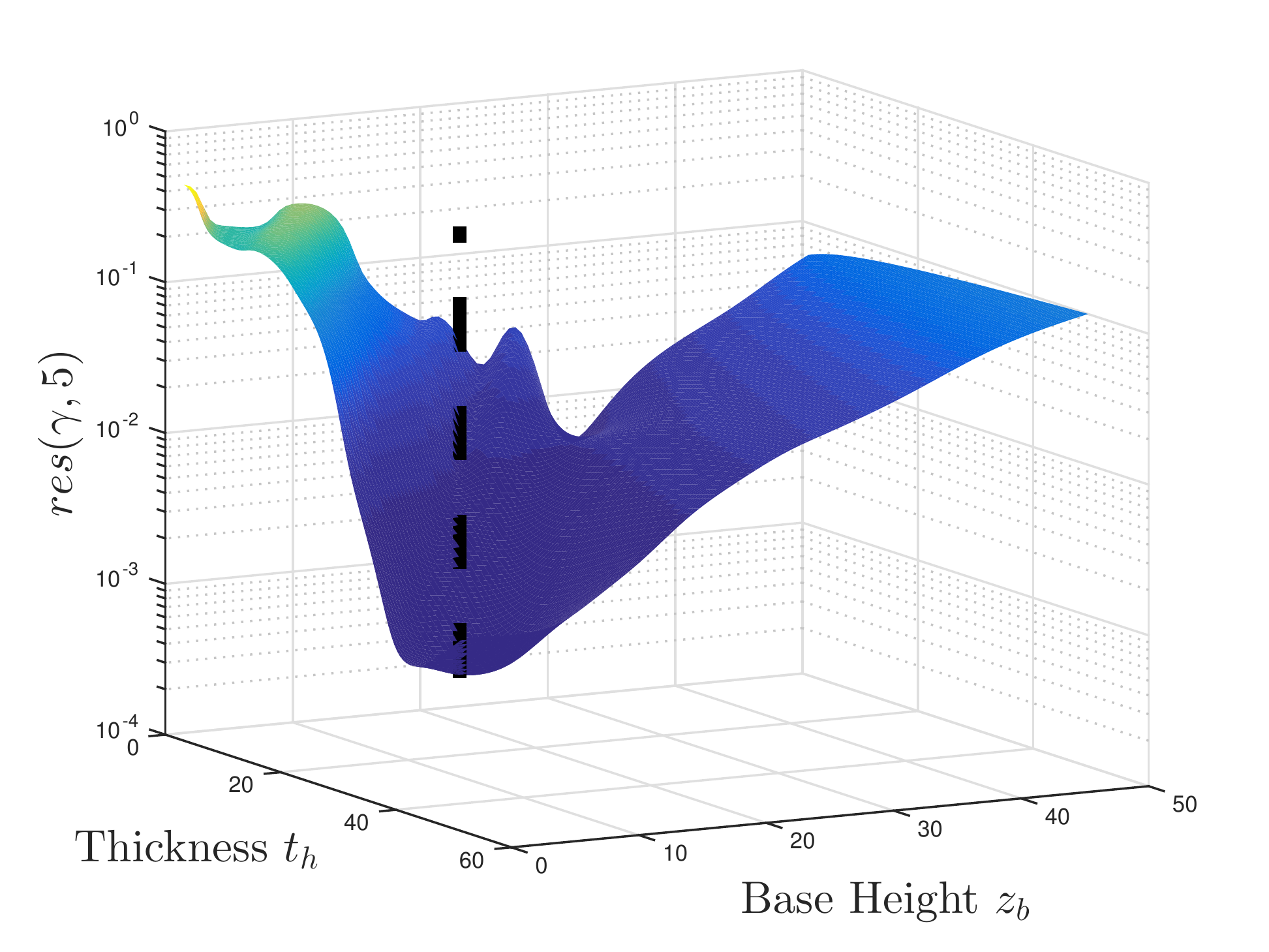}
\put(83,3) {\scriptsize (m)}
\put(11,4) {\scriptsize (m)}
\end{overpic}
  \caption{Objective function of the proposed method plotted in a semi-log scale, defined in equation (\ref{eq:alg1res}). The dotted line indicates the value of the true refractivity profile. }
  \label{fig:alg1objfun}
\end{figure}

It is clear from Fig. \ref{fig:subspaceangle} \& \ref{fig:coordinateplot} that the basis $U(\gamma)$ changes smoothly, whereas the coordinate $c(\gamma)$ of the function causes the oscillatory behavior of the function in (\ref{eq:minl2}). These oscillations are explained by the terms $  \exp{\left( k_{m}^{\gamma}ix_{\text{obs}}\right)}$. Indeed, $x_{\text{obs}}$ is typically large (here, $x_{\text{obs}}= 5 \cdot 10^{4}$m); hence any small change in eigenvalue $k^{\gamma}_m$ caused by a change in $\gamma$ are heavily amplified by the multiplication by $x_{\text{obs}}$, which causes wild oscillations of the term $\exp{\left( k_{m}^{\gamma}ix_{\text{obs}}\right)}$. 

To avoid the multimodal behavior of the objective function (\ref{eq:minl2}) caused by the wild oscillations in coordinates $c(\gamma)$, we propose an alternative objective:
\begin{equation*}
\gamma^{\text{inv}} = \argmin_{\gamma} ||F_{\text{obs}}-\Pi_{U(\gamma)}\left( F_{\text{obs}} \right)|| 
\end{equation*}
where $\Pi_{U(\gamma)}$ (defined precisely in the next section) is a projector onto the low-dimensional space spanned by the propagating modes for the refractivity profile parametrized by $\gamma$. In contrast to (\ref{eq:minl2}), the new objective does not oscillate; see (\ref{fig:alg1objfun}).
\subsection{Proposed inverse solution method} \label{description}

\begin{figure}[!t]
\vspace{0.5cm}
  \centering
\includegraphics[trim={3cm 3cm 3cm 3cm}, width=2.5in]{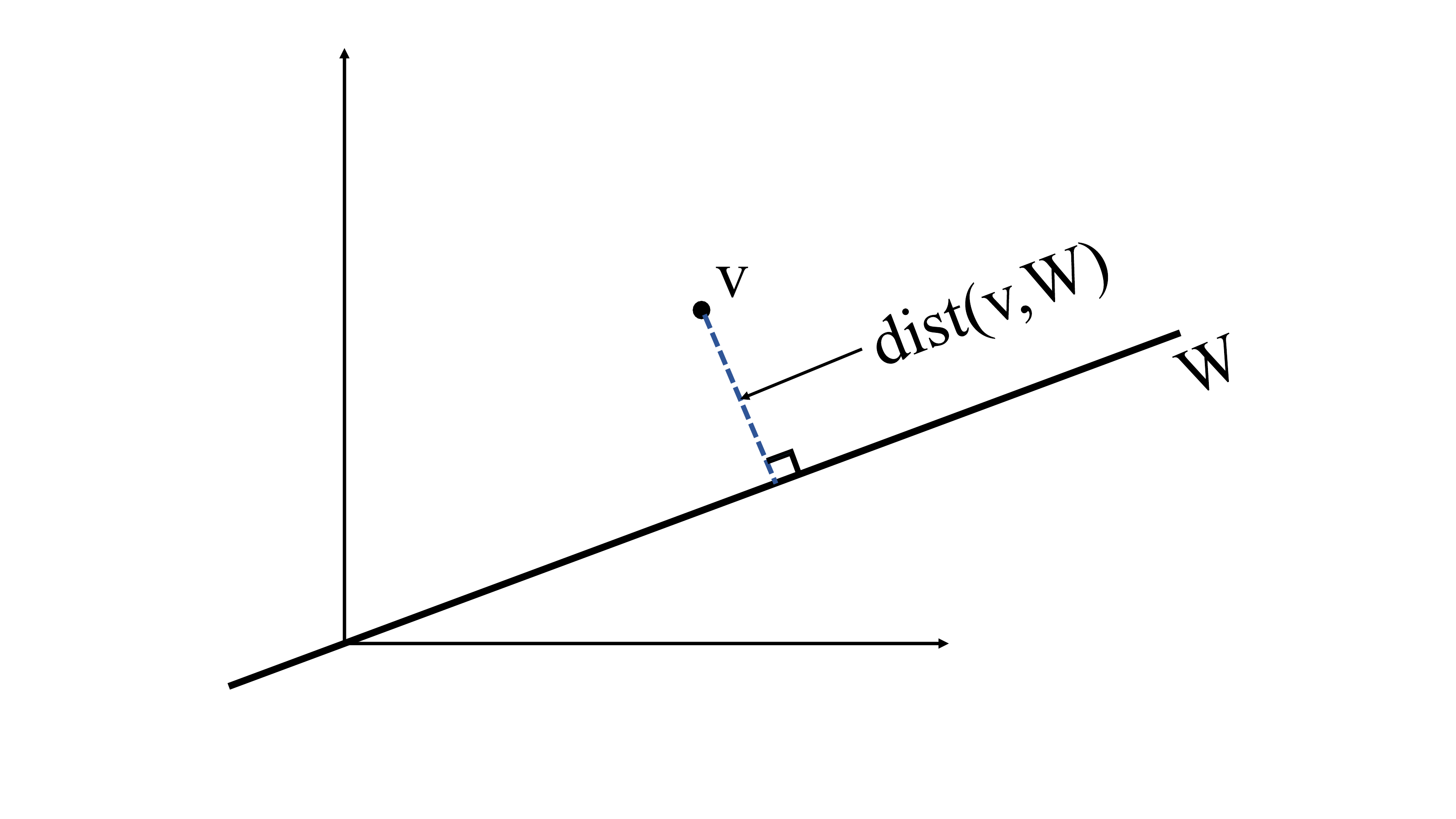}
  \caption{Distance between a vector $v\in \mathbb{R}^2$ and a one dimensional subspace $W \subset \mathbb{R}^2$ (a line) }
  \label{fig:distVW}
\end{figure}

Our algorithm attempts to find a low dimensional subspace that best explains the measurements. 
To achieve this goal, we first need a measure of optimality of a subspace. Let $v \in \mathbb{C}^{n}$, and $\mathcal{W}$ be a $K$ dimensional subspace of $\mathbb{C}^{n}$. It is natural to define the distance between the vector $v$ and the subspace $\mathcal{W}$ as the minimum of the distance between the vector $v$ and any vector $w \in \mathcal{W}$, as in (\ref{eq:subspacedistdefine2}). A graphical representation of the distance between a subspace and a vector in the case where $n=2$ and $K=1$ is shown in figure \ref{fig:distVW}. 
Let the columns of $W \in \mathbb{C}^{n \times K } $ span $\mathcal{W}$, then the squared distance between $v$ and $\mathcal{W}$ is defined as:
\begin{equation}
\label{eq:subspacedistdefine2}
d^2( v,\mathcal{W}) = \min_{w \in \mathcal{W}} \|w-v\|_2^2 = \min_{\phi \in \mathbb{C}^K} \| W\phi - v \|_2^2 
\end{equation}  

Accordingly, we define the distance between $W$ and a collection of $m$ vectors $v_i \in  \mathbb{C}^{n}$, which are the columns of $V \in \mathbb{C}^{n \times m }$ as:

\begin{equation} \label{eq:distVW}
d^2(V,\mathcal{W}) =  \sum_{i=1}^{n} d^2( v_{i}, \mathcal{W}) =   \min_{\Phi\in \mathcal{C}^{K\times m}} ||W\Phi - V||_{F}^2
\end{equation}  
 Equation (\ref{eq:distVW}) provides a way to measure the fit of some collection of measurements $V$ into a subspace $\mathcal{W}$ spanned by the columns of $W$.

In our case, we have $V = F_{\text{obs}}$ (the EM observations),
and $W = U(\gamma)$ (the matrix of propagating modes induced by $\gamma$ sampled at the observation points). Therefore, to find the suspace which best fits the observation, we seek to perform the following minimization:

\begin{equation}
\label{eq:projectioneq}
 \min_{\gamma} d^2 \left( F_{\text{obs}},U \left( \gamma \right) \right) \ .
\end{equation}  

According to our numerical tests, the crucial consideration to make this method successful is to characterize the dimension of the subspace properly. Indeed, it can be observed that the number of modes used, even for a fixed physical setting (i.e. frequency, boundary values, range) depends heavily on the actual refractivity profile that produces the modes. Intuitively, this can be understood in the following way: if the refractivity profile represents a very strong duct, then the trapping increases and therefore more modes are trapped under the duct and propagate in range. In order to automatically choose the number of modes to consider, we propose two strategies:

\subsubsection{Algorithm 1 Minimal subspace dimension}

One strategy is to search over increasingly high dimensional subspaces until we find one that fits enough of the data according to some criterion. The specific criterion we use is: the objective function must be lower than some threshold value of $\tau$, which depends on the noise level\footnote{In section~\ref{experiments} we set $\tau = \eta^2 + 0.03 $ where $\eta$ is the noise level.}. A drawback of this method is that it requires a good estimate of the noise level. We include a regularization term of $\Phi$ into the objective function. We choose the regularization parameter $\alpha$ by optimizing over a representative training set. 
We define a regularized residual function (plotted in Fig. \ref{fig:alg1objfun}):
\begin{equation} \label{eq:alg1res}
\mbox{res}(\gamma,k) =\min_{\Phi} ||U_k(\gamma)\Phi - F_{\text{obs}}||_{F}^2+ \alpha \| \Phi \|^2_{F} \ .
\end{equation}

The algorithm is summarized in Fig. \ref{Alg1}.
\begin{figure}
\begin{algorithmic}[1] 

\STATE \textbf{Input:} EM observations $F_{\text{obs}}$, threshold $\tau$
\STATE $k \leftarrow 1$
\STATE $r \leftarrow \infty$ 
\WHILE{ $r> \tau $ }
\STATE    minimize $\mbox{res}(\gamma,k) $
\STATE$r \leftarrow \min_\gamma \mbox{res}(\gamma,k)$
\STATE $\gamma^{\text{inv}} \leftarrow \argmin_\gamma \mbox{res}(\gamma,k)$
\STATE $k \leftarrow k+1$
\ENDWHILE
\RETURN $\gamma^{\text{inv}}$
\end{algorithmic}
\caption{Algorithm 1: Minimal subspace dimension\label{Alg1}
 }
\end{figure}

\subsubsection{Algorithm 2 Filtering eigenvalues}

Another way to determine the number of eigenvectors to include is to use an \emph{a priori} bound on the eigenvalues to be included. As discussed in section \ref{Background}, we can consider the propagating modes as the eigenvectors associated with eigenvalues which fall within a physically inspired interval and include only such modes. However, using such a hard threshold for included modes within an interval has some drawbacks. One is that at all places of the resulting objective function where a mode is included or excluded, the objective function is discontinuous. The other one is that our characterization of a propagating mode is physically inspired by what we define to be the domain of dependence according to $z_{max}$, but the true domain of dependence is infinite. Instead of a strict cut-off, we use a soft thresholding by defining a filter, $t(\sigma)$, in the following way:
\begin{equation} \label{eq:filterdef}
t^{\gamma}(\sigma ) = \begin{cases}
1, & re(\sigma) > c_2^{\gamma} \\
g^{\gamma}( \sigma ) &c_1^{\gamma} \leq re(\sigma) \leq c_2^{\gamma} \\
0  & re(\sigma ) < c_1^{\gamma} \\
\end{cases}
\end{equation}
$c_1^{\gamma}$ and $c_2^{\gamma}$ are constants that relate to the \emph{a priori} bounds of the eigenvalues and $g^{\gamma}(\sigma)$ is a smooth interpolation. In particular, we set
 \begin{align*}
 k_{max}^{\gamma} &= \max_{z} \{k_0n (z) \mid  0 < z < z_{max}\} ,
  \\ k_{min}^{\gamma} &= \min_{z} \{k_0 n (z) \mid 0 < z < z_{max}\} , \\ c_1^{\gamma} &= k_{min}^{\gamma} \\ c_2^{\gamma} &= 0.9k_{min}^{\gamma} + 0.1{k_{max}}^{\gamma} , 
\\
g^{\gamma}(\sigma) & = \hat{g} \left( \frac{\sigma - c_1^{\gamma}  }{c_2^{\gamma}-c_1^{\gamma} } \right) , \\
\hat{g}(\sigma) & = 6 \sigma^5 - 15\sigma^4 + 10\sigma^3 \ .
 \end{align*}
The modes which have a filtered eigenvalue of 0 are excluded from the objective function.
 The resulting optimization problem is:
 \begin{IEEEeqnarray}{c}
\gamma^{\text{inv}} = \argmin_{\gamma} \left\lbrace  \hat{\Pi}(\gamma) \right\rbrace \IEEEnonumber \\
\hat{\Pi}(\gamma) = \min_\Phi \left\lbrace \| U(\gamma) \Phi -F_{\text{obs}}\|^2_{F} + \alpha \| t(\Sigma (\gamma) )^{-1}\Phi ||^2_{F} \IEEEnonumber  \right. \\ 
\left. +\beta \| t(\Sigma(\gamma)) \|_1 \vphantom{\| EU(\gamma) \Phi -F_{\text{obs}}\|^2_{F} }\right\rbrace
\end{IEEEeqnarray}
where $\Sigma(\gamma)$ is the diagonal matrix of eigenvalues $k_m^{\gamma}$.
We describe each individual term of the objective function:

\begin{itemize}
\item$\|U(\gamma)\Phi-F\|^2_{F}$

This term is the same as defined in (\ref{eq:projectioneq}). As discussed earlier, it models how well the data fits into the subspace spanned by the columns of $U(\gamma)$.

\item $\displaystyle \alpha \|  t(\Sigma(\gamma))^{-1}\Phi \|^2_{F}$

This is the regularization term of $\Phi$. It forces the choice of coordinate in the basis spanned by $U(\gamma)$ to be smaller in the directions associated with eigenvalues that have a small filtered value: that is those that we think \emph{a priori} are less likely to be propagating. $\alpha$ is a regularization parameter is chosen by optimizing over a training set.

\item $ \beta \| t(\Sigma(\gamma)) \|_1$

 The norm used in this term is the sum of the absolute values of the matrix entries. This term is the regularization term on the size of the subspace. Indeed, the objective function without this term would be naturally biased towards refractivity profiles that induce a large number of propagating modes. 
To give intuition as to why this happens, suppose
$\gamma_1$ and $\gamma_2$ are the refractivity profiles with $k$ propagating modes in common, but that $\gamma_2$ admits an additional $(k+1)$st mode that is not propagating for $\gamma_1$. Then, almost any data that fits $\gamma_1$ will fit $\gamma_2$ even better, even if the extra mode only fits model error of measurement noise. Finally,
 $\beta$ is a regularization parameter chosen by optimizing over a training set.
\end{itemize}

It is interesting to note that, unlike an objective function such as the one in ($\ref{eq:minl2}$), the algorithm described in section \ref{description} is agnostic to information about the source or the range of the transmitter. One could imagine that this may be useful if a good model of the source is unavailable, or if the receiver wants to be a passive listener only. The method also combines different observations from different sources or different ranges at virtually no additional computational expense which can increase accuracy. 

\section{ Implementation} \label{implementation}

\subsection{A computational shortcut}

As described in section \ref{Background}, the Sturm-Liouville eigenvalue problem that arises as a result of the separation of variables in the case we are interested in is posed on an infinite domain:

 \begin{equation} \label{eq:SL}
\frac{d^2 \Psi_m(z)}{dz^2}+\left[ k_0^2n(z)^2- k_{rm}^2 \right] \Psi_m(z)=0 \quad  0 < z < \infty
 \end{equation}

 \begin{equation} \label{eq:z0BC}
\beta_1 \frac{\partial \Psi_m(z)}{\partial z }\bigg\rvert_{z=0} + \beta_2  \Psi_m(0) =0
 \end{equation}  
 
  \begin{equation} \label{eq:infBC}
\lim_{z\rightarrow \infty}  \hat{\alpha}_1\frac{\partial \Psi_m(z)}{\partial z } + \hat{\alpha}_2\Psi_m(z) =0 \ .
 \end{equation}  
 
In our case, we have $\beta_1 = 1$, $\beta_2 = \left( 1/\left(2a_e\right) + ik_0\sqrt{\epsilon_s -1}\right) $, $\hat{\alpha}_1 = 1$, $\hat{\alpha}_2 =- ik_0$. One can solve the SL eigenvalue problem with an infinite boundary condition by solving an equivalent problem on a finite domain, but with a boundary condition that is a function of the eigenvalues, for an example of a supporting derivation, see \cite{COA}. This new SL eigenvalue problem takes the form:

 \begin{equation} \label{eq:SL2}
\frac{d^2 \Psi_m(z)}{dz^2}+\left[k_0^2n(z)^2 -k_{rm}^2 \right] \Psi_m(z)=0 \quad  0 < z < D
 \end{equation}

 \begin{equation} \label{eq:z0BC2}
\beta_1 \frac{\partial \Psi_m(z)}{\partial z }\bigg\rvert_{z=0} + \beta_2  \Psi_m(0) =0
 \end{equation}  
 
  \begin{equation} \label{eq:nonlinBC2}
\alpha_1(k_m^2)\frac{\partial \Psi_m(z)}{\partial z }\bigg\rvert_{z=D} + \alpha_2(k_m^2)\Psi_m(D) =0 \ .
 \end{equation}  

For the application at hand, $n(z)$ is assumed to be linear above $D$ and therefore $\alpha_1(k_m^2)$ and $\alpha_2(k_m^2)$ can be expressed in terms of parabolic cylinder functions. One can then discretize this new SL eigenvalue problem and attempt to solve it numerically. This discretized SL eigenvalue problem gives rise to a non-linear algebraic eigenvalue problem. Nonlinear eigenvalue problems are in general expensive to solve, but since the nonlinearity is only in the boundary condition (in other words, in a single entry of the matrix), it is possible to implement fast solvers. For example, in~\cite{kaufman_2006} the author presents an algorithm to solve a similar problem based on Newton's method. However, in our case, this computational expense is unnecessary as we can deal with a standard linear eigenvalue problem instead. Indeed the modes that we are interested in are the so-called trapped modes. The trapped modes are those which are ``trapped" physically low within the domain, and therefore their support lie below a threshold.
Note that if $\Psi_m(z)$ satisfies (\ref{eq:SL2}) and (\ref{eq:z0BC2}), and $ \frac{\partial \Psi_m(z)}{\partial z }\bigg\rvert_{z=D} =0$, $\Psi_m(D) =0$, then  $\Psi_m(z)$ also satisfies ($\ref{eq:nonlinBC2}$), and therefore solves the non-linear eigenvalue problem.
This implies that if we solve the linear eigenvalue problem associated with the SL problem with homogeneous Dirichlet boundary conditions at $z=D$, and the eigenvector's derivative is zero at the upper boundary, then this solution also solves the nonlinear eigenvalue problem. Otherwise, we can use this eigenpair as a first guess in a Newton's method for the non-linear problem as described in \cite{kaufman_2006}. In practice, we find that this step is rarely necessary provided that $D$ is chosen high enough, and thus we only solve the Dirichlet problem to do the inference in the numerical experiments in section~\ref{experiments} in order to save the extra computational expense. 

\subsection{Computation of the modes}

Implementing the algorithm involves solving the Sturm-Liouville eigenvalue problem numerically. We solve this continuous problem by discretizing using finite differences; a treatment and derivation can be found in \cite{COA}. This reduces the problem to one of computing a subset of eigenvectors and eigenvalues of a tridiagonal symmetric matrix, for which optimized solvers can be used. As pointed out in \cite{COA}, one should use 5 to 10 discretization per wavelength in this type of computation. In our case, this induces a discretized system of size approximately $5000 \times 5000$. We have observed the results of the inversion to be insensitive to the discretization size chosen.

\subsection{Inner minimization}

Each function evaluation in the algorithm described above involves a minimization over $\Phi$. However, in that inner minimization, $\gamma$ is fixed. Thus the inner problem is a linear least squares problem which can be solved in closed form. Furthermore, $\Phi$ is a small matrix with a number of rows equal to the number of non-zero filtered eigenvalues (usually fewer than 10) and a number of columns equal to the number of vertical slices of sampled taken (on the order of 30). Therefore the inner minimization may be computed at the cost of a small linear solve.

\subsection{Outer minimization}

The most computationally expensive part of the evaluation of the objective function is the computation of a few eigenvectors. However, the computation of first and second derivatives of the objective function is much cheaper computationally as it only involves matrix multiplications and linear solves of small matrices. This fact, coupled with the small number of local minima of the objective function, motivated the use of derivative-based local optimization method. We use MATLAB's sequential quadratic programming method, described in \cite{nocedal_wright_2006}, and perform multistart. The starting points are chosen by Latin hypercube sampling. Five starting points are used for each subspace dimension in Alg.~1, and ten starting points are used in Alg.~2.

\section{Numerical Experiments} \label{experiments}

\subsection{Error measures}

Defining an error measure is crucial to perform the optimization over the parameters $\alpha$ in Algorithm 1 and $\alpha$ and $\beta$ in Algorithm 2 over a training set, as well as to evaluate the performance of our method. The error measure is defined as the relative normalized $\ell_2$ (RNL2) error between a trial $n(z)$ and a true $n_{\text{true}}(z)$.
First we define the normalized $\ell_2$ error by: 
\begin{equation}
\text{error}_{\ell_2} \left( n(z),n_{\text{true}}(z) \right) = \frac{\int_{\xi=0}^{\xi = 60}  (n(\xi)- n_{\text{true}}(\xi))^2 d\xi }{\int_{\xi=0}^{\xi = 60}  (n_{\text{true}}(\xi ))^2 d\xi }
\end{equation}
We define the RNL2 by dividing the normalized $\ell_2$ error by the expected value of the normalized error of two random refractivity profiles coming from a very large representative set.
\begin{equation}
\text{RNL2}(n(z)) = \frac{\text{error}_{\ell_2}\left( n(z), n_{\text{true}}(z) \right)}
{ \mathbb{E}_{n_i(z),n_j(z)}\left[ \text{error}_{\ell_2} \left( n_i(z), n_j(z) \right)  \right]}
\end{equation}

The expectation is taken over the parameters for which we perform the optimization, and is approximated by averaging $10^4$ trials.
For example, a score of $0.1$ signifies that the algorithm performs $10$ times better than a random guess.

\subsection{Experiment 1: Simulated data originating from a trilinear, horizontally constant index of refraction.}

We simulate electromagnetic wave propagation by solving the partial differential equation in (\ref{eq:Helmholtz}) using the SSFPE method. We set the wavelength $\lambda = 0.1 $ m, use a Gaussian antenna pattern as the source, and sample the field at a fixed range $x_{\text{obs}} = 50$ km and heights of $z_{\text{obs},j} = j $ for $j\in \{1,2,...,30\}$.
Therefore, each measurement $F_{\text{obs},i}$ is a vector of length 30, whose entry $j$ corresponds to the field at range $x=x_{\text{obs}}$ and altitude $z=j$. We simulate 5 such measurements for each test case, where different measurements are obtained by varying the  height (between $20$~m and $30$~m) and tilt (between $-0.5$ and $0.5$ degrees off of horizontal) of the antenna (represented as a Gaussian source):
\begin{align*}
\text{tilt} \sim U(-0.5, 0.5), \qquad \text{height} \sim U(20, 30)
\end{align*}
where $U(a,b)$ denotes the uniform distribution on $[a,b]$.
 This forms a matrix of measurements $F_{\text{obs}}$ of dimension $30 \times 5$.
We then contaminate the data with Gaussian white noise of standard deviation $0.3 \| F_{\text{obs}} \|_2$. 
The observations are then normalized to have unit norm in order to keep the different terms of the objective function scaled relative to each other.

\begin{figure*}[!t]
\centering
\subfloat[$ \text{RNL2} =  \left(0.28 ; 0.14 \right) $ ]{\includegraphics[width=0.23\textwidth]{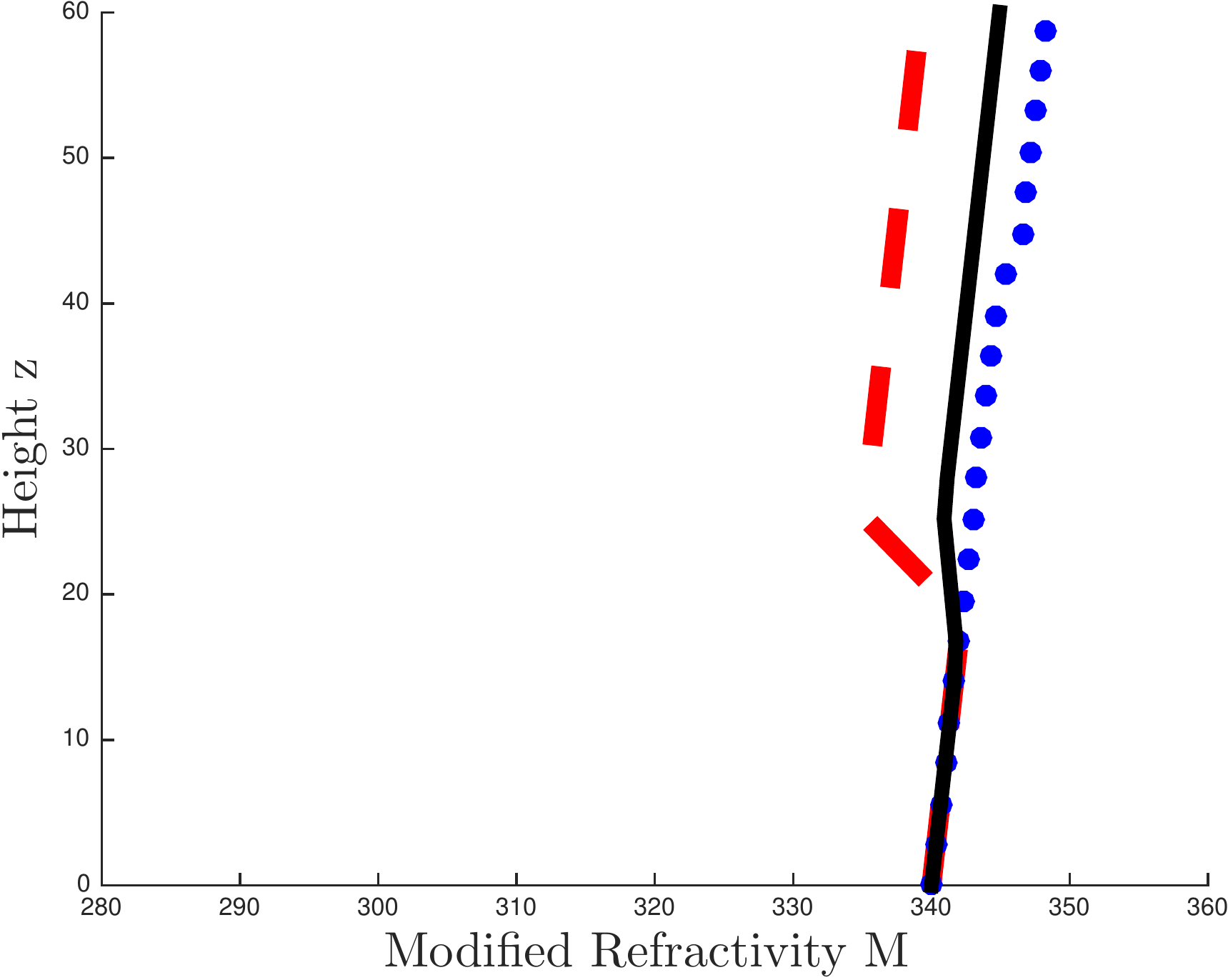}
\label{fig_first_case}}
\hfil
\subfloat[$ \text{RNL2} =  \left(0.25 ; 0.37 \right) $ ]{\includegraphics[width=0.23\textwidth]{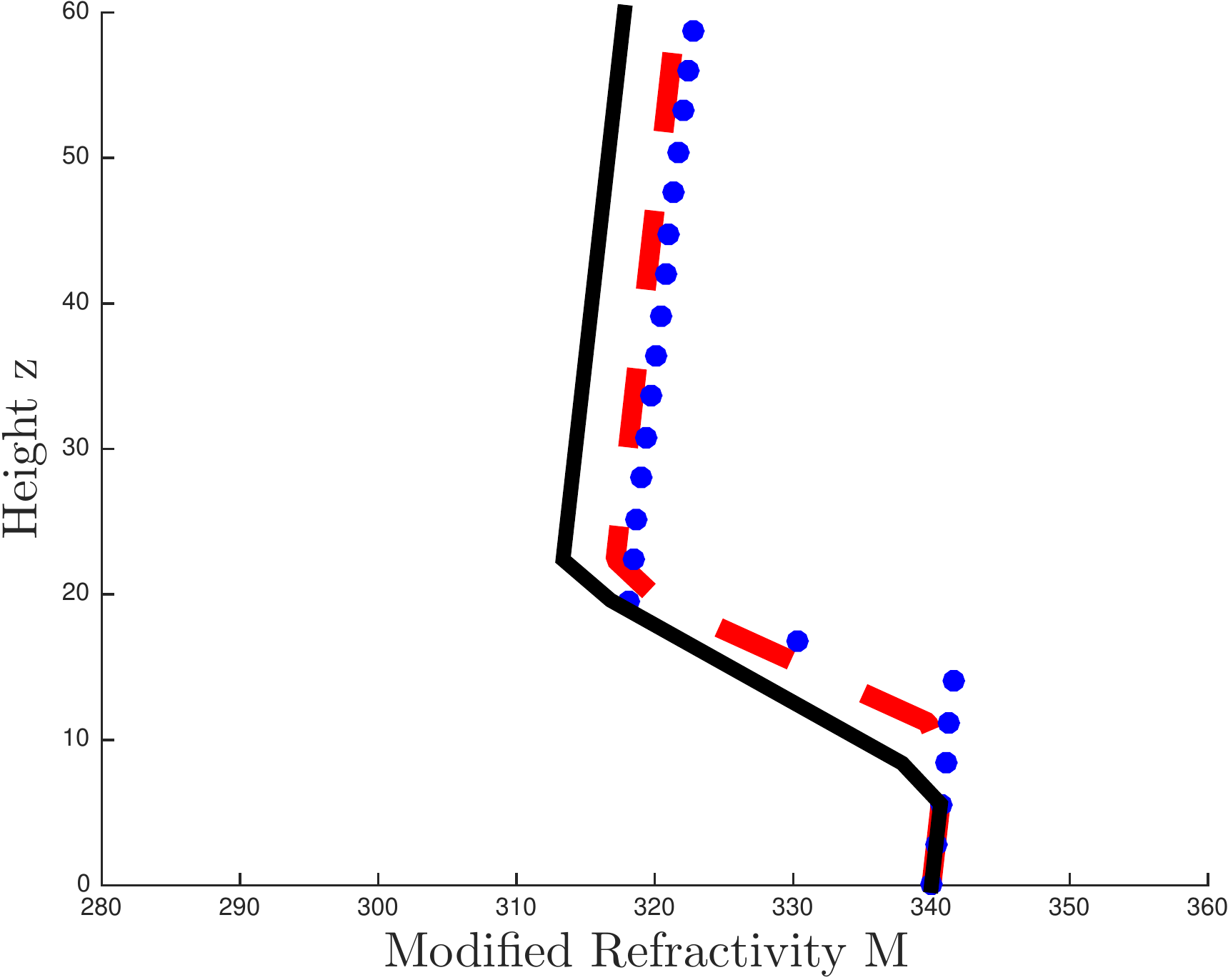}
\label{fig_second_case}}
\hfil
\subfloat[$ \text{RNL2} =  \left(0.38 ; 0.37\right) $ ]{\includegraphics[width=0.23\textwidth]{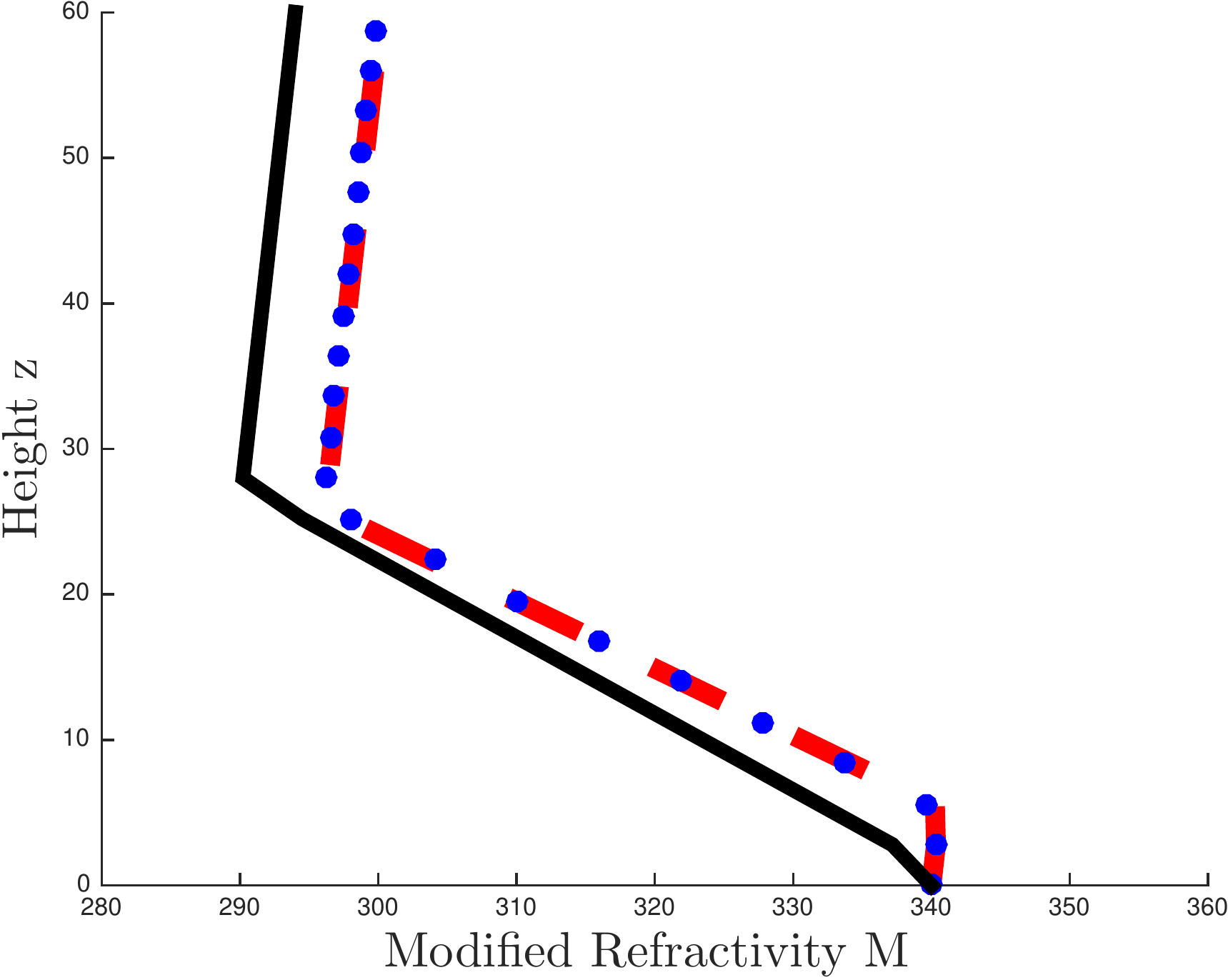}
\label{fig_second_case}}
\hfil
\subfloat[$ \text{RNL2} =  \left(0.46 ; 0.15\right) $ ]{\includegraphics[width=0.23\textwidth]{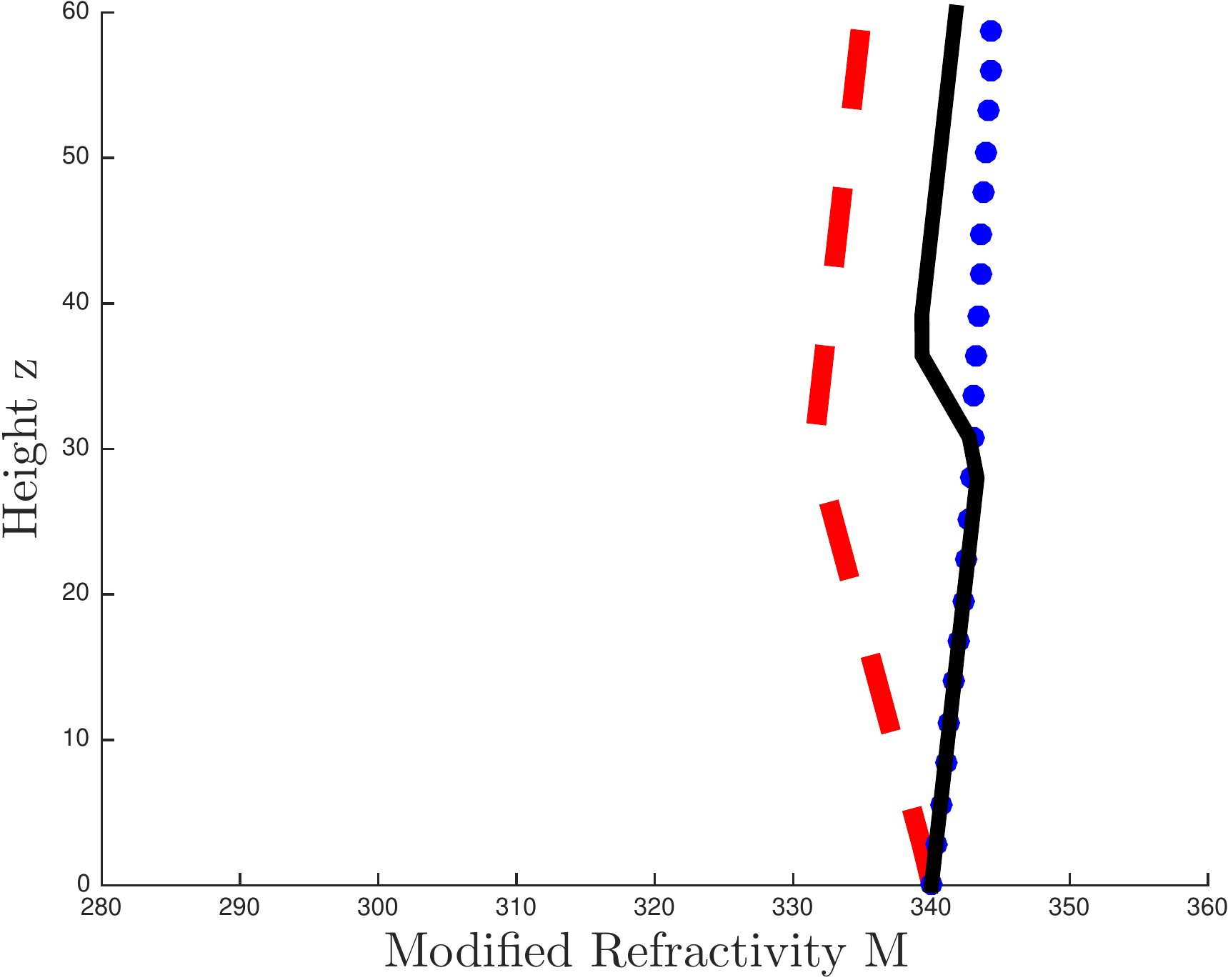}
\label{fig_second_case}}
\vfil

\subfloat[$ \text{RNL2} =  \left(0.05 ; 0.41\right) $ ]{\includegraphics[width=0.23\textwidth]{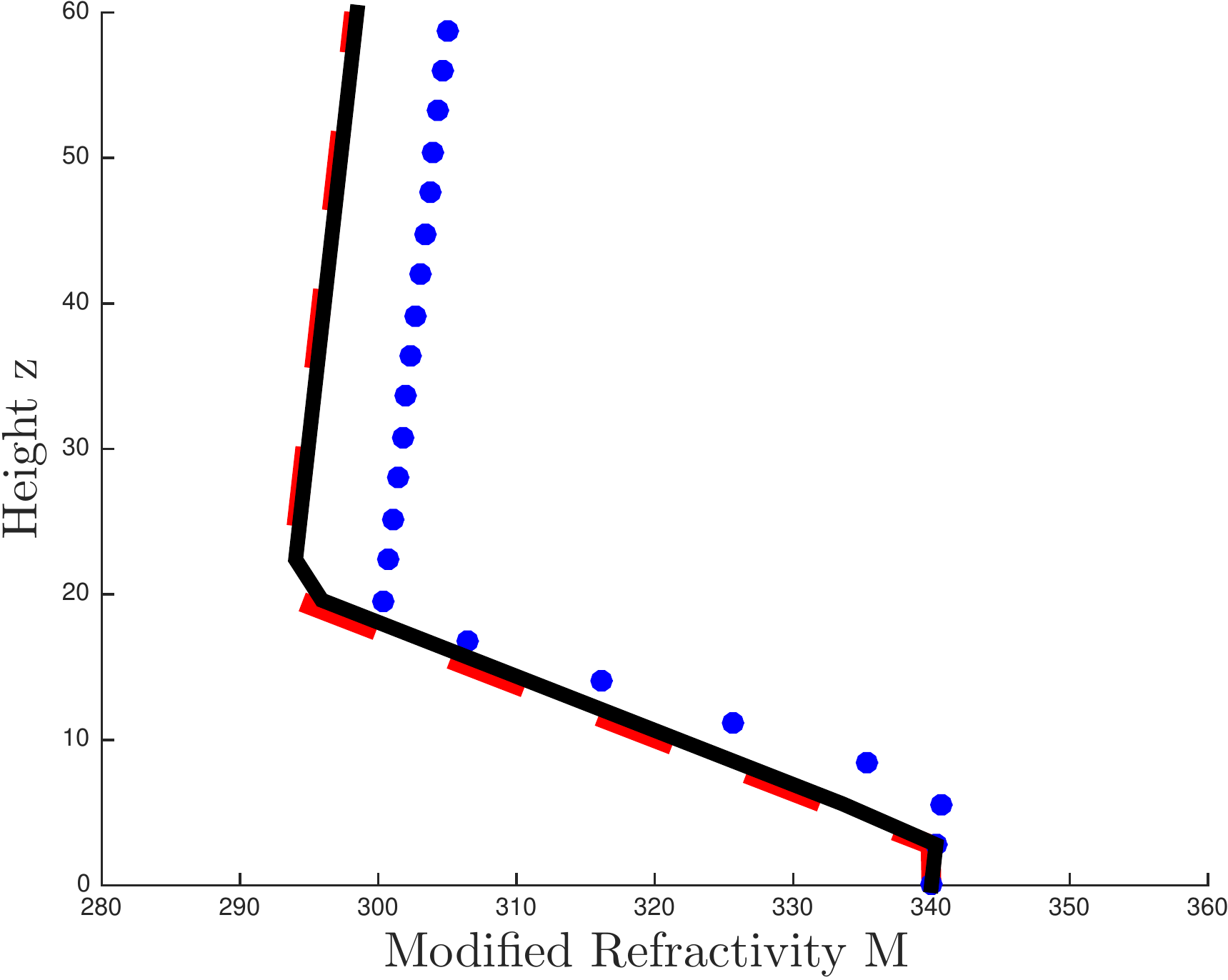}
\label{fig_first_case}}
\hfil
\subfloat[$ \text{RNL2} =  \left(0.07 ; 0.05 \right) $ ]{\includegraphics[width=0.23\textwidth]{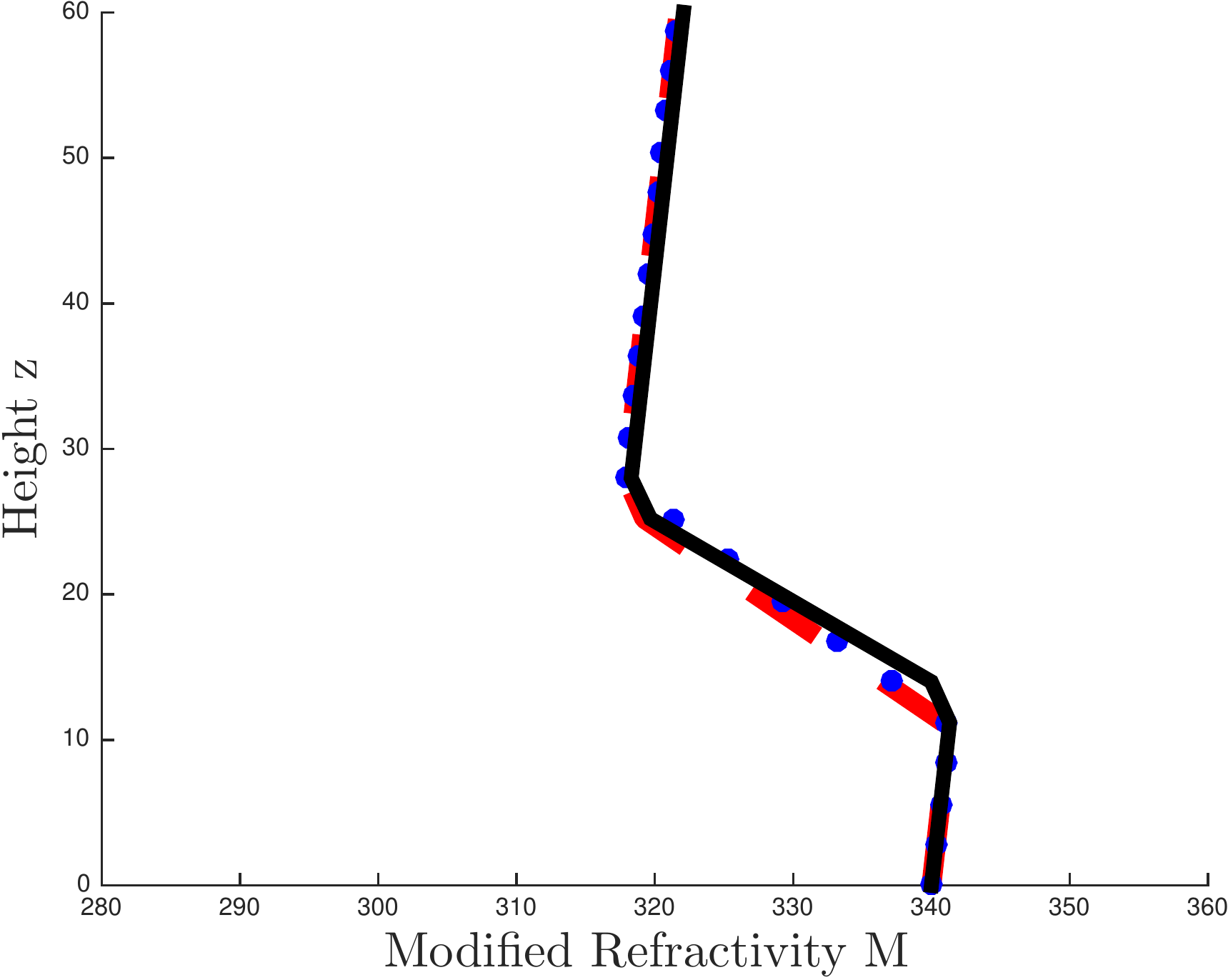}
\label{fig_second_case}}
\hfil
\subfloat[$ \text{RNL2} =  \left(0.44 ; 0.21\right) $ ]{\includegraphics[width=0.23\textwidth]{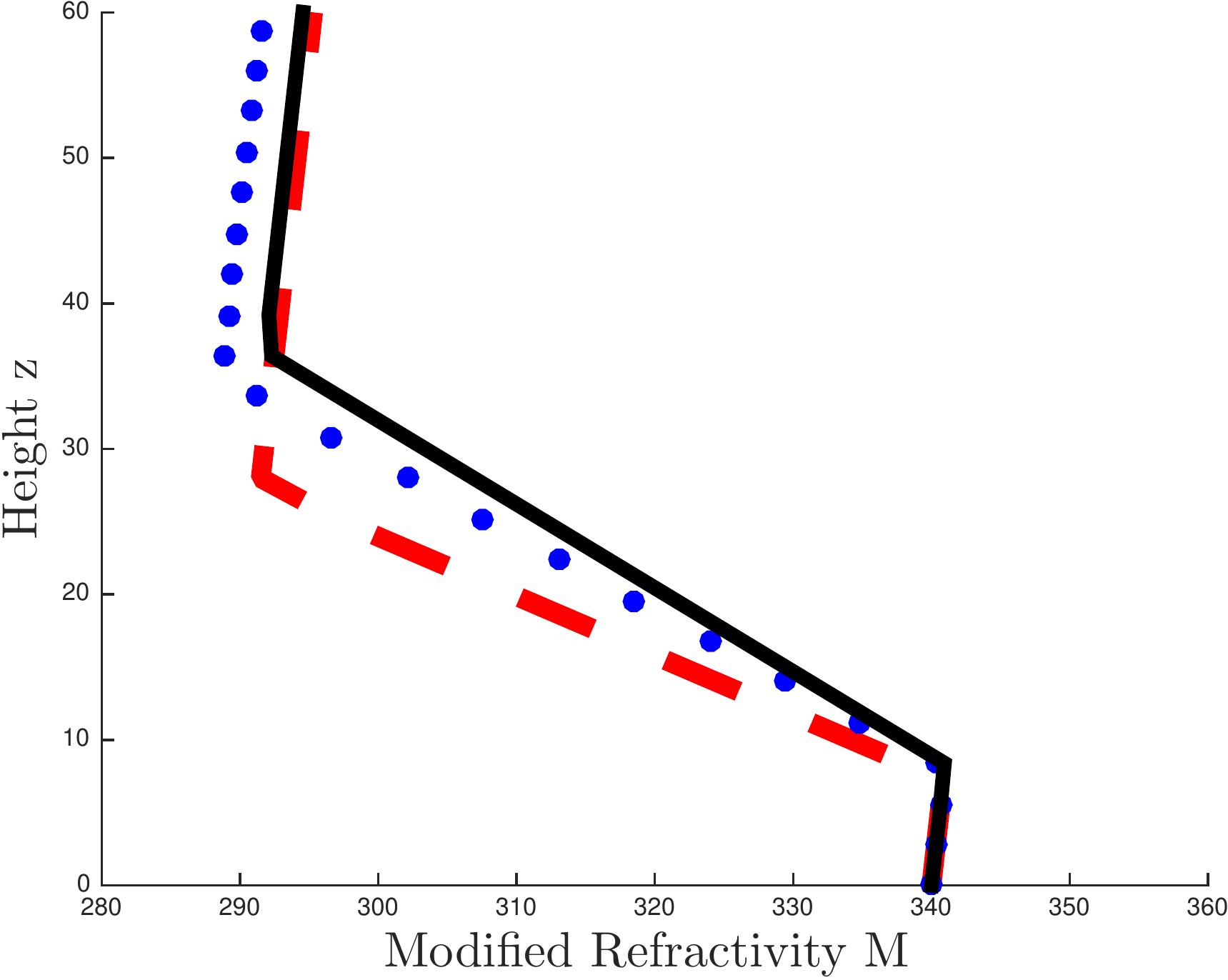}
\label{fig_second_case}}
\hfil
\subfloat[$ \text{RNL2} =  \left(0.17 ; 0.30\right) $ ]{\includegraphics[width=0.23\textwidth]{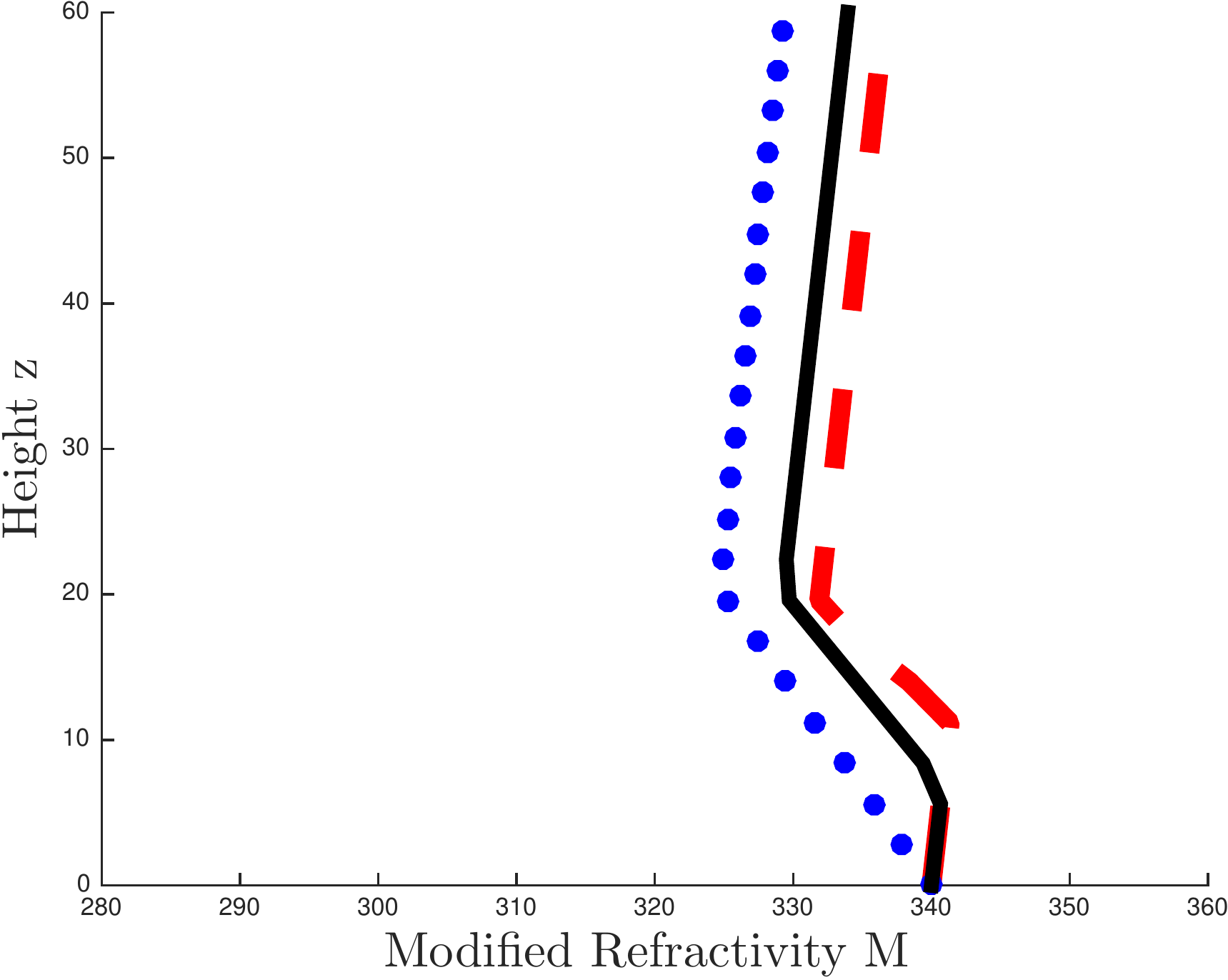}
\label{fig_second_case}}
\vfil

\subfloat[$ \text{RNL2} =  \left(0.39 ; 0.39\right) $ ]{\includegraphics[width=0.23\textwidth]{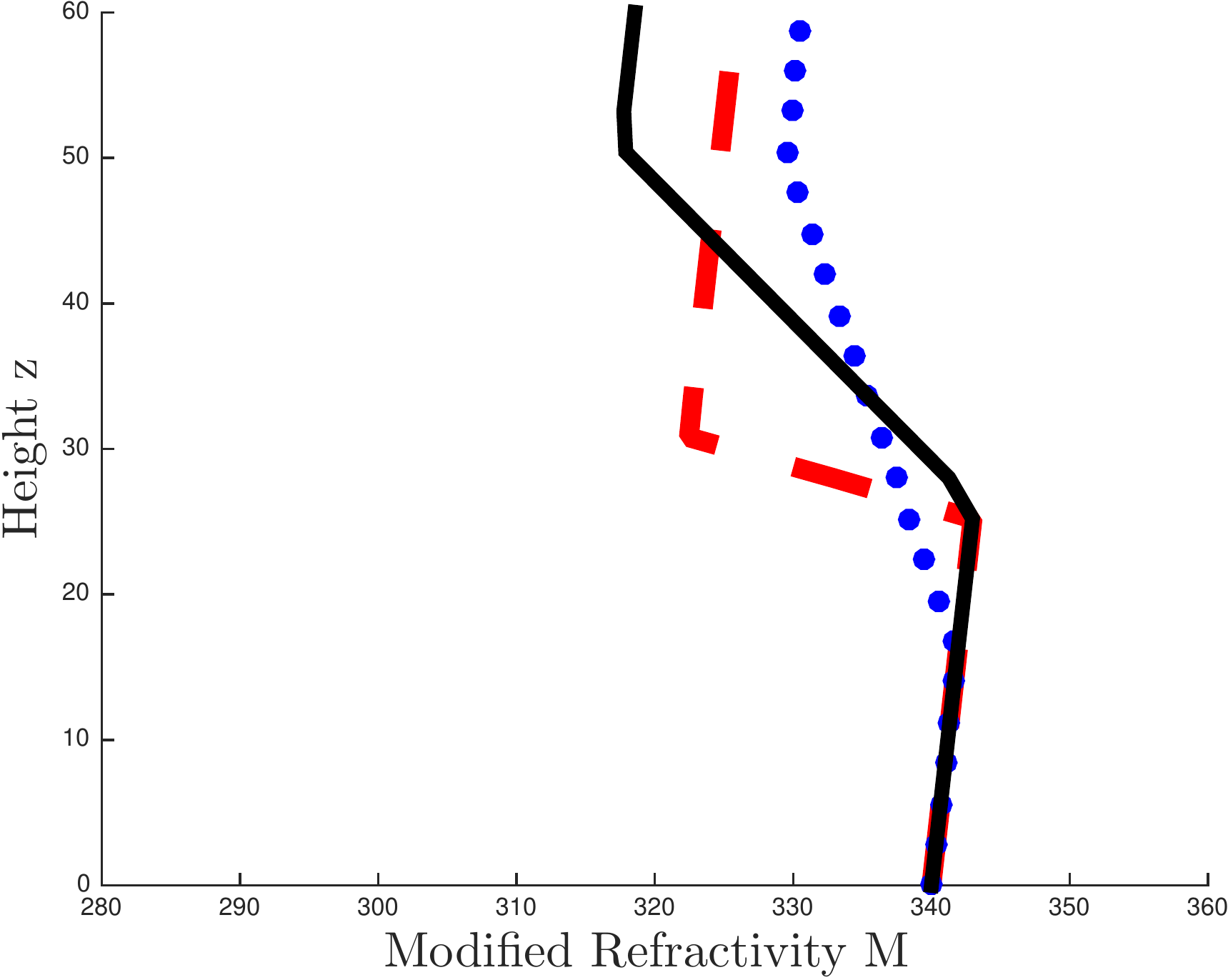}
\label{fig_first_case}}
\hfil
\subfloat[$ \text{RNL2} =  \left(0.04 ; 0.05\right) $ ]{\includegraphics[width=0.23\textwidth]{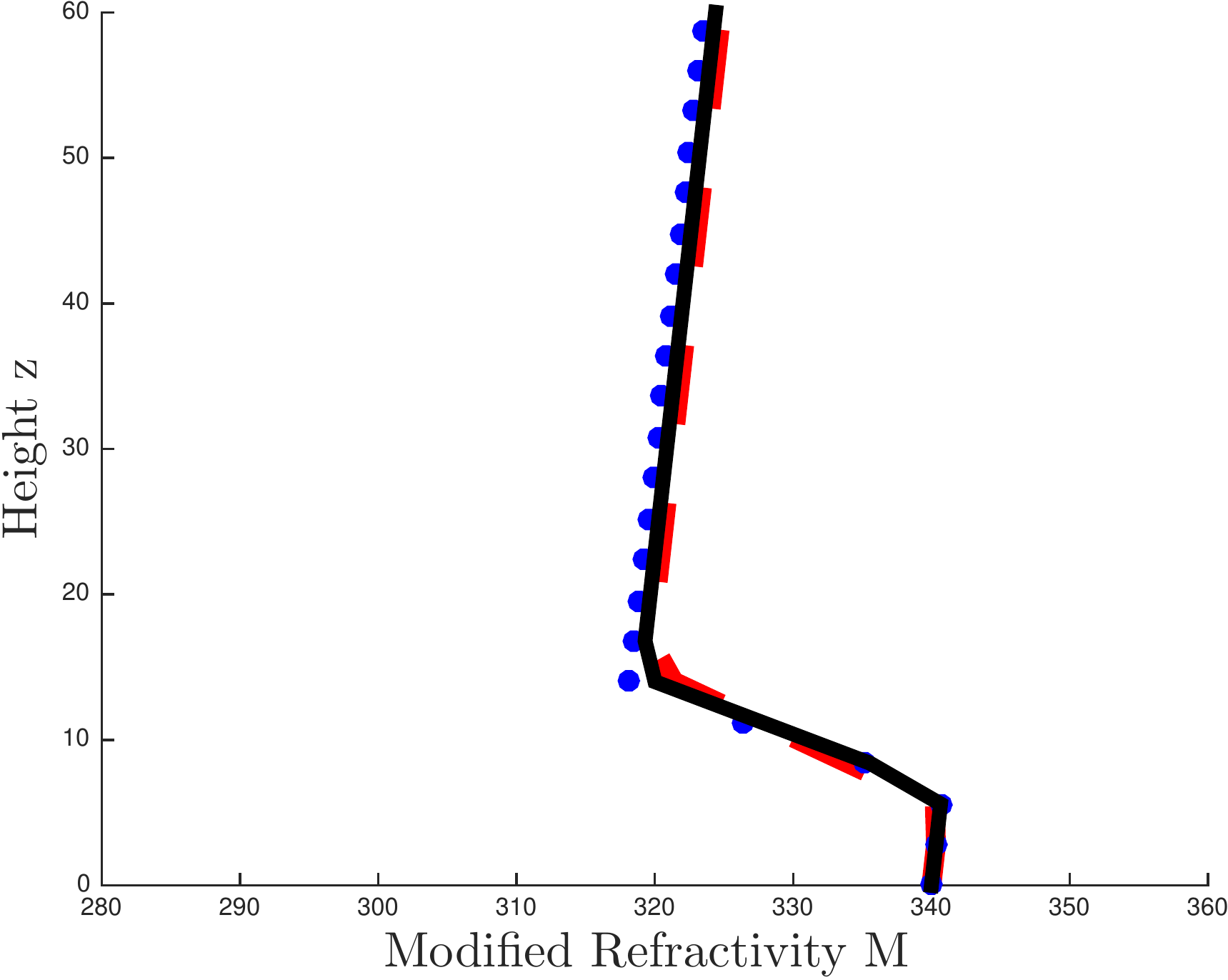}
\label{fig_second_case}}
\hfil
\subfloat[$ \text{RNL2} =  \left(1.12 ; 0.12\right) $ ]{\includegraphics[width=0.23\textwidth]{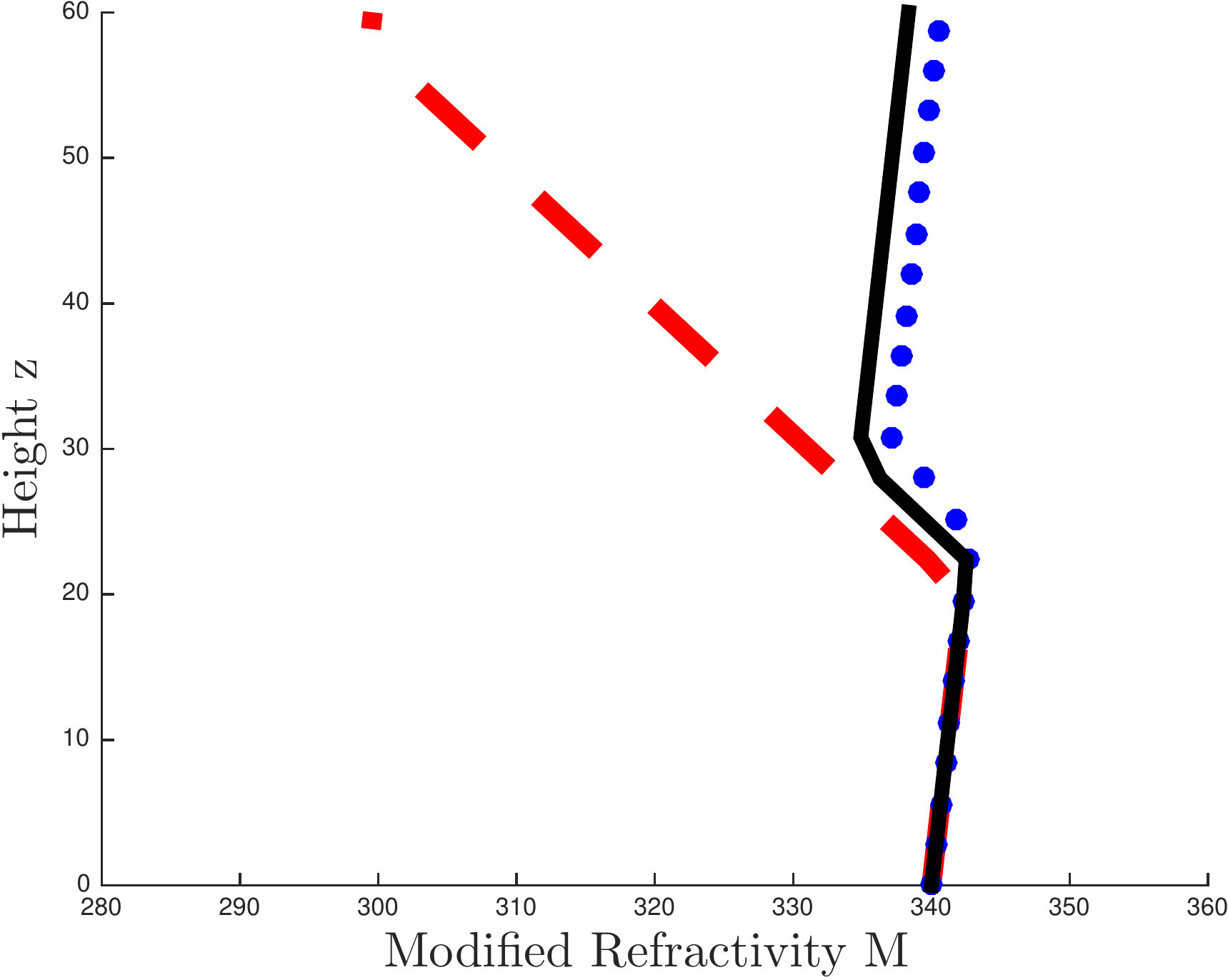}
\label{fig:1k}}
\hfil
\subfloat[$ \text{RNL2} =  \left(0.45 ; 0.17\right) $ ]{\includegraphics[width=0.23\textwidth]{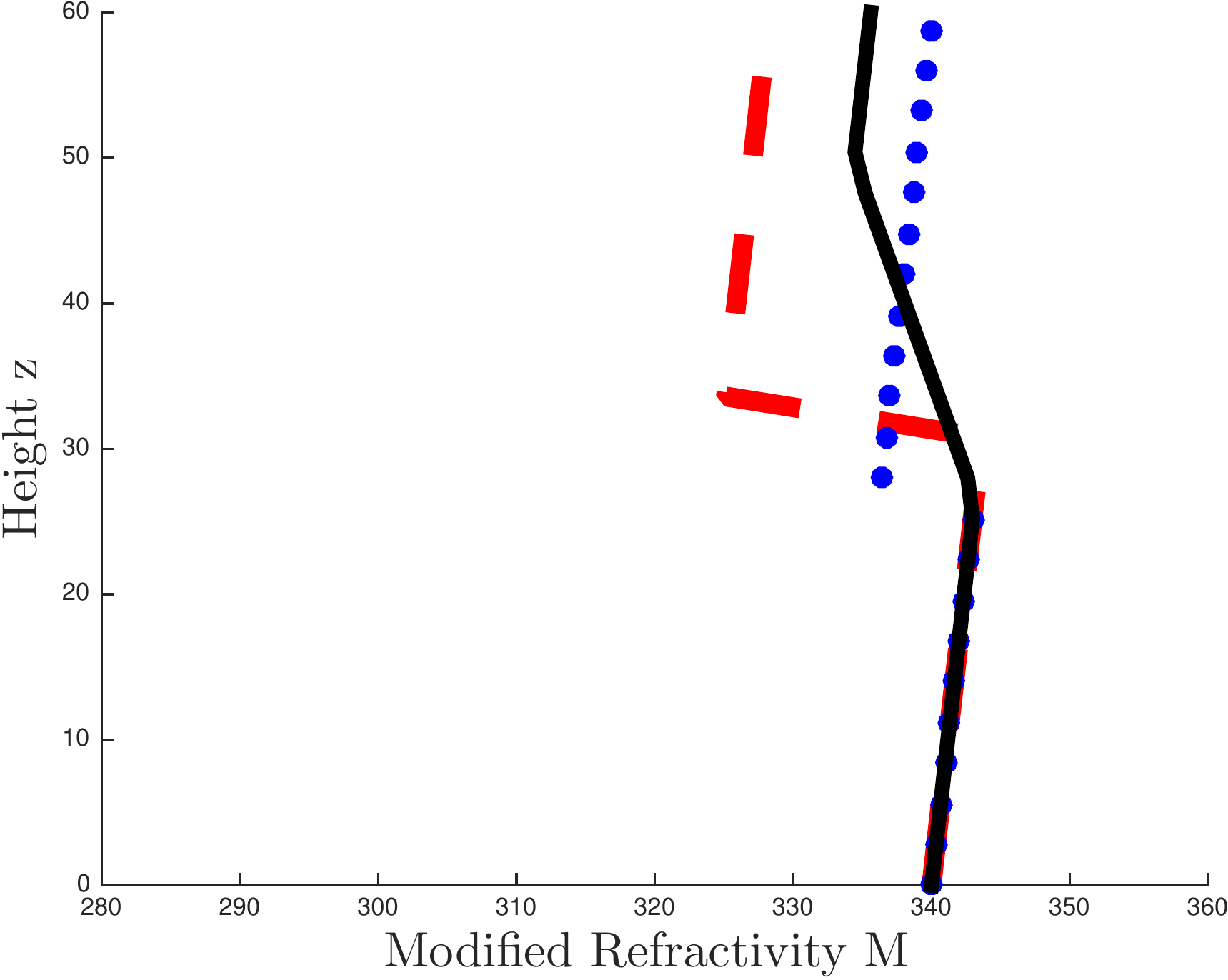}
\label{fig_second_case}}
\vfil

\subfloat[$ \text{RNL2} =  \left(0.76 ; 0.16\right) $ ]{\includegraphics[width=0.23\textwidth]{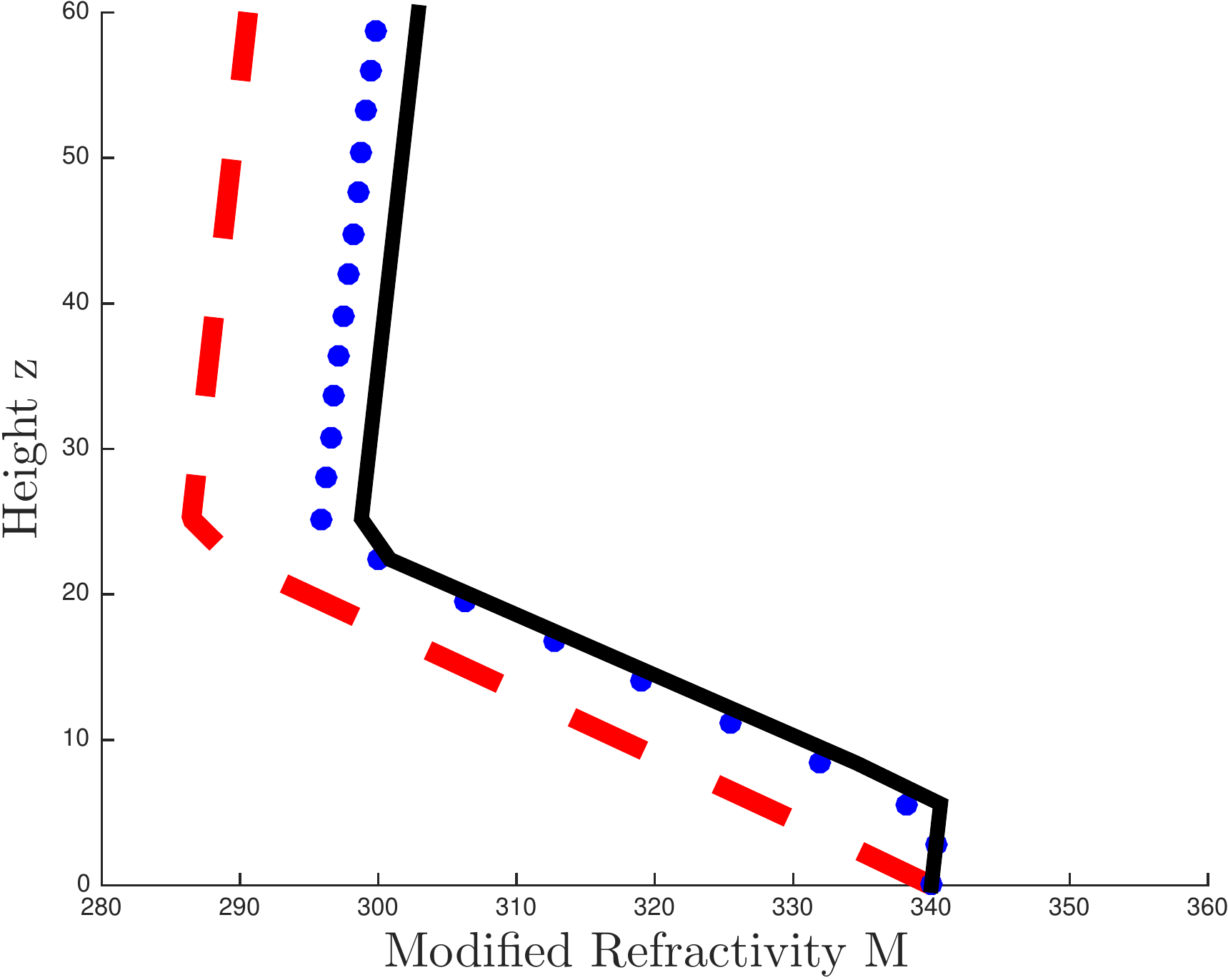}
\label{fig_first_case}}
\hfil
\subfloat[$ \text{RNL2} =  \left(0.58 ; 0.34 \right) $ ]{\includegraphics[width=0.23\textwidth]{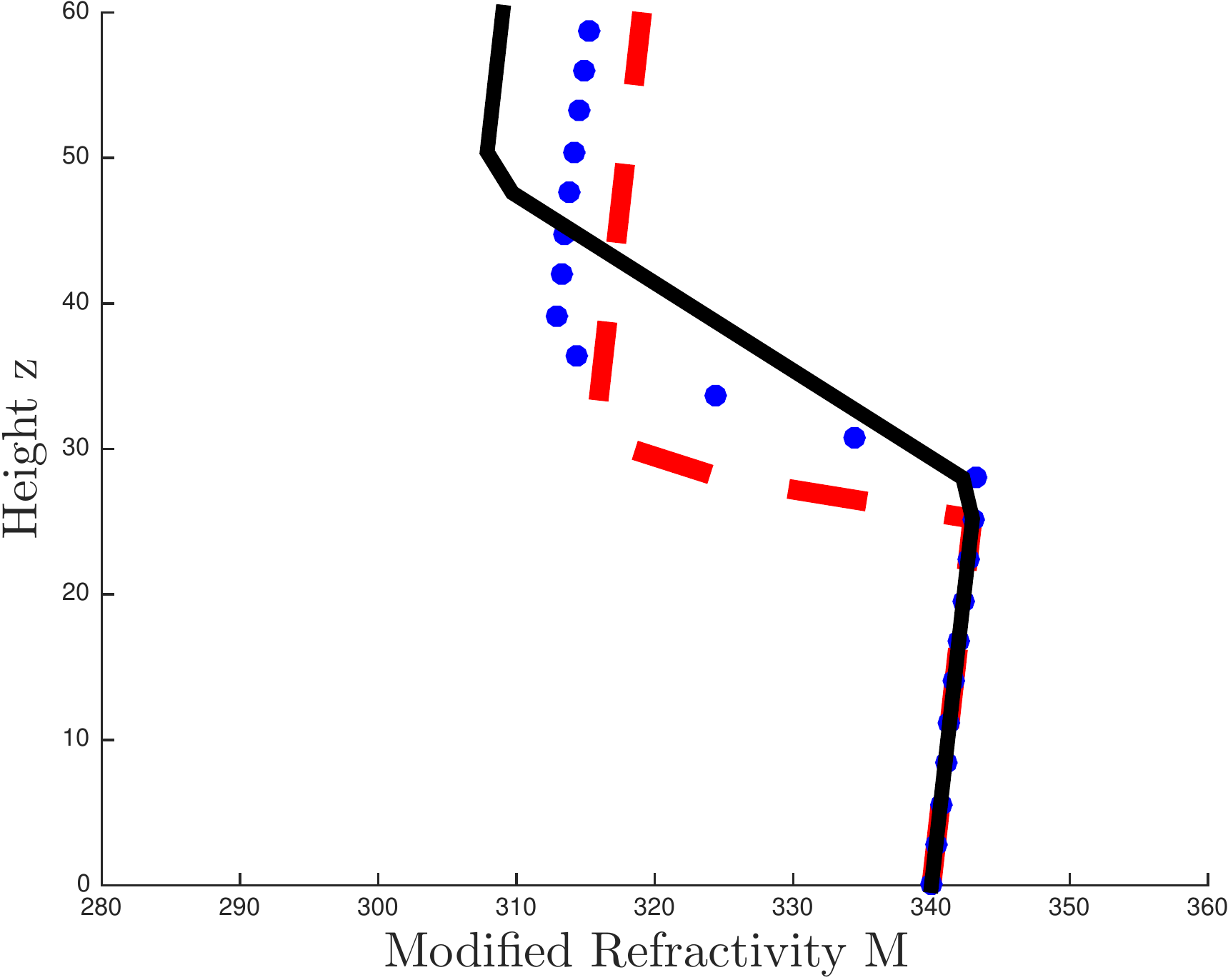}
\label{fig_second_case}}
\hfil
\subfloat[$ \text{RNL2} =  \left(0.10 ; 0.11 \right) $ ]{\includegraphics[width=0.23\textwidth]{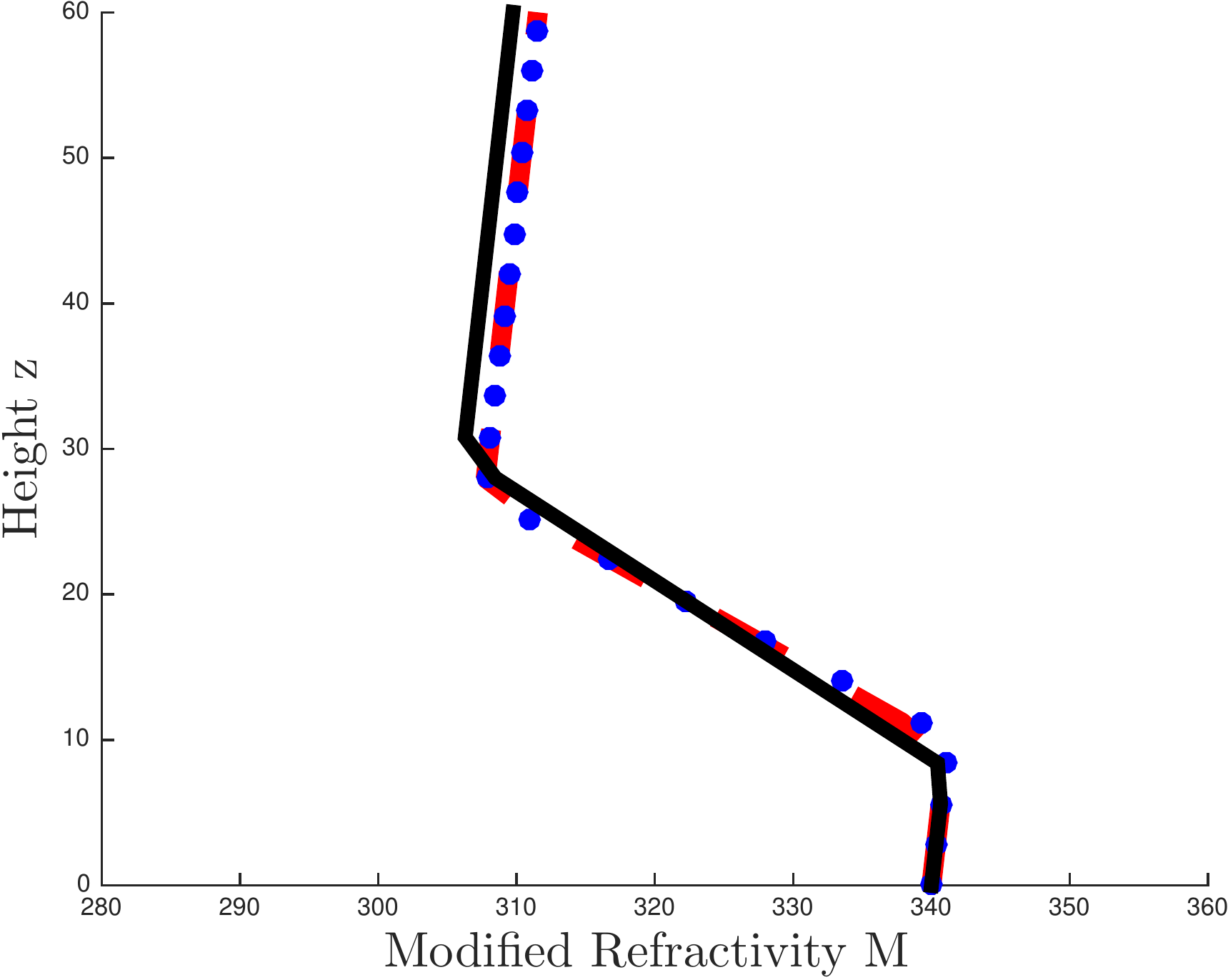}
\label{fig_second_case}}
\hfil
\subfloat[$ \text{RNL2} =  \left(0.02 ; 0.08\right) $ ]{\includegraphics[width=0.23\textwidth]{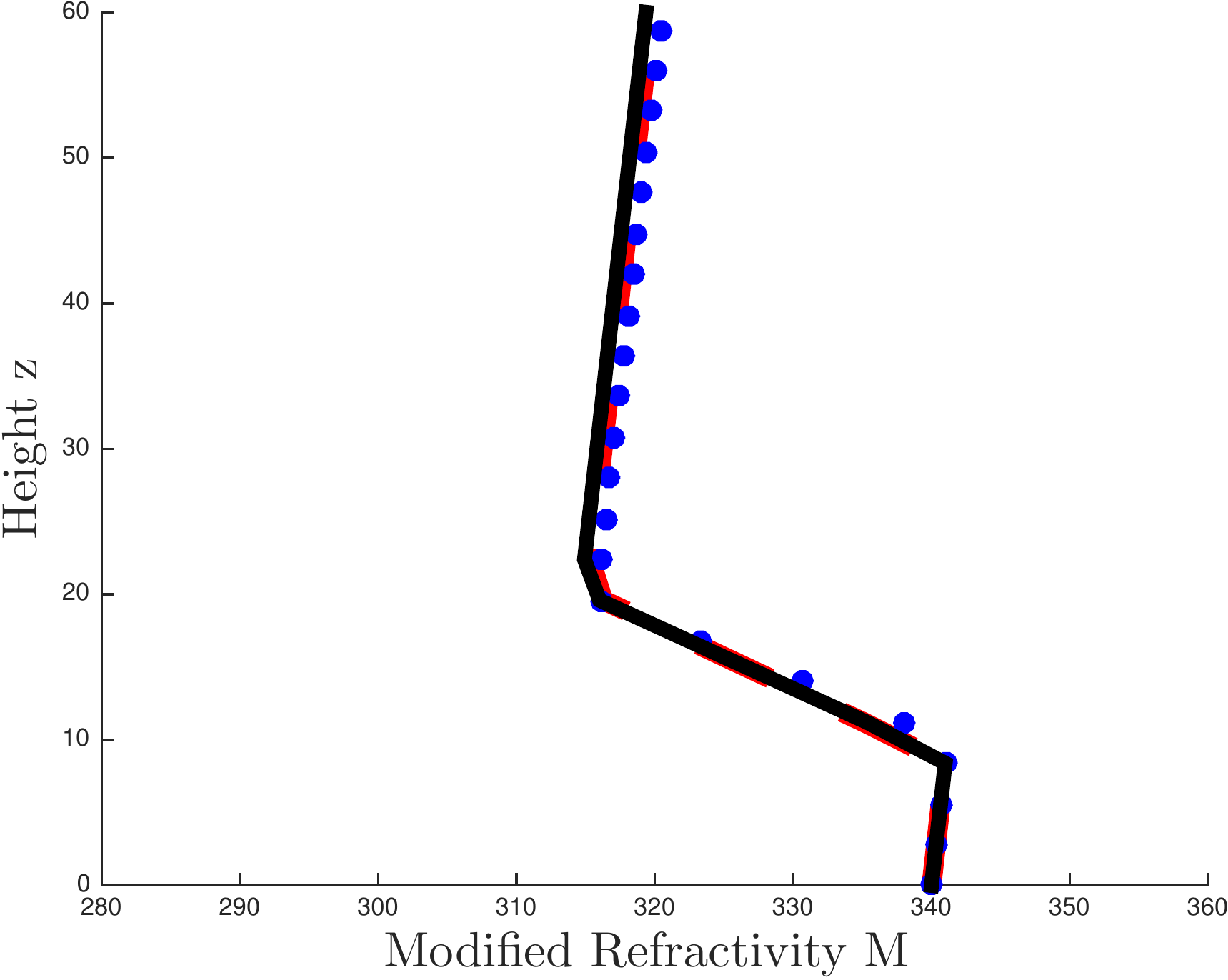}
\label{fig_second_case}}
\vfil

\subfloat[$ \text{RNL2} =  \left(0.31 ; 0.19\right) $ ]{\includegraphics[width=0.23\textwidth]{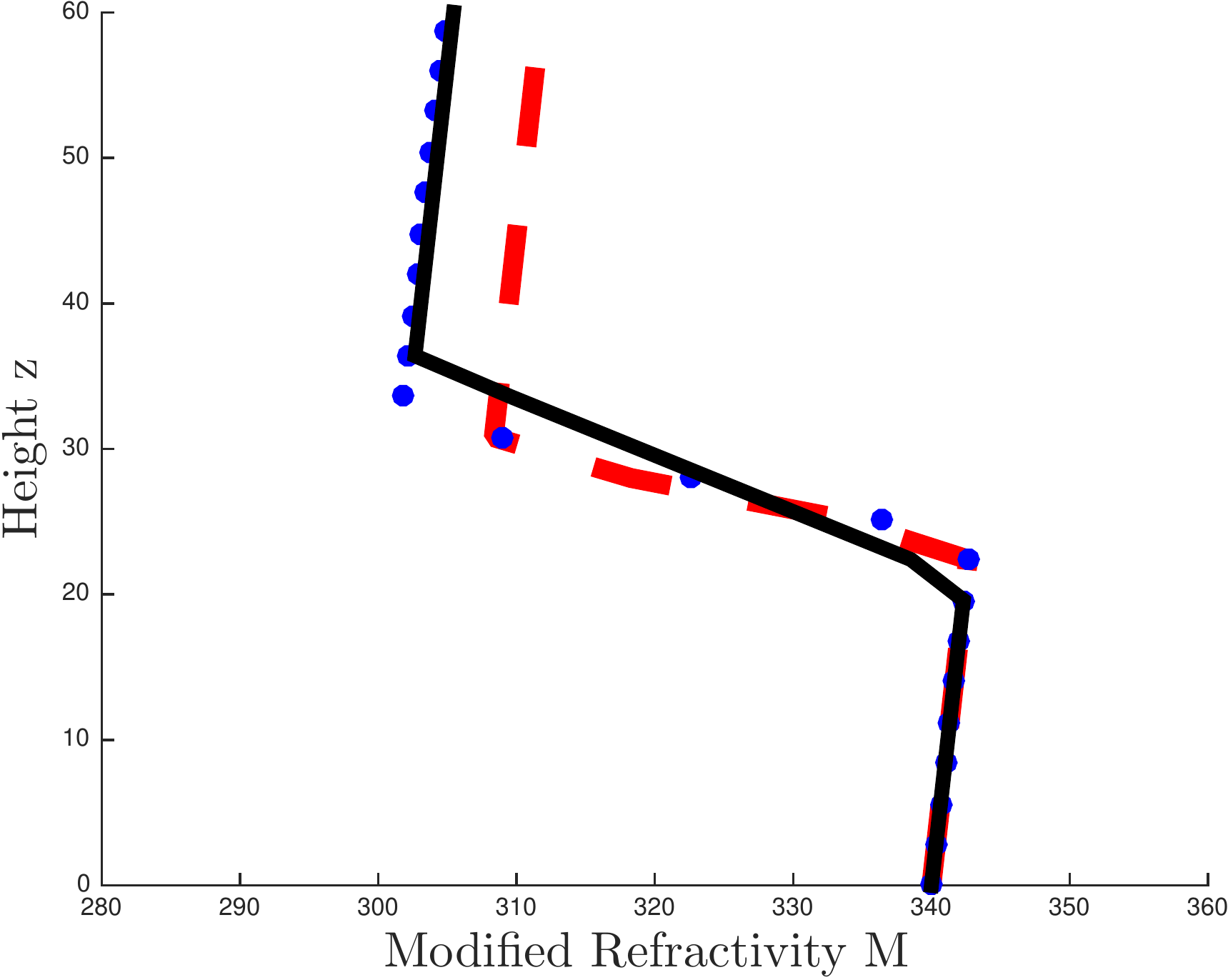}
\label{fig_first_case}}
\hfil
\subfloat[$ \text{RNL2} =  \left(0.07 ; 0.26\right) $ ]{\includegraphics[width=0.23\textwidth]{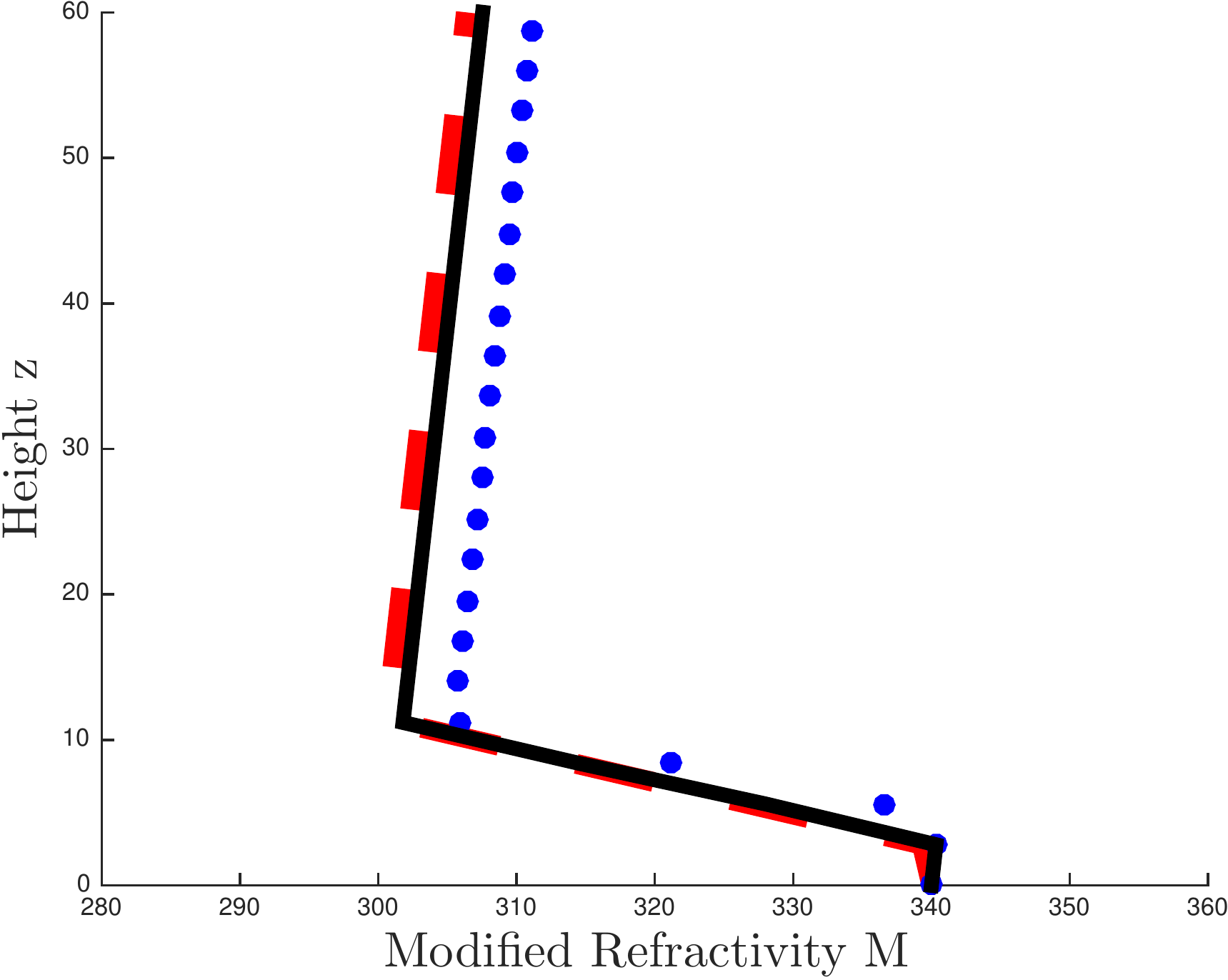}
\label{fig_second_case}}
\hfil
\subfloat[$ \text{RNL2} =  \left(0.12 ; 0.25\right) $ ]{\includegraphics[width=0.23\textwidth]{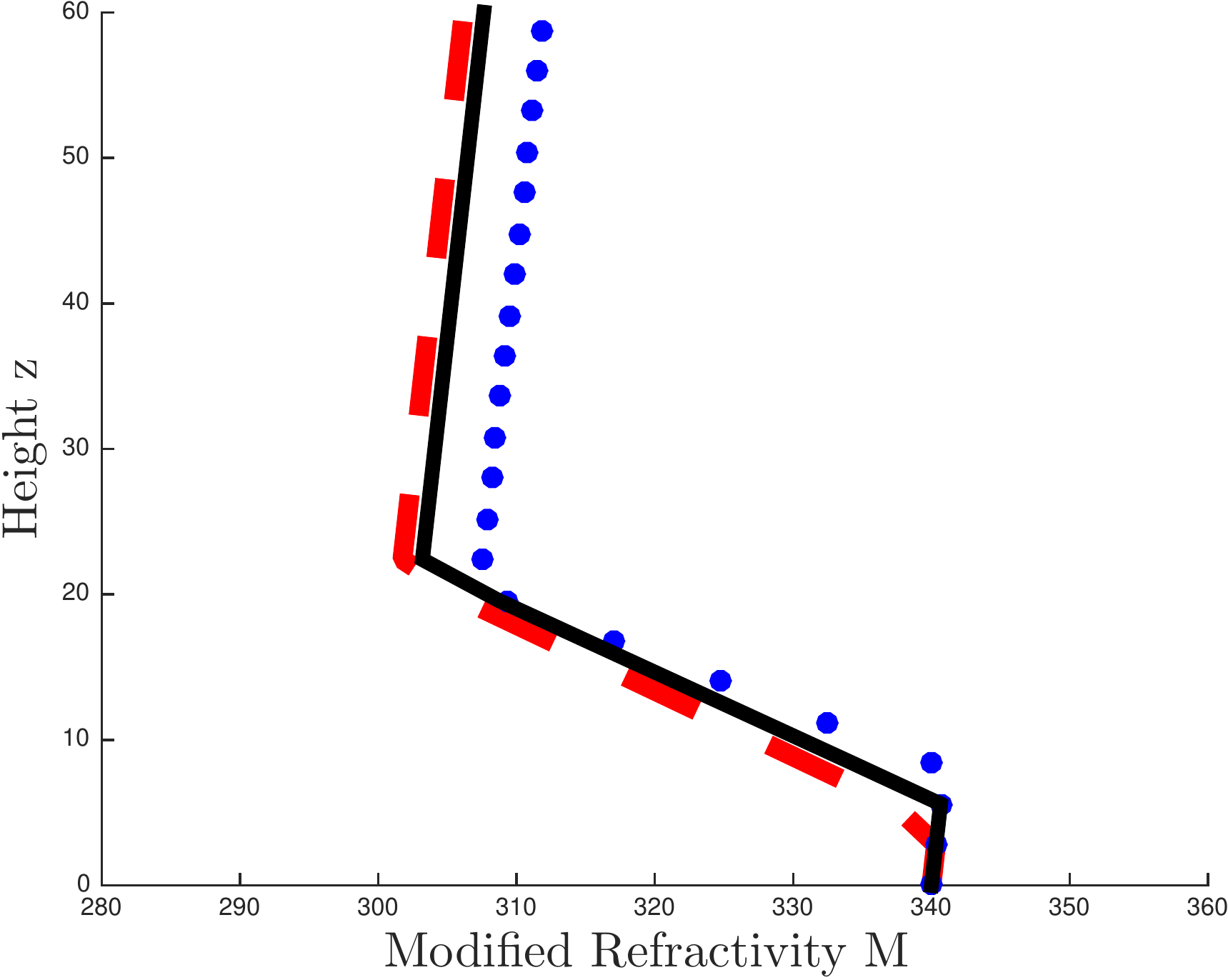}
\label{fig_second_case}}
\hfil
\subfloat[$ \text{RNL2} =  \left(0.52 ; 0.49\right) $ ]{\includegraphics[width=0.23\textwidth]{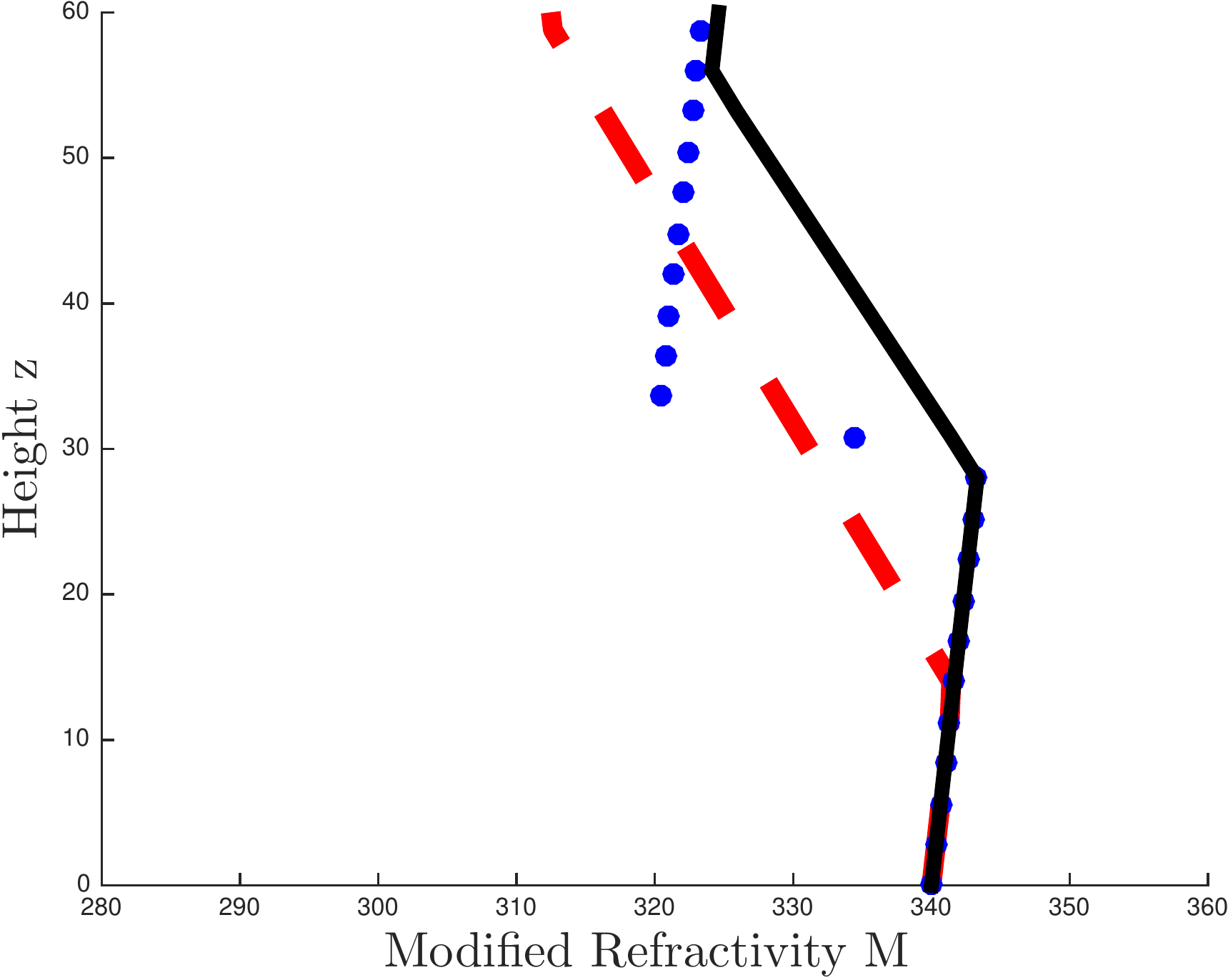}
\label{fig_second_case}}
\vfil

\caption{ Results for algorithm 1 and 2 on experiment 1.
The solid back lines are the true profile that generated the data: $n_{\text{true}}(z)$, the red dashed plot are the inverted profiles obtained using algorithm 1, and the dotted blue lines are the inverted profile obtained using algorithm 2. For each case, the RNL2 score is given below the plots. The first number is the RNL2 score of algorithm 1, and the second is the score of algorithm 2.}
\label{fig:sim1}
\end{figure*}

For the test cases, we fix $s_1=0.118$ M-unit/m\cite{gingras1997electromagnetic}, which is consistent with the mean over the whole of the United States, and has been observed to have very little variability~\cite{rogers1998demonstration}. In order to produce unbiased test cases, we generate twenty synthetic refractivity profiles by randomly sampling the parameters in the following way:
\begin{align*}
z_b \sim U(0, 30), && M_d \sim U(0, 50), && t_h\sim U(0, 30) \ .
\end{align*}
These parameters are consistent with low altitude surface-based ducts. The realizations are shown in Fig.~\ref{fig:sim1}. 
In terms of physical domain, this means that we are observing data from $0$ to $30$ m in altitude, and are trying to invert parameters that define non-standard refractivity profiles from $0$ to $60$ m. 
We attempt the inverse problem of identifying the refractivity profiles from the observational data. We use the algorithms described in section \ref{description}. For algorithm 1, we set $\alpha = 10^{-4}$. For algorithm 2, we set $\alpha = 3 \times 10^{-4}$, and $\beta =3  \times 10^{-3}$. These parameters were obtained by minimizing the RNL2 score over a separate training set and were found to be insensitive to the noise level. The result of Alg.~1 and 2 on the twenty refractivity profiles are shown in Fig.~\ref{fig:sim1}. 

\subsection{Experiment 2: Simulated data originating from a trilinear, horizontally varying index of refraction.}
We present results on simulated data originating from a horizontally varying index of refraction. 
This experiment violates the assumption that $n(x,z)$ is horizontally constant, thus one cannot expect in general to find an approximate solution of the form~(\ref{eq:approx}) which induces the low-rank decomposition of the solution of~(\ref{eq:Helmholtz}). It is this low-rank structure that is exploited by Alg.~1 and~2 and therefore we expect that their performance would degrade in this experiment.

\begin{figure*}[!t]
\centering
\subfloat[ $ \text{RNL2} =  \left( 0.29 ; 1.65 \right) $  ]{\includegraphics[width=0.23\textwidth]{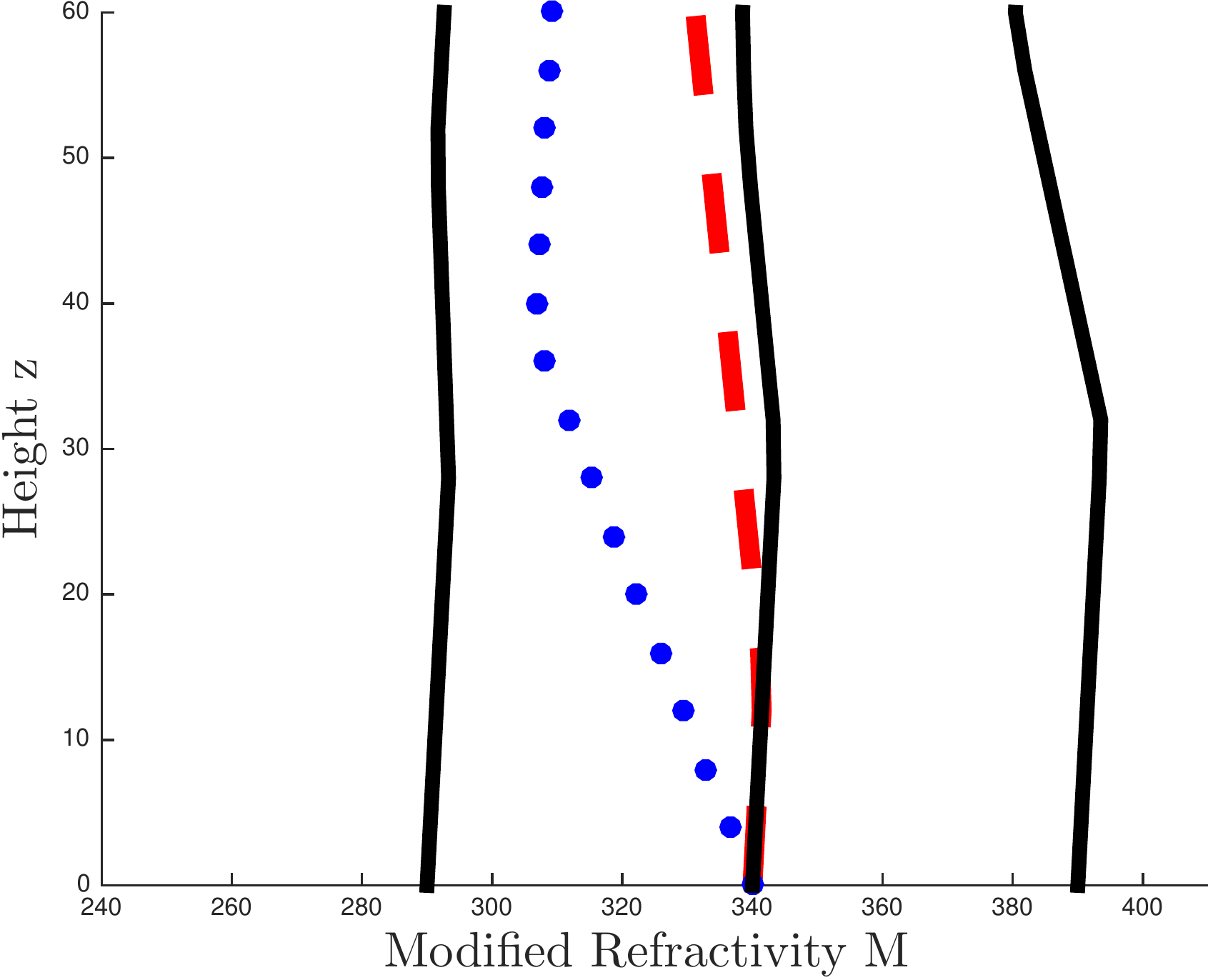}
\label{fig:2a}}
\hfil
\subfloat[$ \text{RNL2} =  \left(0.12; 0.14 \right)$ ]{\includegraphics[width=0.23\textwidth]{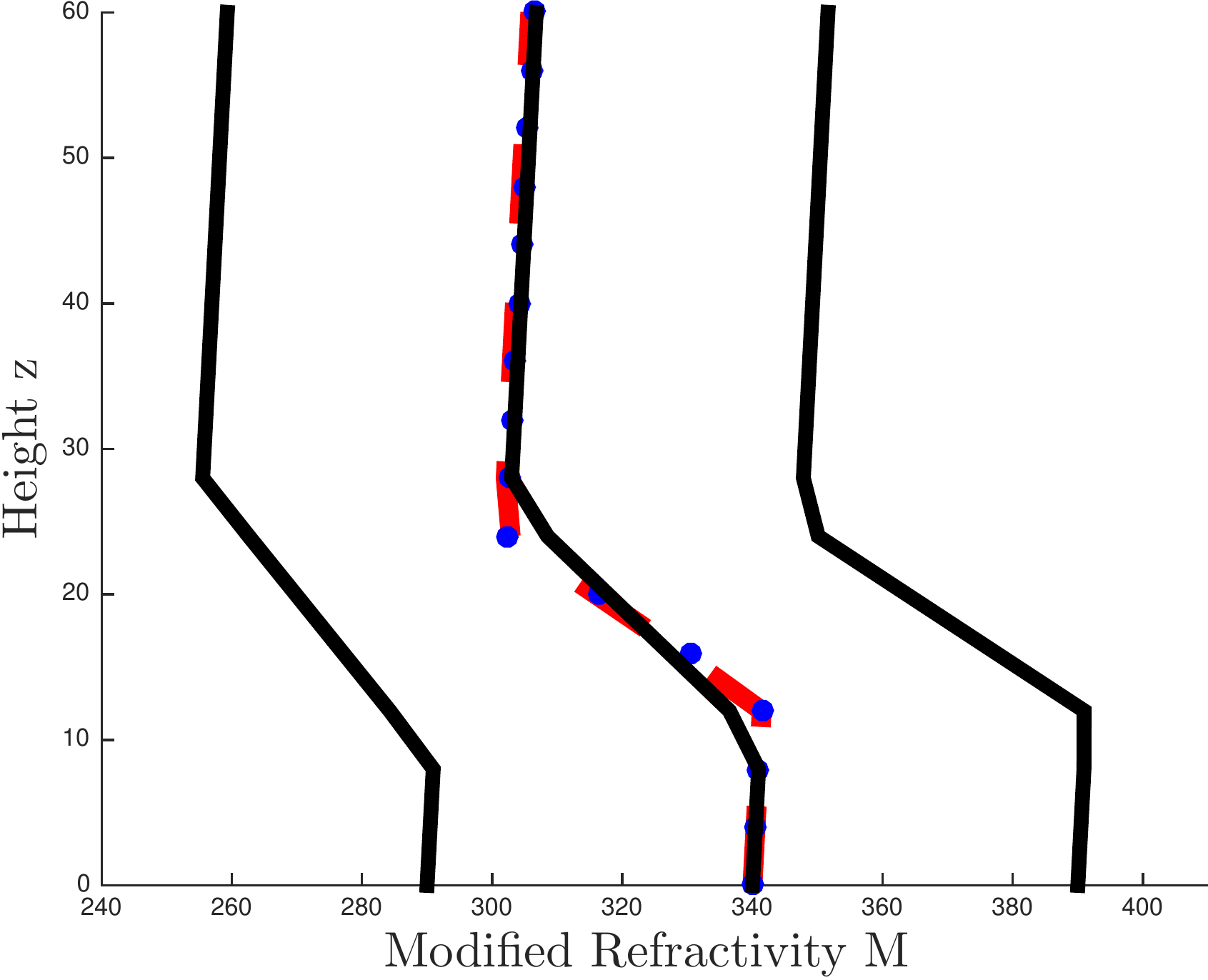}
\label{fig_second_case}}
\hfil
\subfloat[$ \text{RNL2} =  \left(0.27; 0.23\right)$]{\includegraphics[width=0.23\textwidth]{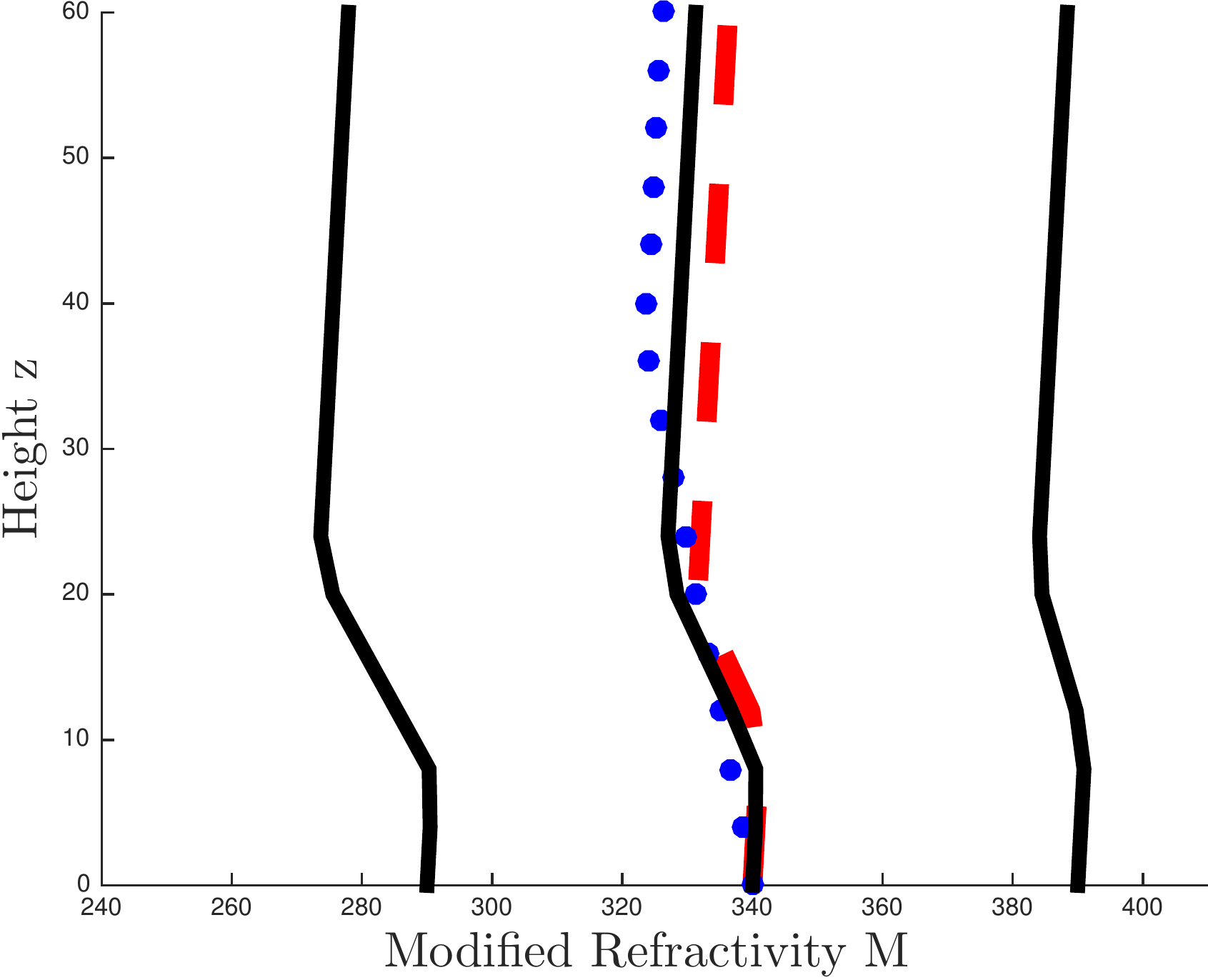}
\label{fig_second_case}}
\hfil
\subfloat[$ \text{RNL2} =  \left(0.06; 0.38 \right)$]{\includegraphics[width=0.23\textwidth]{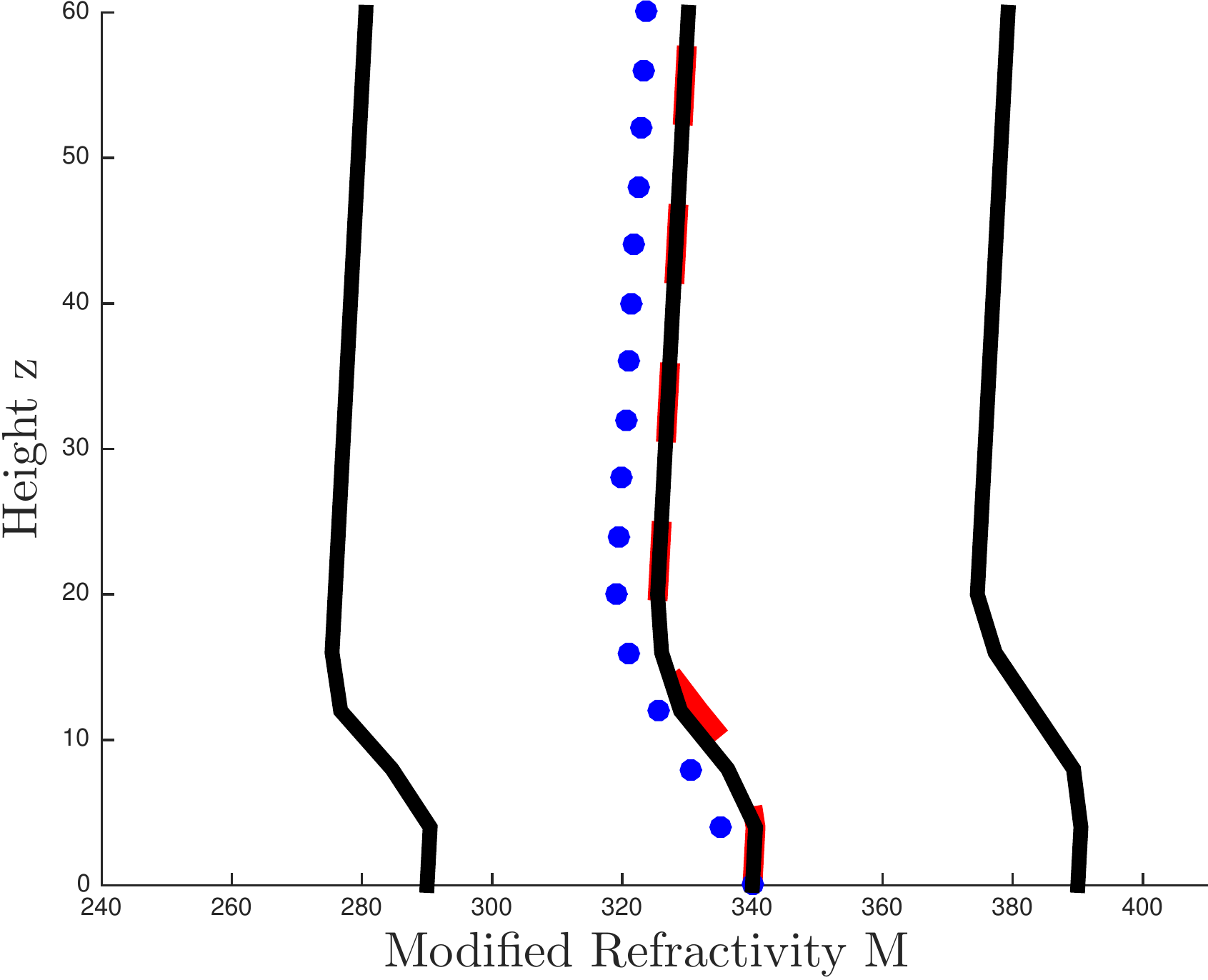}
\label{fig_second_case}}
\vfil

\subfloat[$ \text{RNL2} =  \left(0.81; 0.40 \right)$]{\includegraphics[width=0.23\textwidth]{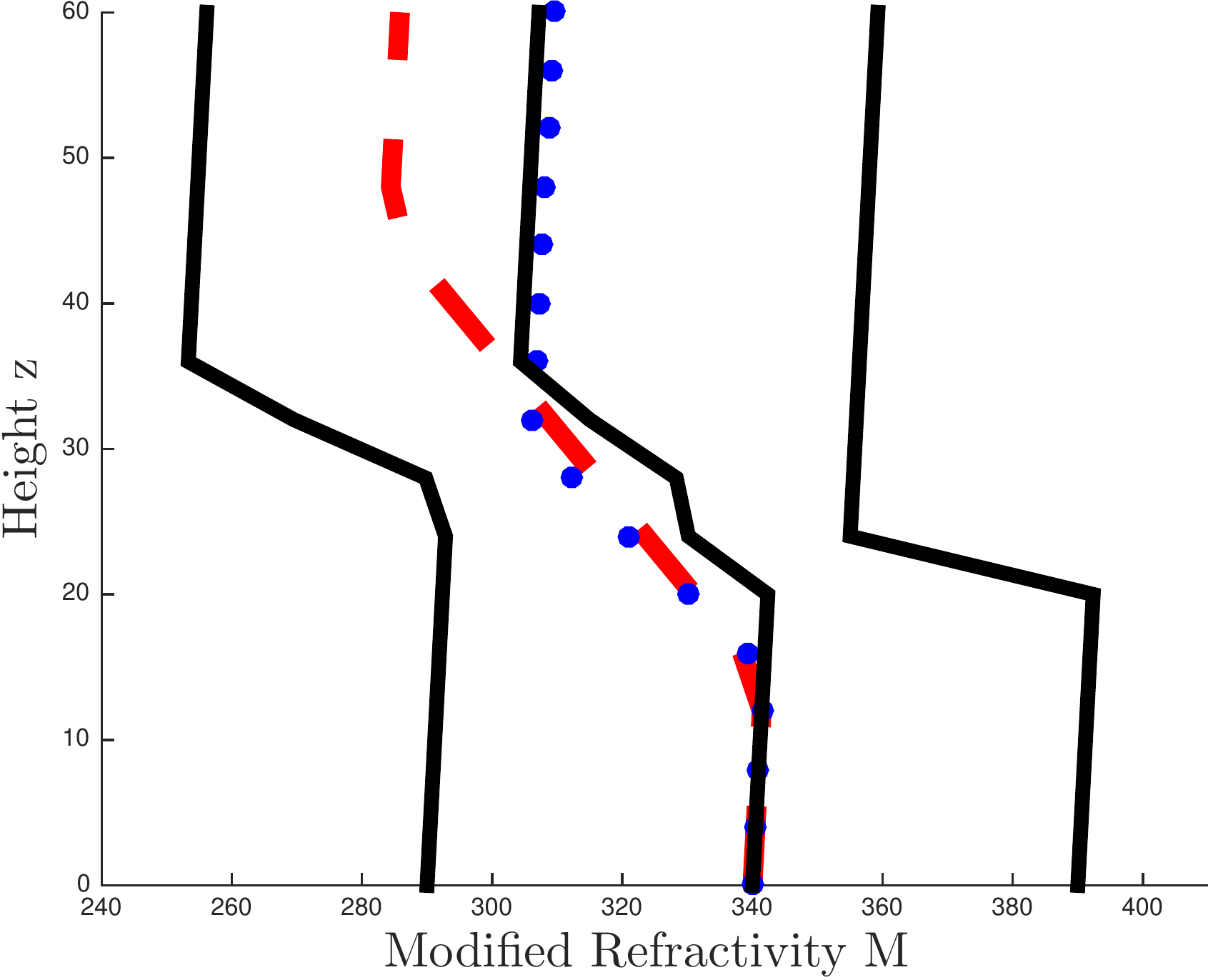}
\label{fig_first_case}}
\hfil
\subfloat[$ \text{RNL2} =  \left(0.52 ; 0.73 \right)$ ]{\includegraphics[width=0.23\textwidth]{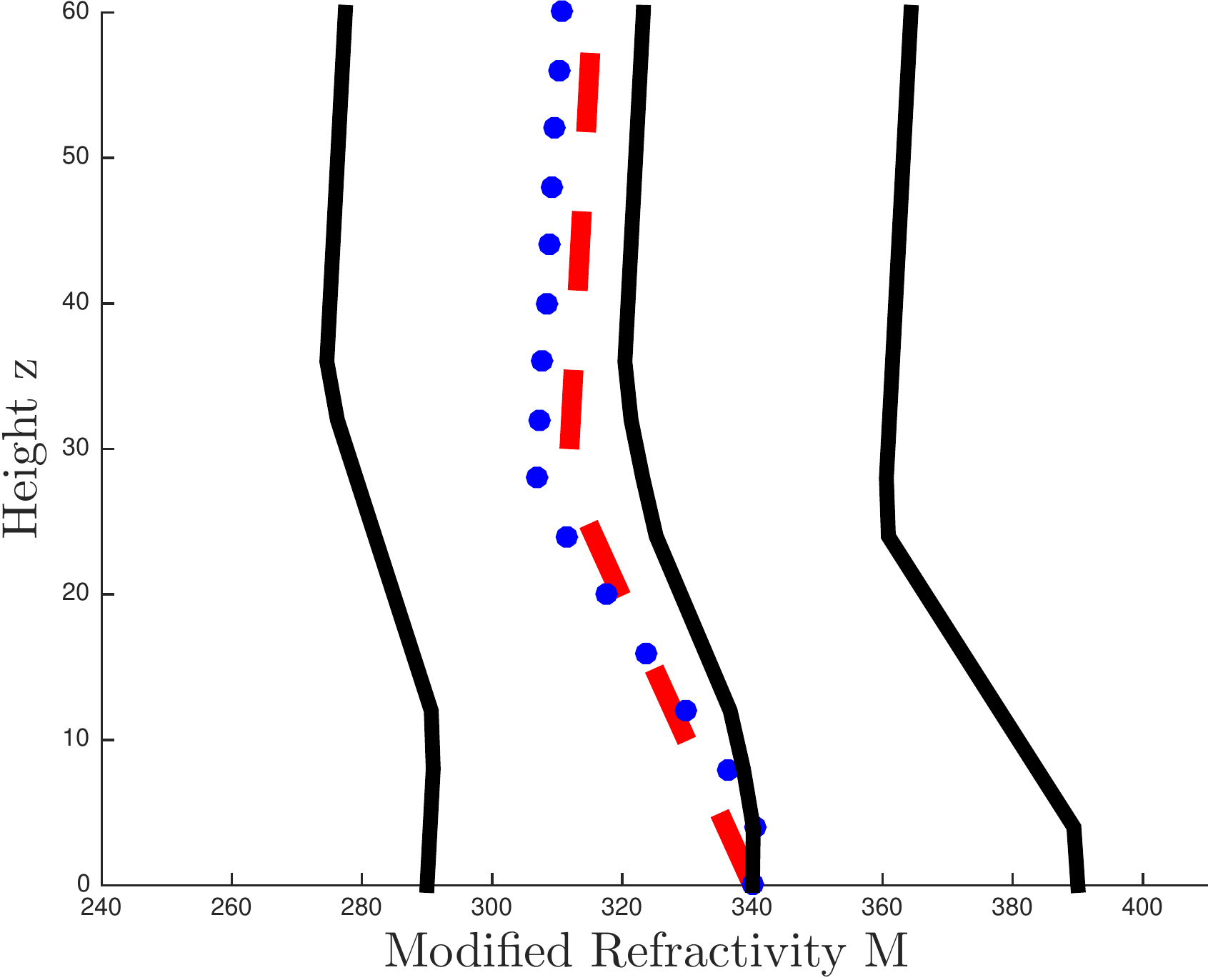}
\label{fig_second_case}}
\hfil
\subfloat[$ \text{RNL2} =  \left(0.40 ; 0.18 \right)$]{\includegraphics[width=0.23\textwidth]{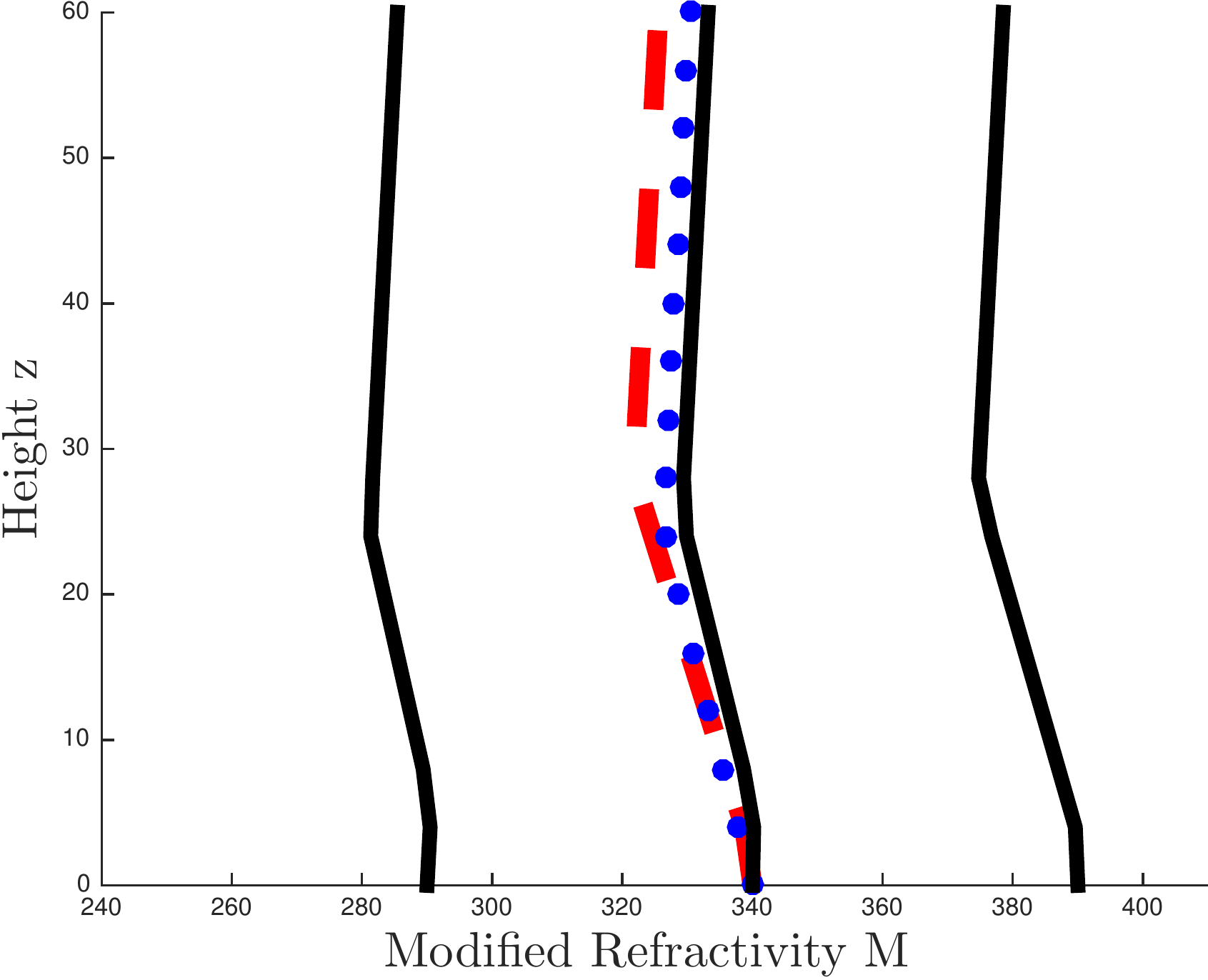}
\label{fig_second_case}}
\hfil
\subfloat[$ \text{RNL2} =  \left(0.11 ; 0.17 \right)$]{\includegraphics[width=0.23\textwidth]{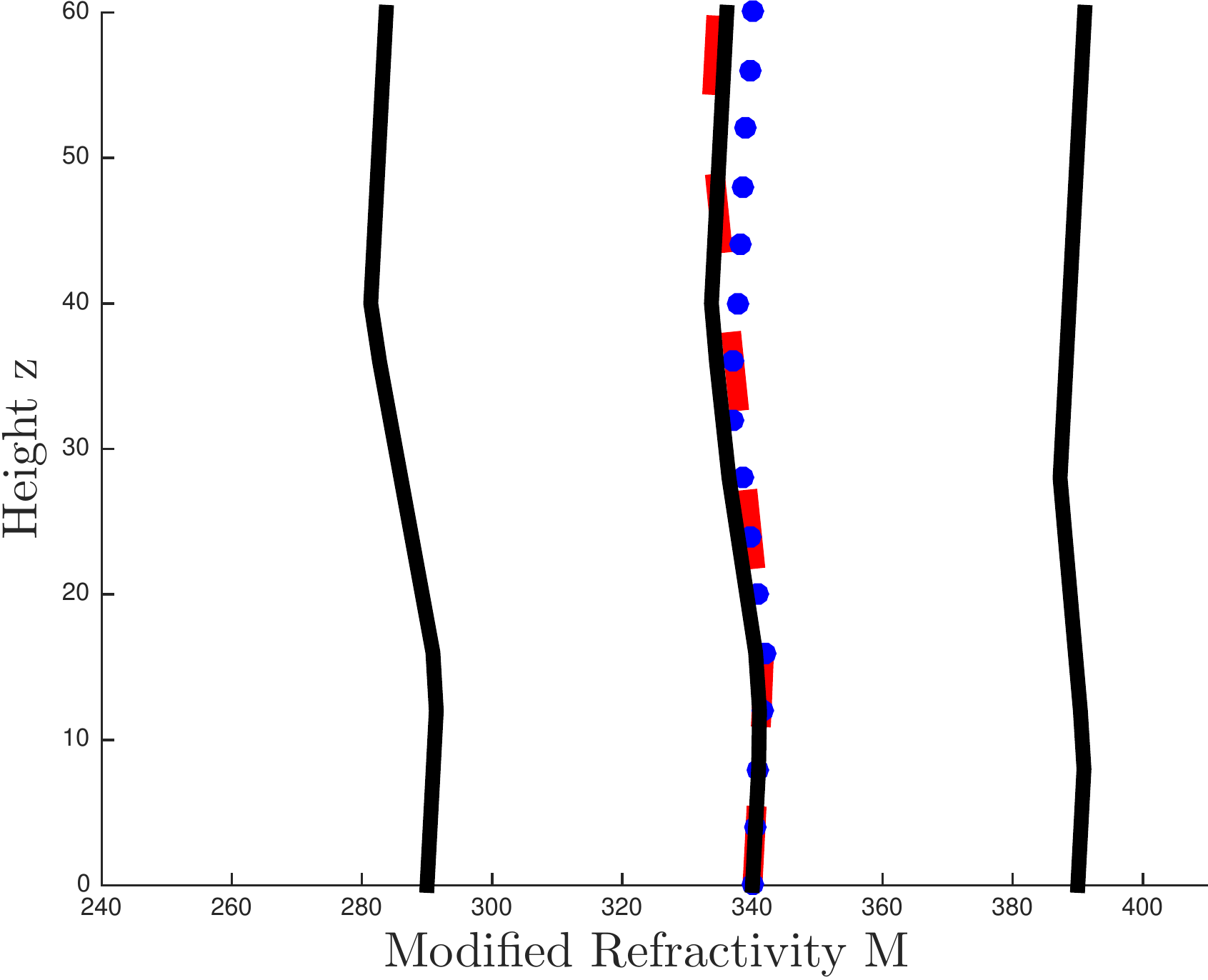}
\label{fig_second_case}}
\vfil

\subfloat[$ \text{RNL2} =  \left(0.34 ; 0.07 \right)$]{\includegraphics[width=0.23\textwidth]{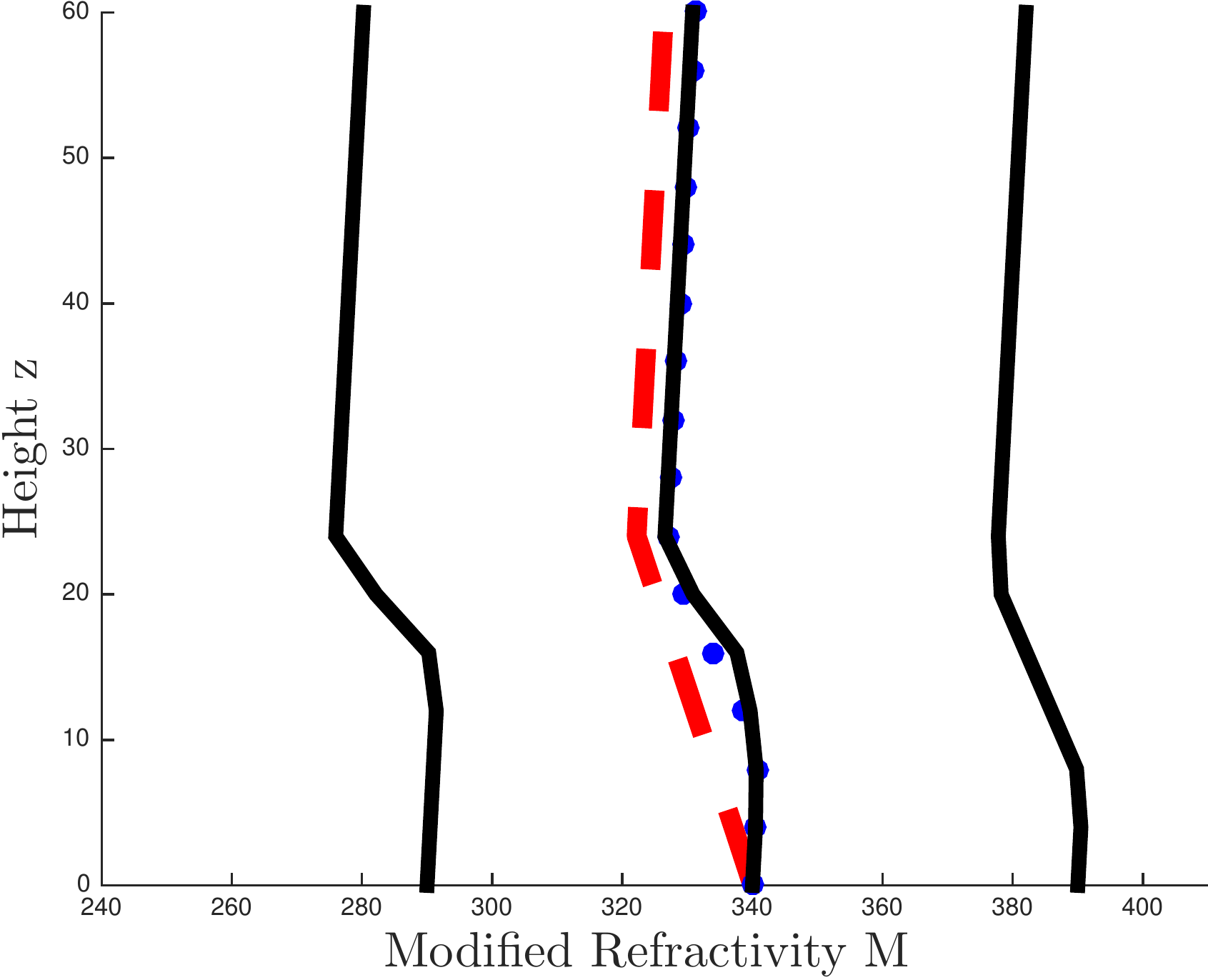}
\label{fig_first_case}}
\hfil
\subfloat[$ \text{RNL2} =  \left( 0.10 ; 0.11 \right)$ ]{\includegraphics[width=0.23\textwidth]{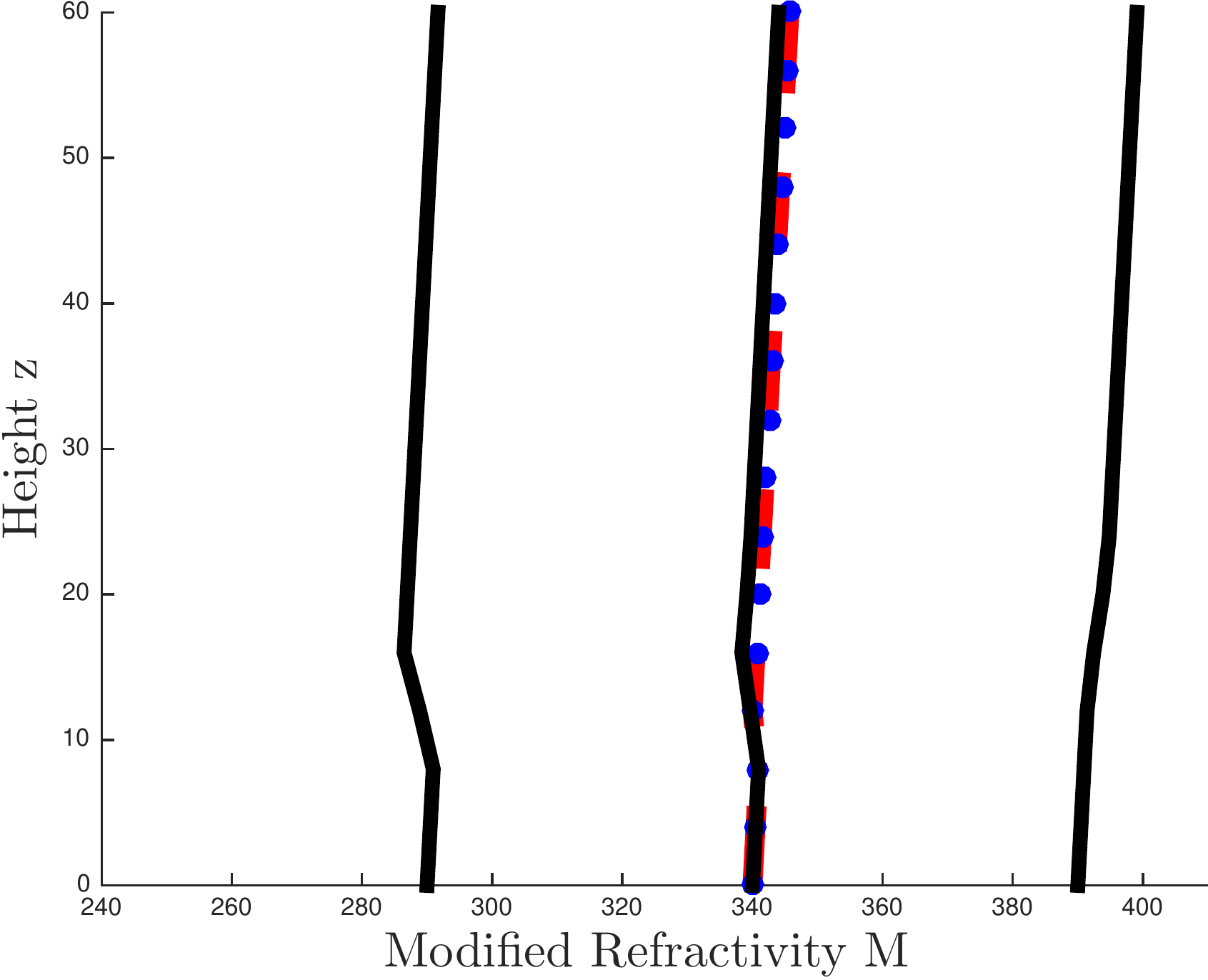}
\label{fig_second_case}}
\hfil
\subfloat[$ \text{RNL2} =  \left(0.10 ; 0.67\right) $ ]{\includegraphics[width=0.23\textwidth]{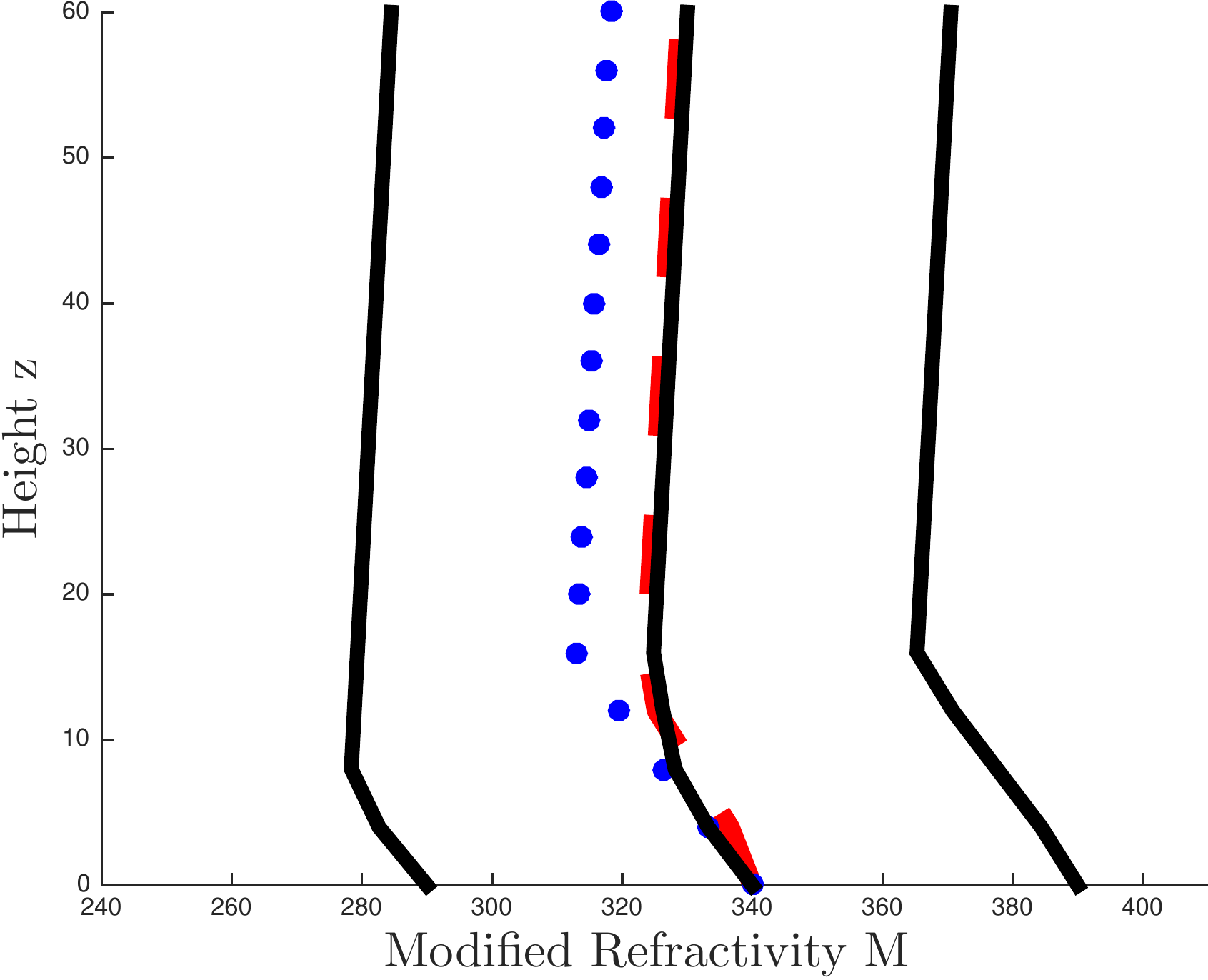}
\label{fig_second_case}}
\hfil
\subfloat[$ \text{RNL2} =  \left(0.23 ; 0.17\right) $ ]{\includegraphics[width=0.23\textwidth]{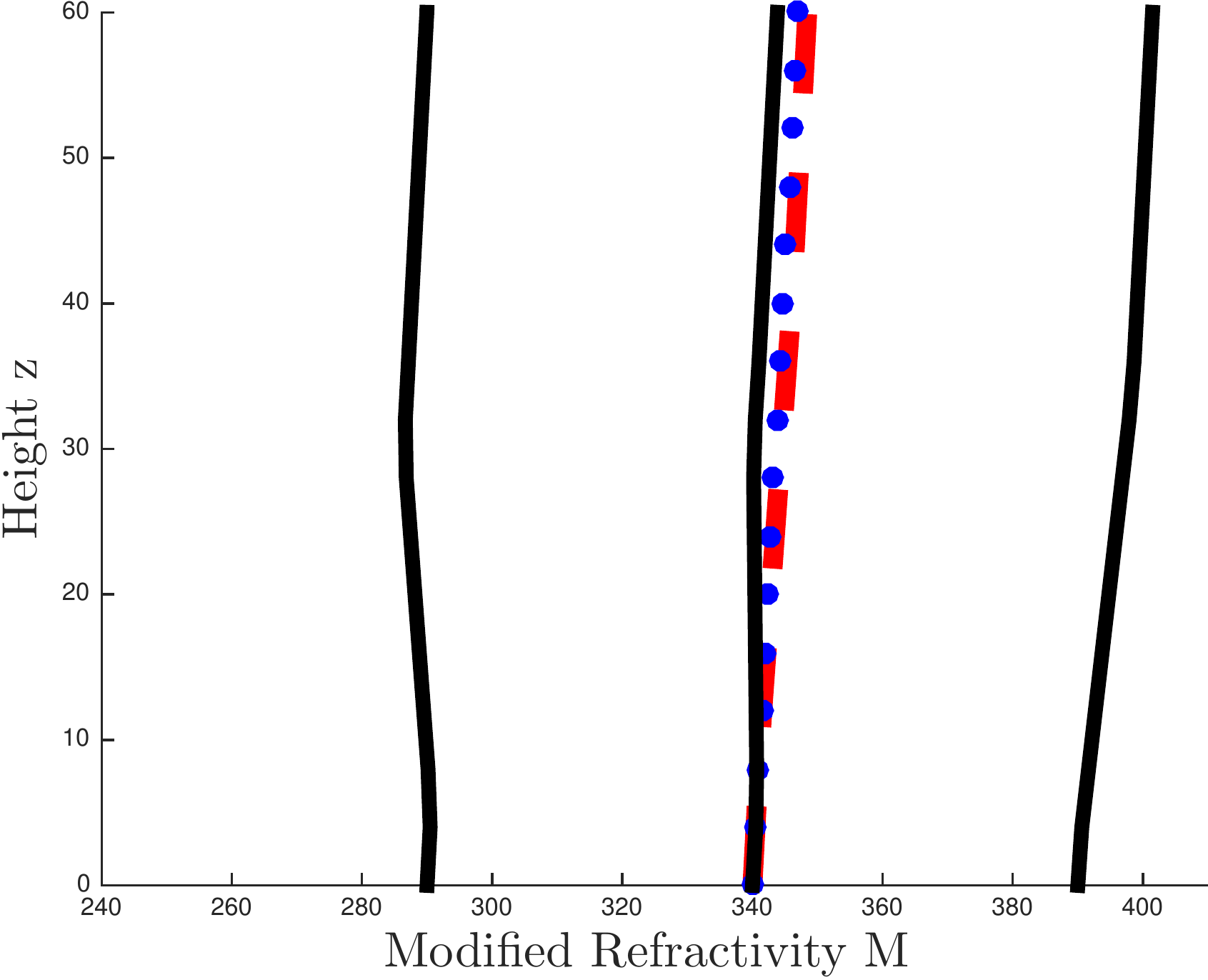}
\label{fig_second_case}}
\vfil

\subfloat[$ \text{RNL2} =  \left(0.97 ; 0.52\right) $ ]{\includegraphics[width=0.23\textwidth]{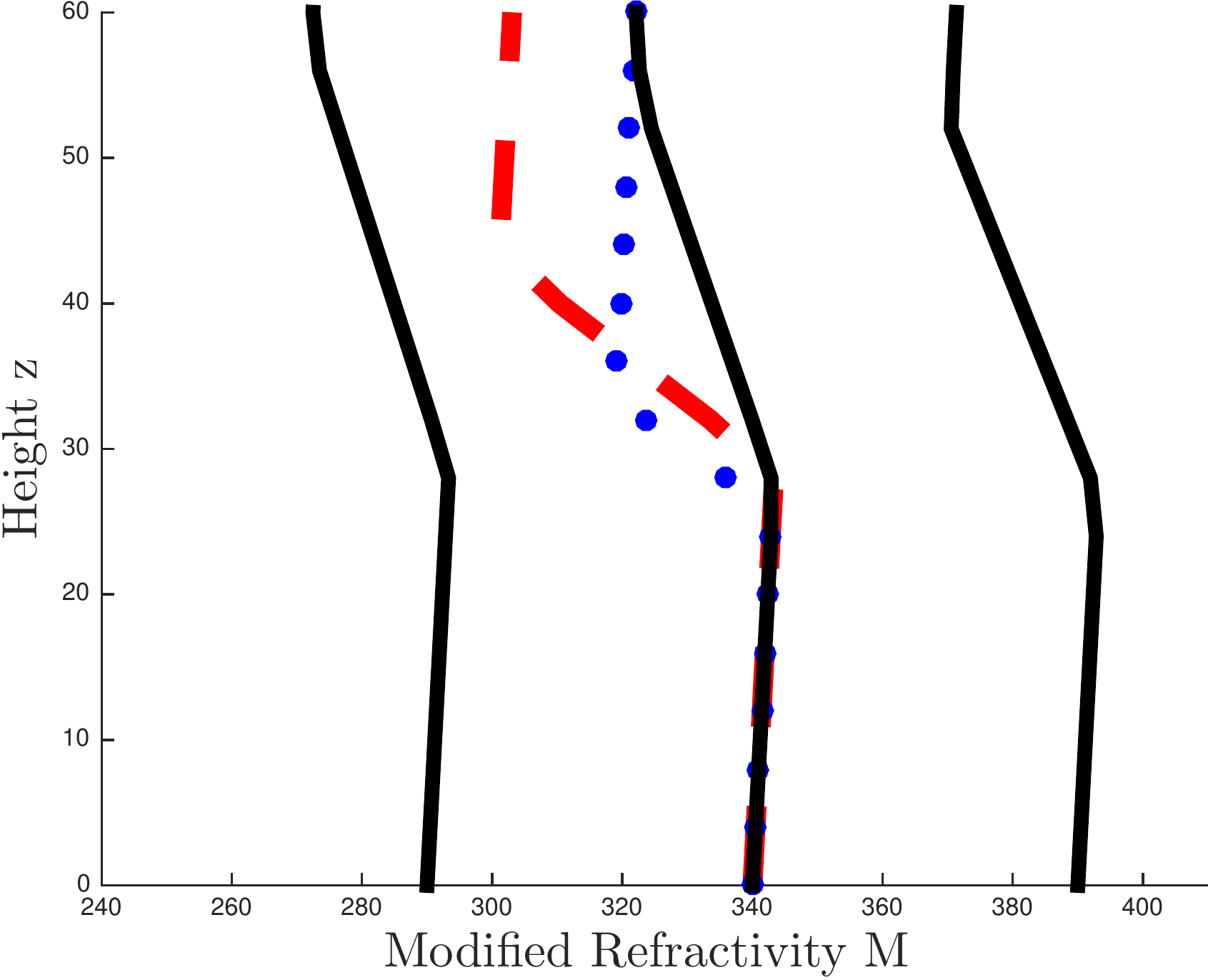}
\label{fig_first_case}}
\hfil
\subfloat[$ \text{RNL2} =  \left(0.18 ; 0.07 \right) $ ]{\includegraphics[width=0.23\textwidth]{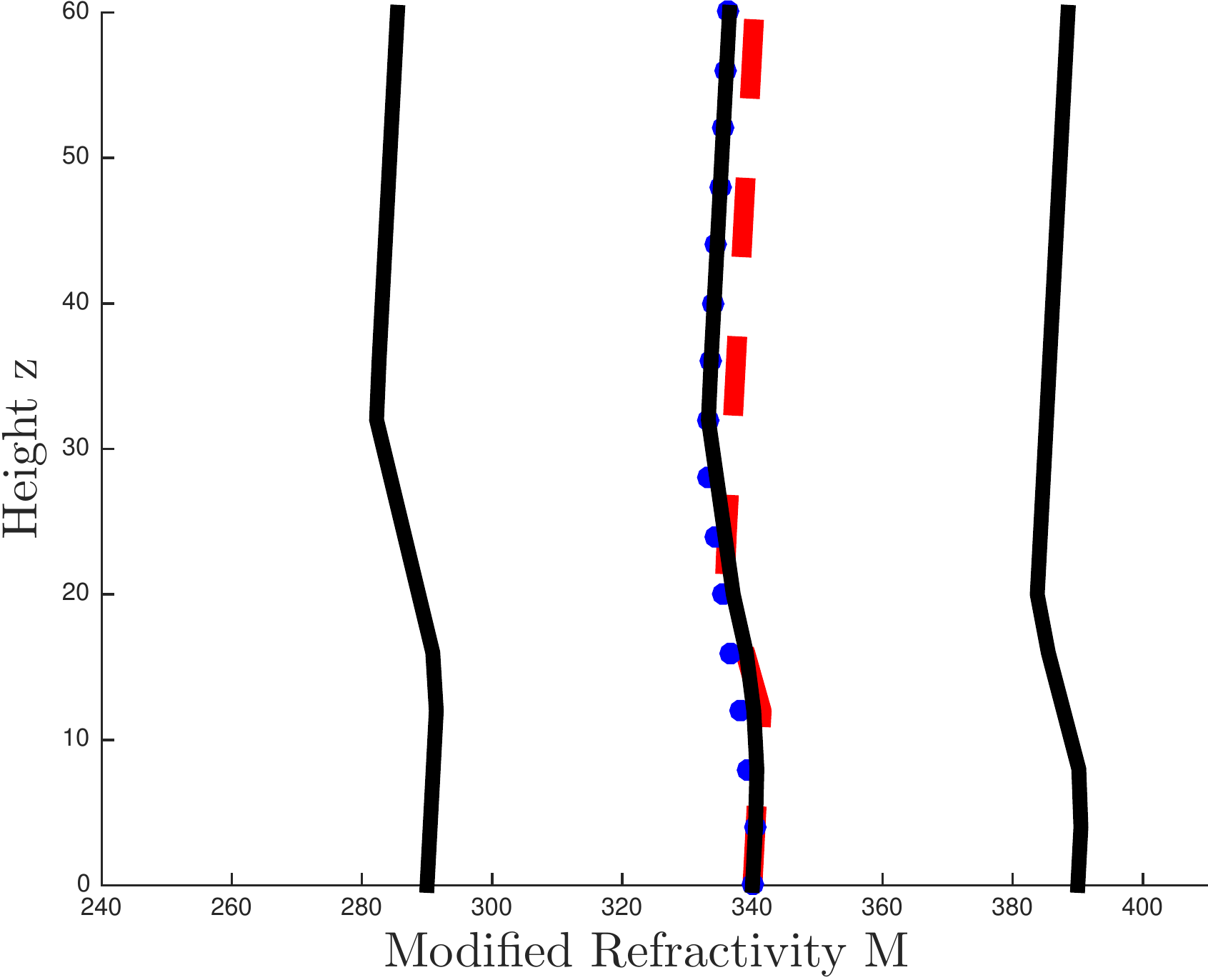}
\label{fig_second_case}}
\hfil
\subfloat[$ \text{RNL2} =  \left(0.86 ; 0.05 \right) $ ]{\includegraphics[width=0.23\textwidth]{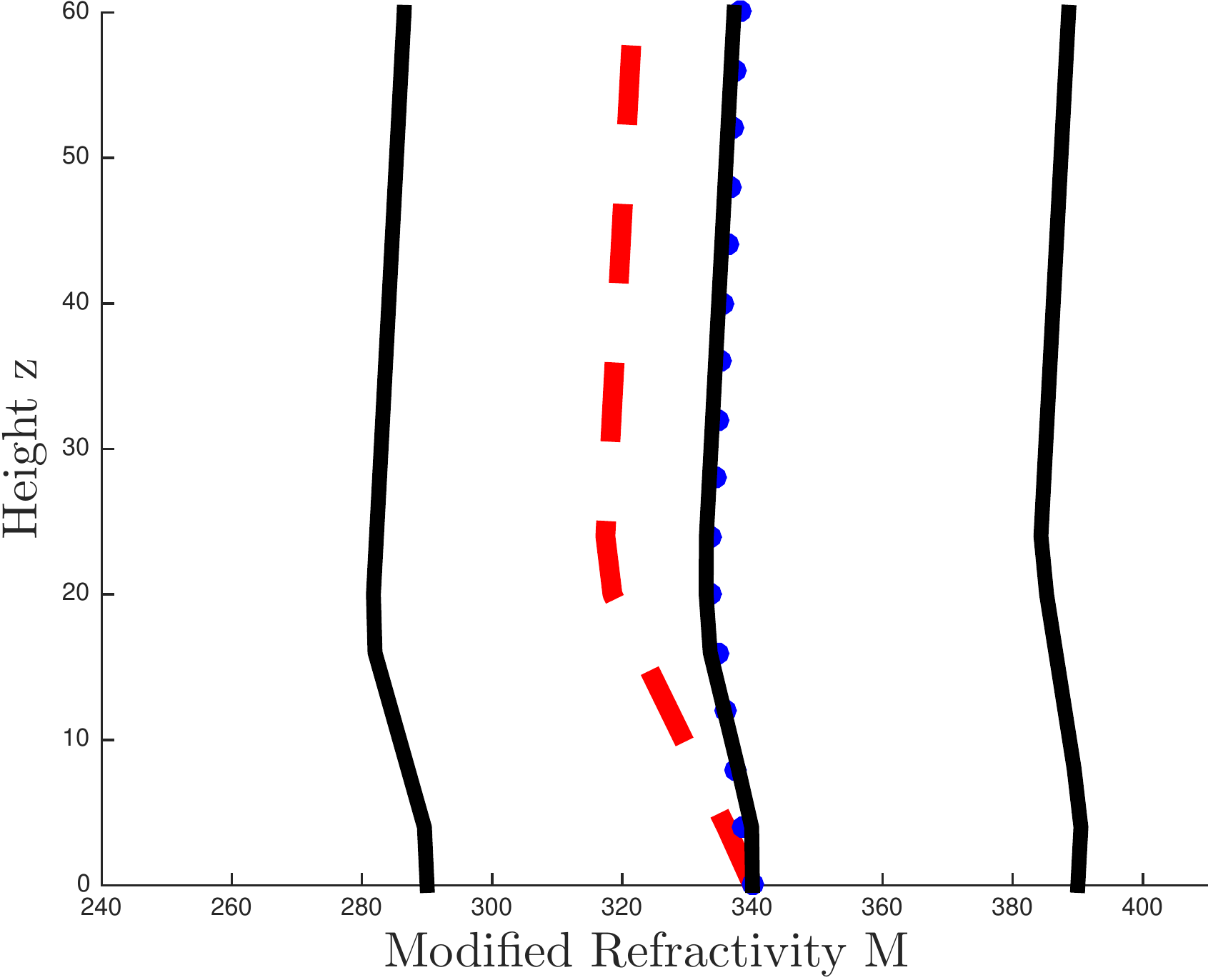}
\label{fig_second_case}}
\hfil
\subfloat[$ \text{RNL2} =  \left(0.55 ; 0.27\right) $ ]{\includegraphics[width=0.23\textwidth]{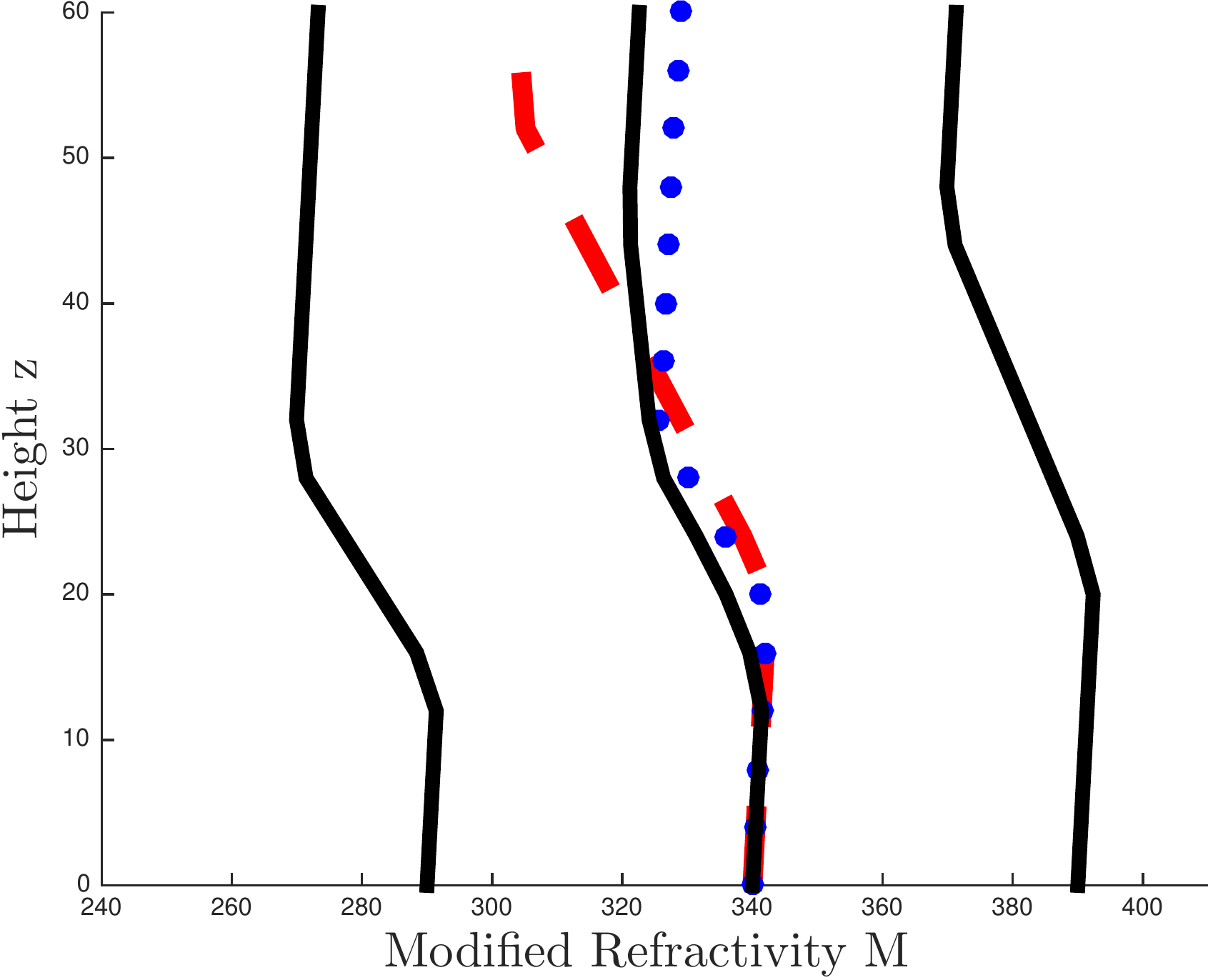}
\label{fig_second_case}}
\vfil

\subfloat[$ \text{RNL2} =  \left(0.55 ; 0.21\right) $ ]{\includegraphics[width=0.23\textwidth]{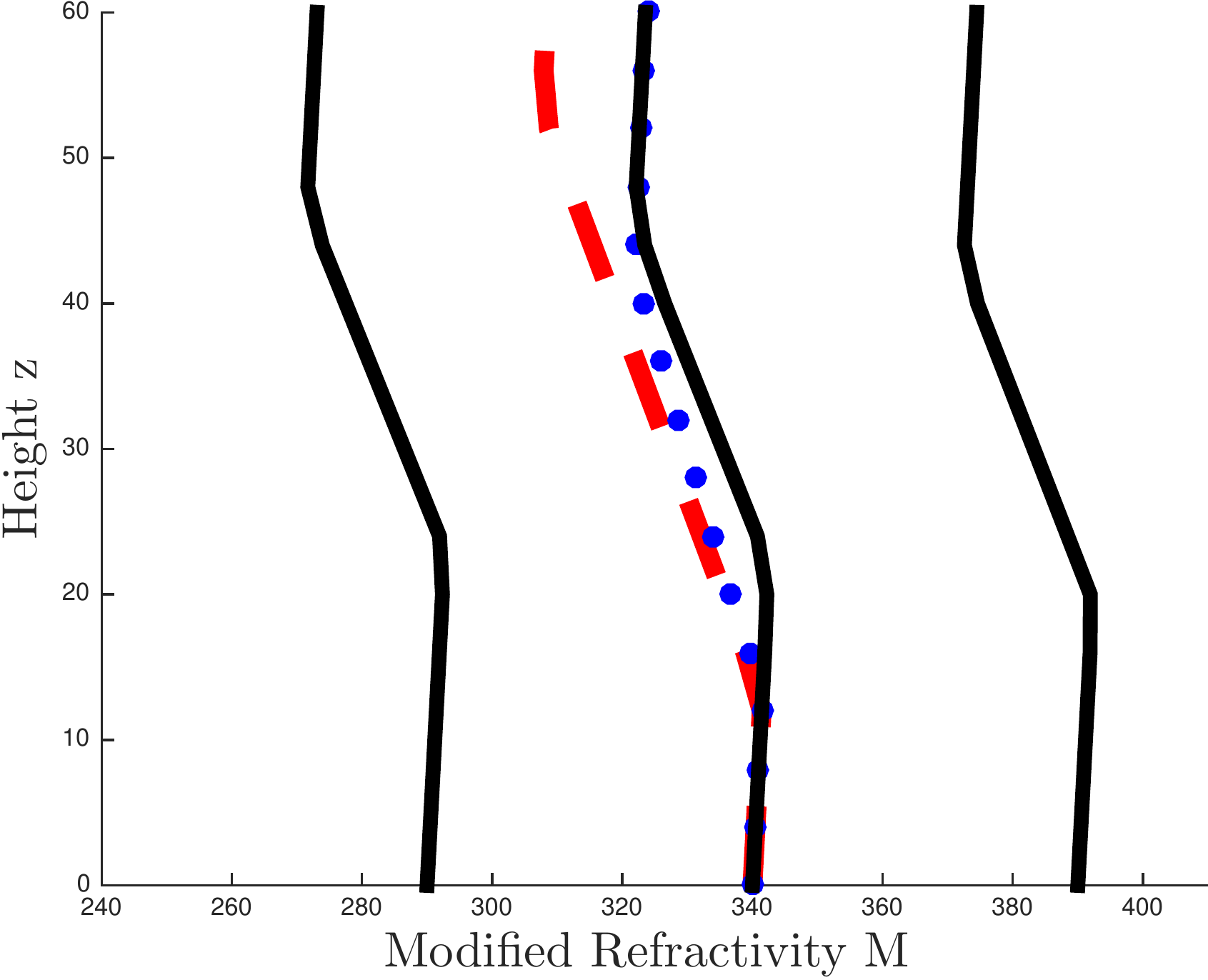}
\label{fig_first_case}}
\hfil
\subfloat[$ \text{RNL2} =  \left(0.80 ; 0.19\right) $ ]{\includegraphics[width=0.23\textwidth]{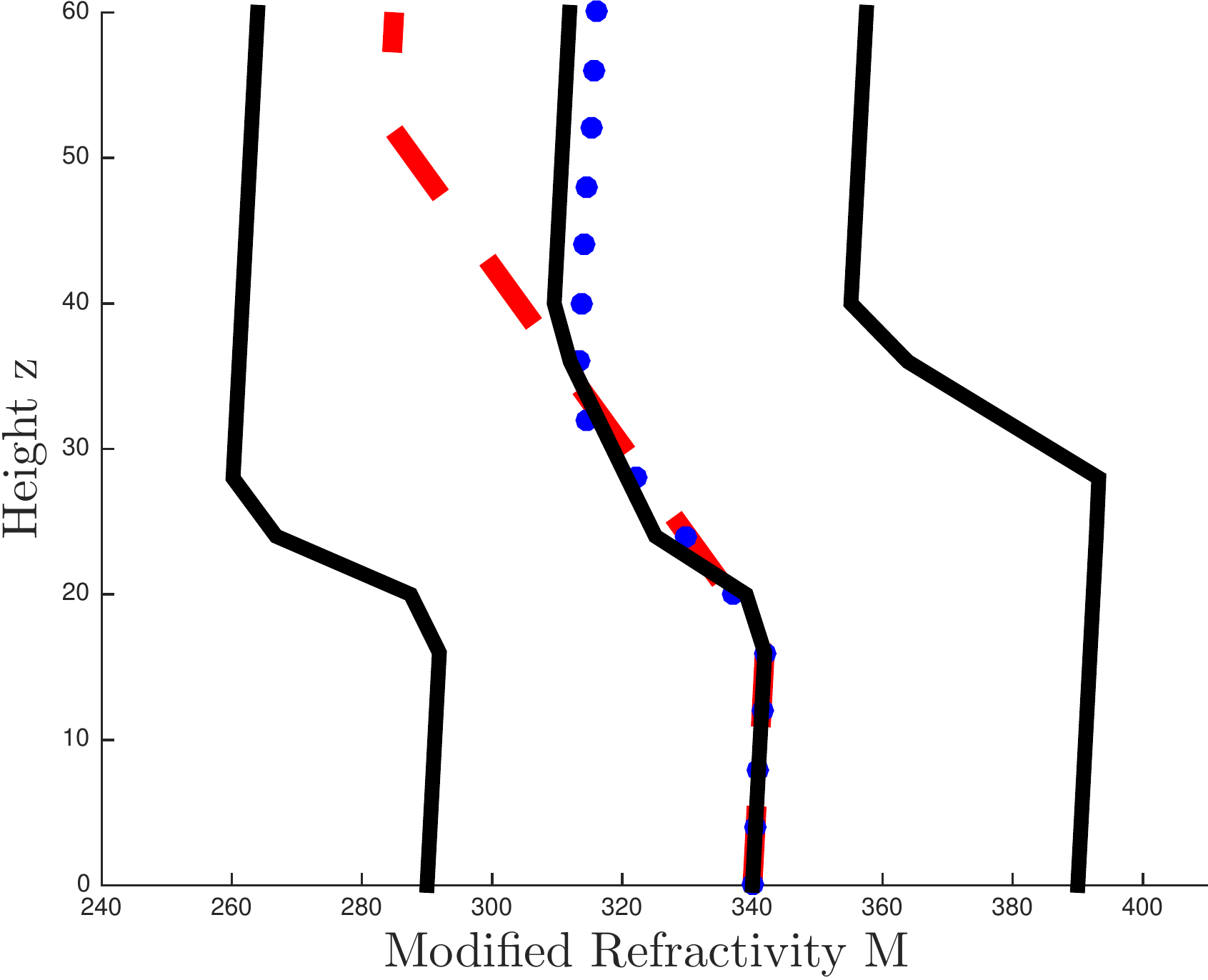}
\label{fig_second_case}}
\hfil
\subfloat[$ \text{RNL2} =  \left(0.25 ; 0.22\right) $ ]{\includegraphics[width=0.23\textwidth]{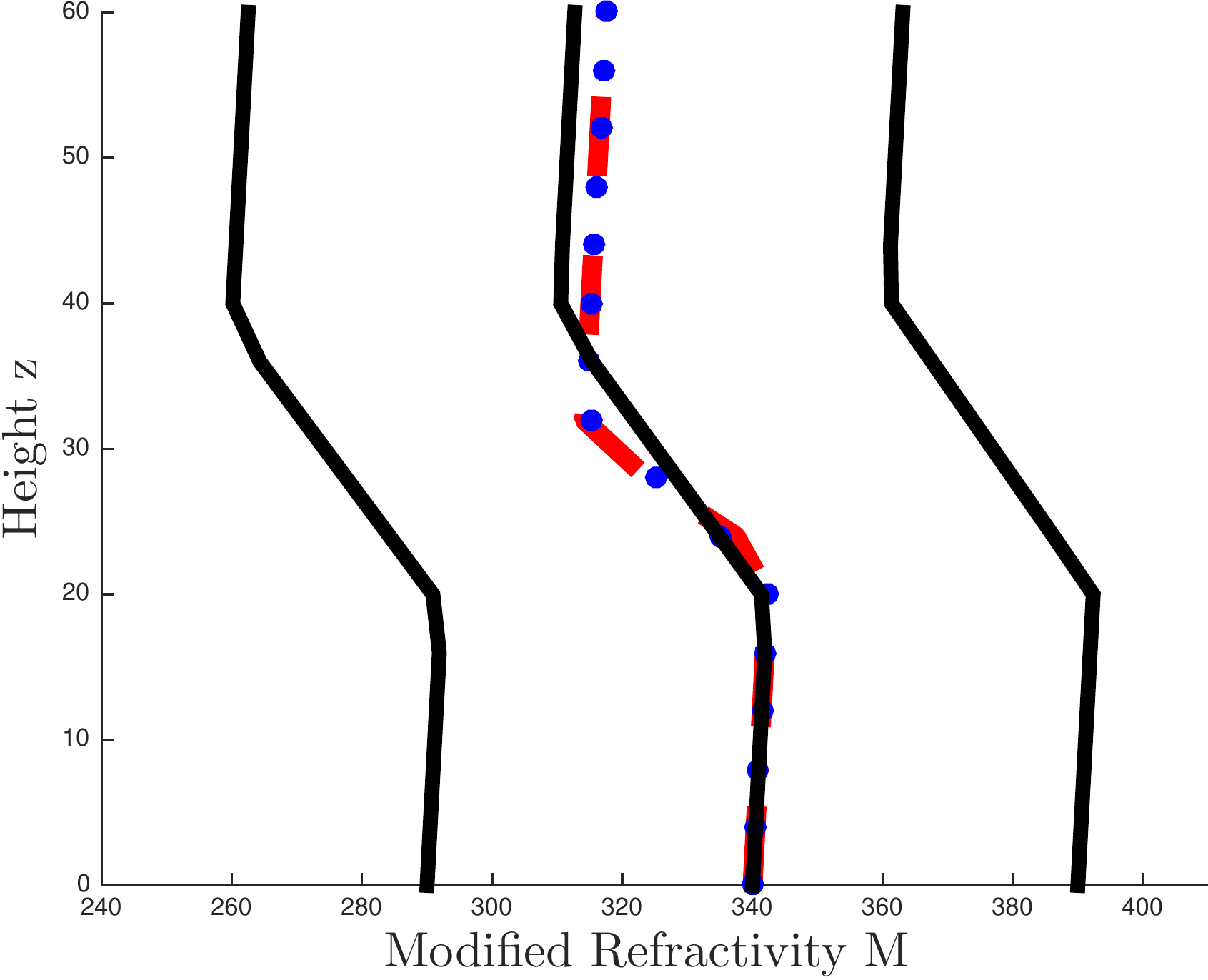}
\label{fig_second_case}}
\hfil
\subfloat[$ \text{RNL2} =  \left(0.29 ; 0.35\right) $ ]{\includegraphics[width=0.23\textwidth]{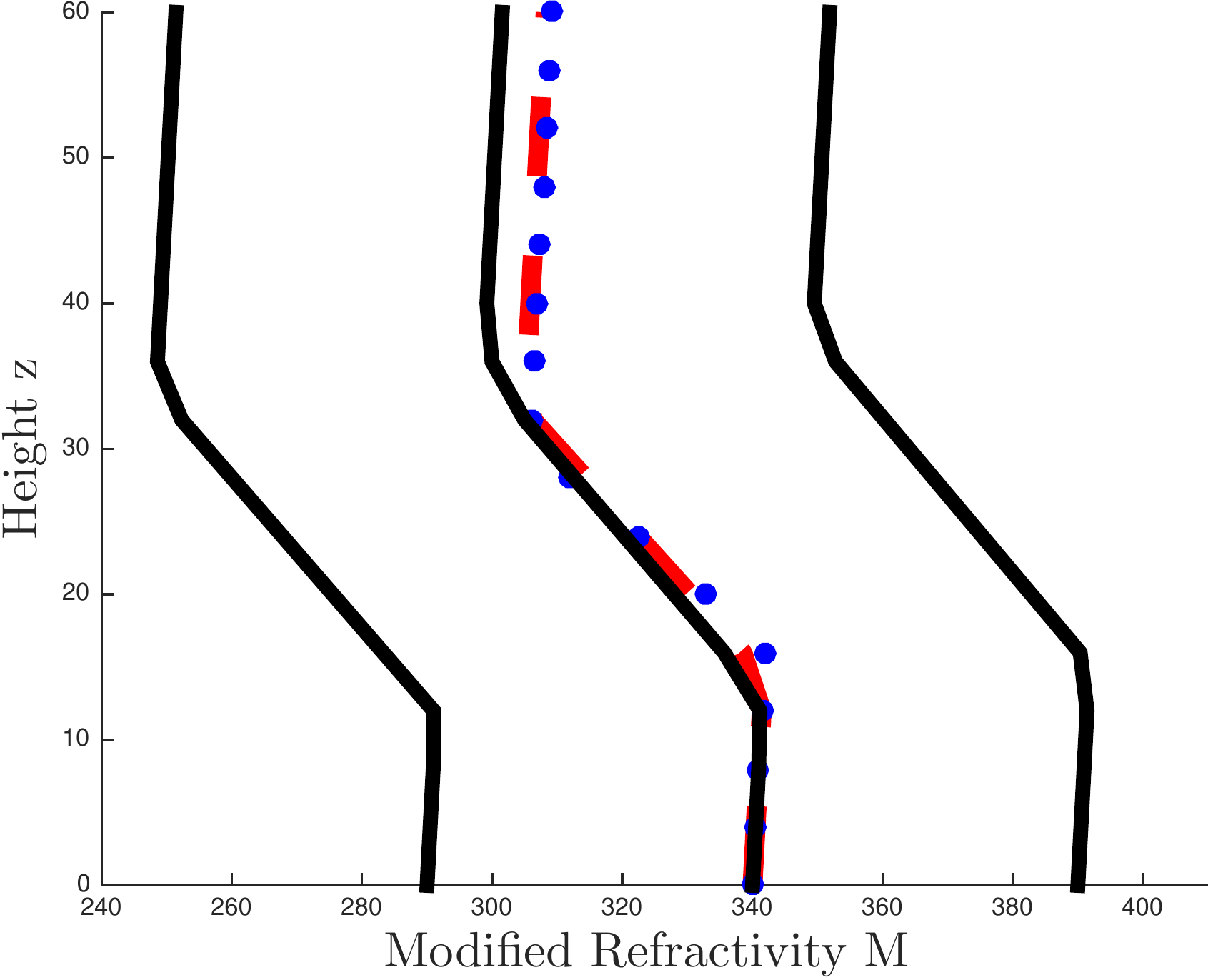}
\label{fig_second_case}}
\vfil

\caption{ Results for Alg.~1 and~2 on experiment 2.
 The solid black lines on the left are $n(0,z)$, the  solid black lines the right are $n(80,z)$, the solid black lines in the middle are $n_{\text{true}}(z)$,  the red dashed plot are the inverted profiles obtained using Alg. 1, and the dotted blue lines are the inverted profile obtained using Alg. 2. For each case, the RNL2 score is given below the plots. The first number is the RNL2 score of Alg. 1, and the second is the score of Alg. 2.}
\label{fig:sim2}
\end{figure*}

Aside from the way that the refractivity profiles are generated, the setup is the same as experiment 1. The horizontal variation in the refractivity profiles is achieved by interpolating two trilinear refractivity profiles. That is, we set $n(0,z)$ to some trilinear refractivity profile, and $n(80km,z)$ to some other trilinear refractivity profile. We then pointwise linearly interpolate the two refractivity profiles (that is, not the parametrizations) between $n(0,z)$, $n(80km,z)$ to produce refractivity profiles on $n(x,z)$ for all $x \in \left[0, 80 \right]$~km.  
As a result, the interpolated refractivity profiles are not trilinear. We define the ``true" refractivity profile as the average of the refractivity profiles between the transmitter and receiver:  $n_{\text{true}}(z) = 1/r \int_{x=0}^{x=r} n(x,z) dx$ where $r$ is the range of the receiver.
 We set the range of the receiver to $r = 50$ km. Note that this average will also not be trilinear. 
We generate twenty synthetic refractivity profiles by randomly sampling the parameters $\left( z_b^0, M_d^0, t_h^0 \right)$ in the following manner:
\begin{align*}
z_b^0 \sim U(0, 30)  && t_h^0 \sim U(0, 30) && M_d^0 \sim U(0, 50)
\end{align*}
 and the parametrization $\left( z_b^{80}, M_d^{80}, t_h^{80} \right)$  of $n(80,z)$ by sampling 
 \begin{align*}z_b^{80} \sim U(z_b^0 - 10, z_b^0+10)  \\ t_h^{80} \sim U(t_h^0 - 10, t_h^0+10)
 \\ M_d^{80} \sim U(M_d^0-15, M_d^0+15) \ .
 \end{align*}
This allows for a variation in the refractivity profile on the order of $20\%$ of the maximal value of each parameter along the propagation path.
We then contaminate the data with Gaussian white noise of standard deviation $0.3 \| F_{\text{obs}} \|_2$. The result of Alg 1~and~2 on these twenty synthetic test cases are shown in Fig.~\ref{fig:sim2}. 

\section{Discussion}

\begin{table}[tbhp]
{\footnotesize
  \caption{Error statistics of Alg.~1 and 2 on the datasets 1 and 2.\label{tab:2Doperations}}
\begin{center}
  \begin{tabular}{c c c c c}  \toprule
   Alg. &  Dataset & Mean error & Median error & Standard deviation \\ \midrule 
1  & 1 & 0.331 & 0.296 & 0.2808 \\
2 &1  & 0.231 & 0.1982 & 0.129\\
1  & 2& 0.441 & 0.370 & 0.2917 \\
2 & 2 & 0.274 & 0.200&  0.208\\
\bottomrule
  \end{tabular}
\end{center}
}
\end{table}

We observe that in all but two cases (Alg. 1 in Fig. \ref{fig:1k} and Alg. 2 in Fig. \ref{fig:2a}) the RNL2 scores are significantly below one, indicating that the algorithm performed far better than a random guess. Qualitatively, the height and structure of the inverted ducts and true ducts are in most cases similar.
Overall, Alg. 2 performs better than Alg. 1 and does not require an estimate of the noise level, thus Alg.~2 is preferable to Alg.~1.
The median RNL2 score indicates that the output of the Alg~2. is $5$ times more accurate than a random guess. The addition of horizontal variation does not seem to greatly affect the quality of the inference (especially for Alg. 2, where the median RNL2 score is 0.198 and 0.200 for the horizontally constant and varying case respectively).

The two cases where the algorithms produce a RNL2 score greater than one are attributed to a failure in the minimization of the objective function and could be resolved by using more starting points in the local optimization algorithm at the cost of a higher computational expense. However, we note that the ten starting points used in Alg.~2 seem sufficient in the vast majority of cases to allow the use of local optimization algorithms. Thus, we conclude that the issue of multimodality pointed out in~\ref{sec:analysis} is largely resolved. 

The timings shown in Table~\ref{tab:time}  were performed on a single Intel i7 core processor operating at 3.60GHz. We note that the timings should be interpreted as lower bounds as the algorithms were implemented in MATLAB, and were not fully optimized. \

The most important feature of this method is that it is able to overcome the highly multimodal behavior associated with the physics of EM wave propagation. This allows the use a local optimization method instead of a global optimization method which is typically used in the literature (such as genetic algorithms in, for example,~\cite{vasilis,gerstoft2003inversion,douvenot2008,penton2018rough, zhao2012evaporation}).

As this method is able to cheaply find an estimate of the refractivity profile using local optimization, it could also be used to warm-start a different method which would typically require a global optimization method.
That is, in a first step, one could use this method to find a good first guess. Then, in a second step, a local optimization method started at that initial guess could be used to minimize a multimodal, but perhaps more accurate objective function such as the ones in \cite{gerstoft2003inversion,douvenot2008,penton2018rough, zhao2012evaporation}.
This should allow for more accurate prediction while still benefiting from the lower computational cost associated with local optimization algorithms.

\begin{table}
\caption{Average running time on experiments 1 and 2} \label{tab:time} 
\begin{center}

\begin{tabular}{ c c c  }
\toprule
Algorithm  & Experiment  & Avg. run time \\
\midrule
1 &  1 & 3 min 15 sec \\
1 &  2 & 2 min 59 sec  \\
2 &  1 & 5 min 12 sec  \\
2 &  2 & 5 min 19 sec  \\
\bottomrule
\end{tabular}
\end{center}
\end{table}

\section{Conclusion}

We presented a new method for characterizing the refractivity profile in the MABL which relies on the low-rank structure of the field within parts of the domain, inherited by the governing Helmholtz equation. This low-rank structure allows us to formulate the inverse problem in terms of a minimization of a new objective function. The objective function performs a projection operation of the data onto the subspace spanned by the eigenvectors that form the low-rank approximation of an electromagnetic field induced by a particular refractivity profile. Performing an optimization on this well behaved objective function allows us to accurately solve the inverse problem in around five minutes, allowing for real-time characterization of the MABL.

We conducted two numerical experiments to demonstrate the efficacy of the method. The first experiment was conducted on noisy simulated data computed with horizontally constant refractivity profiles, and the second experiment was performed on noisy simulated data computed with horizontally varying refractivity profiles. We observed that in both setups, the method is able to infer the refractivity profile using local optimization algorithms with few starting points thanks to the small number of local minima of the objective function.

\section*{Acknowledgment}

The authors wish to thank Dr. Frank Ryan, of Applied Technology, Inc., for his advice and many helpful suggestions during this research work. The authors gratefully acknowledge ONR Division 331 \& Dr. Steve Russell for the financial support of this work through grant N00014-16-1-2077.

\ifCLASSOPTIONcaptionsoff
  \newpage
\fi



\bibliographystyle{IEEEtran}
\bibliography{mypaperbib.bib}
%

%




\end{document}